\documentclass[journal, 11 pt, onecolumn]{IEEEtran}
\makeatletter
\makeatletter
\let\IEEEproof\proof
\let\IEEEendproof\endproof
\let\proof\@undefined
\let\endproof\@undefined
\makeatother

\usepackage{amsthm,amssymb,euscript,amscd,amsbsy,psfrag,cite}
\usepackage{graphicx}
\usepackage{graphics}
\usepackage{epsf,subfigure}

\newtheorem{coro}{Corollary}
\newtheorem{lemma}{Lemma}

\newtheorem{prop}{Proposition}
\newtheorem{definition}{Definition}
\newtheorem{remark}{Remark \textrm}

\newcommand{\disp}{\displaystyle}

\newcommand{\BI}{\textnormal{BI}}
\newcommand{\E}{\mathbf{E}}
\newcommand{\trace}{\textnormal{trace}}

\newcommand{\tr}{\textnormal{tr}}
\newcommand{\diag}{\textnormal{diag}}

\newcommand{\MMSE}{\textnormal{MMSE}}
\newcommand{\MSE}{\textnormal{MSE}}
\newcommand{\CMMSE}{\textnormal{CMMSE}}
\newcommand{\PMMSE}{\textnormal{PMMSE}}
\newcommand{\CRB}{\textnormal{CRB}}

\newcommand{\eqa}{\stackrel{\textnormal{(a)}}{=}}
\newcommand{\eqb}{\stackrel{\textnormal{(b)}}{=}}

\newcommand{\iid}{\stackrel{\textnormal{\small{i.i.d.}}}{\large{\sim}}}

\newcommand{\bfA}{\pmb{A}}
\newcommand{\bfB}{\pmb{B}}
\newcommand{\bfC}{\pmb{C}}
\newcommand{\bfD}{\pmb{D}}
\newcommand{\bfF}{\pmb{F}}
\newcommand{\bfJ}{\pmb{J}}
\newcommand{\bfK}{\pmb{K}}
\newcommand{\bfS}{\pmb{S}}
\newcommand{\bfX}{\pmb{X}}

\newcommand{\bfphi}{\pmb{\phi}}

\newcommand{\bfGa}{\pmb{\Gamma}}

\newcommand{\bfI}{\pmb{I}}
\newcommand{\bfM}{\pmb{M}}

\newcommand{\bfP}{\pmb{P}}
\newcommand{\bfzero}{\pmb{0}}

\newcommand{\uy}{\underline{y}}
\newcommand{\cy}{\check{y}}
\newcommand{\ucy}{\underline{\check{y}}}
\newcommand{\uN}{\underline{N}}
\newcommand{\uu}{\underline{u}}

\newcommand{\ue}{\underline{e}}
\newcommand{\ur}{\underline{r}}
\newcommand{\uv}{\underline{v}}
\newcommand{\uZ}{\underline{Z}}
\newcommand{\uB}{\underline{B}}

\newcommand{\uG}{\underline{G}}
\newcommand{\uH}{\underline{H}}
\newcommand{\uC}{\underline{C}}
\newcommand{\uL}{\underline{L}}
\newcommand{\uW}{\underline{W}}

\newcommand{\uzero}{\underline{0}}
\newcommand{\ubary}{\underline{\bar{y}}}
\newcommand{\ubars}{\underline{\bar{s}}}

\newcommand{\us}{\underline{s}}
\newcommand{\ux}{\underline{x}}

\newcommand{\bbC}{\underline{\mathbb{C}}}
\newcommand{\bbA}{\pmb{\mathbb{A}}}
\newcommand{\bbX}{\underline{\mathbb{X}}}
\newcommand{\bbR}{\mathbb{R}}
\newcommand{\bbD}{\underline{\mathbb{D}}}

\newcommand{\bbS}{\mathbb{S}}

\newcommand{\tbbC}{\tilde{\underline{\mathbb{C}}}}
\newcommand{\tbbD}{\tilde{\underline{\mathbb{D}}}}
\newcommand{\tbbA}{\tilde{\pmb{\mathbb{A}}}}
\newcommand{\tSi}{\tilde{\pmb{\Sigma}}}

\newcommand{\cbbC}{\check{\underline{\mathbb{C}}}}
\newcommand{\cbbA}{\check{\pmb{\mathbb{A}}}}
\newcommand{\cSi}{\check{\pmb{\Sigma}}}

\newcommand{\tildex}{\underline{\tilde{x}}}

\newcommand{\utildey}{\underline{\tilde{y}}}
\newcommand{\tildes}{\underline{\tilde{s}}}
\newcommand{\hatx}{\underline{\hat{x}}}
\newcommand{\hatW}{\underline{\hat{W}}}

\newcommand{\hats}{\underline{\hat{s}}}
\newcommand{\hatG}{\pmb{\widehat{\mathcal{G}}}}
\newcommand{\hatr}{\hat{r}}

\newcommand{\Si}{\pmb{\Sigma}}

\newcommand{\calF}{{\mathcal{F}}}
\newcommand{\calG}{{\pmb{\mathcal{G}}}}

\newcommand{\calM}{{{\mathcal{M}}}}
\newcommand{\calN}{{\mathcal{N}}}
\newcommand{\calP}{{\mathcal{P}}}
\newcommand{\calR}{{\mathcal{R}}}

\newcommand{\calB}{{\pmb{\mathcal{B}}}}
\newcommand{\calI}{{\pmb{\mathcal{I}}}}
\newcommand{\calT}{{\mathcal{T}}}

\newcommand{\calZ}{{\mathcal{Z}}}
\newcommand{\bcalZ}{{\pmb{\calZ}}}
\newcommand{\ep}{{\epsilon}}
\newcommand{\la}{{\lambda}}

\newcommand{\ee}{\end{equation}}
\newcommand{\be}{\begin{equation}}
\newcommand{\ba}{\begin{array}}
\newcommand{\ea}{\end{array}}

\renewcommand{\QED}{\QEDopen}

\let \proof \IEEEproof
\let \endproof \IEEEendproof

\begin{document}

\title{ \large \bf
Convergence of Fundamental Limitations in Feedback Communication, Estimation, and Feedback Control over Gaussian Channels}

\author{ \normalsize {Jialing Liu and Nicola Elia
\thanks{This research was supported by NSF under Grant ECS-0093950.  The material in
this paper was presented in part at the 43rd Annual Allerton 
Conference on Communication, Control, and Computing, Monticello, IL, 
September 2005, and the 45th {IEEE} Conference on Decision and Control, San Diego, CA, December 2006.}
\thanks{J. Liu was with the Department of Electrical
and Computer Engineering, Iowa State University, Ames, IA 50011
USA.  He is now with Motorola Inc., Libertyville, IL 60048 USA (e-mail:jialingliu@motorola.com). }
\thanks{N. Elia is with the Department of Electrical
and Computer Engineering, Iowa State University, Ames, IA 50011
USA (e-mail: nelia@iastate.edu). }
}}

\maketitle

\small \vspace{-18pt}
\begin{abstract}

In this paper, we establish the connections of the fundamental limitations in feedback communication, estimation, and feedback control over Gaussian channels, from a unifying perspective for information, estimation, and control.  The optimal feedback communication system over a Gaussian necessarily employs the Kalman filter (KF) algorithm, and hence can be transformed into an estimation system and a feedback control system over the same channel.  This follows that the information 
rate of the communication system is alternatively given 
by the decay rate of the Cramer-Rao bound (CRB) of the  
estimation system and by the Bode integral (BI) of the control system.  Furthermore, the optimal tradeoff between the channel input power and information rate in feedback communication is alternatively characterized by the optimal tradeoff between the (causal) one-step prediction mean-square error (MSE) and (anti-causal) smoothing MSE (of an appropriate form) in estimation, and by the optimal tradeoff between the regulated output variance with causal feedback and the disturbance rejection measure (BI or degree of anti-causality) in feedback control.  All these optimal tradeoffs have an interpretation as the tradeoff between causality and anti-causality.  Utilizing and motivated by these relations, we provide several new results regarding the feedback codes and information theoretic characterization of KF.  Finally, the extension of the finite-horizon results to infinite horizon is briefly discussed under specific dimension assumptions (the asymptotic feedback capacity problem is left open in this paper).

\end{abstract}



\section{Introduction} \label{sec:intro}

Communication systems in which the transmitters have access to 
noiseless feedback of channel outputs have been widely studied.  The \emph{fundamental limitations} in these systems, i.e. the feedback capacities, and the capacity-achieving codes, have been a central focus in the information theoretic literature.  As 
one of the most important case, the single-input single-output 
Gaussian channels with noiseless feedback have attracted 
considerable attention; see 
\cite{kailath1,kailath2,omura68,butman-1969,butman-1976,cover-pombra-1989, 
ozarow90:2,yanagi92,ordentlich, 
feder:isit04,tati:capI,kavcic_it07,sahai:phd,elia_c5,kim04} and 
references therein for the capacity characterization and coding scheme 
design for these channels. 
There exist different approaches in addressing the fundamental limitations for such channels, categorized roughly (by no means strict as the approaches are intrinsically related) as follows: 1) \emph{Estimation theory related approaches}, which utilizes concepts such as maximum likelihood (ML) or minimum mean-square error (MMSE) estimates in constructing the coding schemes (cf. e.g. \cite{kailath1,kailath2,butman-1969,butman-1976,ozarow90:2,ihara:book}); 2) \emph{Information theoretic approaches}, most notably the Cover-Pombra formulation based on the asymptotic equipartition (AEP) property and the mutual information between the message and the channel outputs (cf. e.g. \cite{cover-pombra-1989, 
ordentlich,kim04}), and the directed information formulation based on the input-output characterization of the channels \footnote{The directed information in feedback communication systems may be viewed 
as the causal counterpart of mutual information used in  
communication systems without feedback, the supremum of which (under applicable constraints, if any) 
is the capacity. See also Appendix \ref{appsub:directI}.} (cf. e.g. \cite{massey:dmi,tati:phd,tati:capI}); and 3) \emph{Control theory related approaches}, which regards the feedback communication problems as optimal control problems (cf. e.g. \cite{omura68,tati:capI,sahai:phd,kavcic_it07,elia_c5}).

In particular, Schalkwijk and Kailath 
\cite{kailath1,kailath2} proposed the Schalkwijk-Kailath (SK) codes for 
additive white Gaussian noise (AWGN) channels, achieving the 
asymptotic feedback capacity (i.e. the infinite-horizon feedback 
capacity, denoted $C_\infty$, which is the highest information rate over the time spans between 0 
and infinity, subject to an average power constraint) and greatly reduce the coding 
complexity and coding delay.  The SK codes were suggested by the Robbins-Monro stochastic approximation and recursive ML algorithm which have an \emph{estimation} theoretic flavor.  Along the line of 
\cite{kailath1,kailath2}, Butman, Ozarow, and numerous other 
researchers have proposed extensions of the SK codes to Gaussian 
feedback channels with memory and obtained tight capacity bounds, 
see e.g. \cite{butman-1969,butman-1976,ozarow90:2}.

Cover and Pombra 
\cite{cover-pombra-1989} introduced a general coding 
structure (called the Cover-Pombra structure, or the CP structure 
for short) to achieve the finite-horizon feedback capacity (denoted 
$C_T$, the highest information rate over the time span between 0 and $T$ subject to an average power constraint) 
for Gaussian channels with memory, based on classical \emph{information} theoretic concepts.  Their development builds on the mutual information between the message and the channel outputs (hence circumventing the causality issue pointed out by Massey \cite{massey:dmi} without appealing to directed information) and AEP for arbitrary Gaussian processes.  The CP structure was initially 
regarded to have prohibitive computation complexity if the coding 
length $(T+1)$ is large (see, however, Section \ref{appsub:cpcolor} for more detailed discussion), and efforts have been made to reduce the complexity and 
to refine the CP structure.  By exploiting the special properties of 
a moving-average Gaussian channel with feedback, Ordentlich 
\cite{ordentlich} discovered the finite rank property of the innovations 
in the CP structure, which reduces the computation complexity. 
Shahar-Doron and Feder \cite{feder:isit04} reformulated the CP 
structure along this direction, and obtained an SK-based coding 
scheme to achieve $C_T$ with reduced computation complexity.   
Furthermore, utilizing the CP structure as a starting point, Kim \cite{kim04} proved that a closed-form expression \footnote{This expression was initially identified by Elia \cite{elia_c5} and Yang 
\emph{et al} \cite{kavcic_it07} and has been conjectured to be 
$C_\infty$; however, a rigorous proof was not available until Kim 
\cite{kim04}. } of the asymptotic capacity $C_\infty$ for an first-order moving-average Gaussian channel with feedback, and obtained an SK-based coding 
scheme to achieve $C_\infty$. This is the first Gaussian channel 
with memory (except for the degenerated case of AWGN channel) \footnote{By Gaussian channels with memory,
researchers normally refer to frequency-selective Gaussian channels, including Gaussian channels with inter-symbol interference (ISI) and channels with colored Gaussian noise, a convention also adopted in this paper (although some other Gaussian channels may also have memory). The Gaussian channels with memory may sometimes be referred to as general Gaussian
channels (in contrast to the specific AWGN channels), or even simply as Gaussian channels.} that 
has an established asymptotic feedback capacity and available 
capacity-achieving codes, to the best of our knowledge.  On the 
other hand, Vandenberghe \emph{et al} \cite{maxdet} showed that the 
computation of $C_T$ based on the CP structure can be reformulated 
as a convex optimization problem.

Tatikonda and Mitter \cite{tati:phd,tati:capI} provided an extensive study of 
feedback communication systems and their capacities. They extended the notion of \emph{directed information }
proposed in~\cite{massey:dmi} and proved that its supremum equals the operational capacity; reformulated the problem of computing $C_T$ as a 
\emph{stochastic control optimization problem}; and proposed a dynamic 
programming based solution and characterized the sufficient statistics required for encoding and decoding. This idea was further explored in 
\cite{kavcic_it07} by Yang \emph{et al}, which uncovered the Markov 
property of the optimal input distributions for Gaussian channels 
with memory, established a class of refined, finite-dimensional optimal input distributions, and eventually reduced the finite-horizon stochastic 
control optimization problem to a manageable size (with complexity $O(T)$). Moreover, under a 
\emph{stationarity conjecture} that $C_\infty$ can be achieved by a stationary input process, $C_\infty$ is 
given by the solution of a finite-dimensional optimization problem. 
This is the first computationally efficient \footnote{Here we do not 
mean that their optimization problem is convex.  The computation 
complexity associated with the optimization problem is determined mainly on the channel order which does not grow to infinity as the time horizon increases to infinity.} method to calculate the feedback capacity in infinite horizon for general Gaussian channels.  A Kalman filter (KF) was used in \cite{kavcic_it07} to generate the sufficient statistics of the output feedback.

Omura \cite{omura68} identified a stochastic \emph{optimal control problem} for feedback communication systems.  Omura showed that the solution to the control problem is optimal for AWGN channels in the sense of achieving the capacity; however, how this approach might be extended to achieve the capacities of more general channels remained to be seen \footnote{Rather than showing the feedback capacity problem can be posed as a control problem as Tatikonda and Mitter did, Omura formulated the control problem to minimize MMSE. Whether this may yield information theoretic optimality was not explored by Omura \cite{omura68} except for the AWGN case.  Later works such as \cite{tati:capI,sahai:main,elia_c5,kavcic_it07} and the present paper have established results on the intrinsic relationship between communication and control within a more general framework.}.  Sahai and Mitter \cite{sahai:phd,sahai:main} 
investigated the problem of tracking unstable sources over a channel 
and introduced the notion of \emph{anytime capacity} to capture the 
fundamental limitations in that problem, which again reveals connections 
between communication and control and brings various new insights to 
feedback communication problems. Furthermore, Elia \cite{elia_c5} 
established the \emph{equivalence} between reliable communication 
and stabilization over Gaussian channels with memory, 
showed that the achievable transmission rate is given by the Bode 
sensitivity integral of the associated control system, and presented 
an optimization problem based on robust control to compute lower 
bounds of $C_{\infty}$. These lower bounds can be achieved by generalized SK codes that have an 
interpretation of tracking unstable sources over Gaussian channels.  For a time-varying 
fading AWGN channel whose fade is modelled as a Markov 
process with channel output feedback and channel 
state information (CSI), a \emph{control-oriented} coding scheme multiplexing across multiple subsystems according to CSI was 
constructed by Liu \emph{et al} \cite{liu:markov} to achieve the 
ergodic capacity, and it is shown to be an extension of the SK codes 
to time-varying channels with appropriate channel state information.  
For a recent survey of various topics on feedback communication, see e.g. 
\cite{kim04,liu:phd} and references therein.  

As we have seen, different approaches have been shown useful in addressing the Gaussian feedback communication problem.  This paper attempts to present a converging point:  We study the Gaussian channels with feedback from a perspective that unifies information, estimation, and control, which encompasses many of the existing approaches scattered in the literature.  We demonstrate that the feedback communication 
problem over a Gaussian channel can be reformulated as an optimal estimation 
problem or an optimal control problem.  In fact, we show that the existing coding structures either necessarily contain Kalman filters or are reformulations of Kalman filters: The CP structure necessitates a KF in order to be optimal, the SK code can be easily obtained or extended by transforming a KF, and the control-oriented schemes can be derived from a KF by the duality between control and estimation \cite{kailath:book}. As a result, \emph{the fundamental limitations in feedback 
communication, estimation, and feedback control coincide}. 

Particularly, the achievable rate of the feedback communication 
system is alternatively given by the decay rate of the Cramer-Rao 
bound (CRB) for the associated estimation system as well as the Bode 
integral (BI) of the associated control system. 
In addition, the fundamental limitations in terms of the \emph{optimal tradeoffs in feedback communication, estimation, and feedback control coincide, all of which may be interpreted as the tradeoff between causality and anti-causality}. In feedback communication, this fundamental limitation is the optimal tradeoff between the input power and information rate.   Alternatively in the associated estimation system, it can be  characterized by the optimal tradeoff between the (causal) one-step prediction and (anti-causal) smoothing, or in the associated control system by the optimal tradeoff between the variance of a regulated output (generated using causal feedback) and the BI (or degree of anti-causality or instability).  That is, the optimal pairs $(P,R)$, $(\PMMSE_T, (\log \det \MMSE_T^{-1})/(2T+2))$, and  $(P_u,\log DI)$ correspond to each other, where $P$ is the average channel input power and $R$ is the average information rate in the communication system; $\PMMSE_T$ is the time average of the \emph{one-step prediction MMSE} of the to-be-estimated process in the estimation system and $\MMSE_T$ is the \emph{anti-causal smoothing MMSE} of the initial state of the process; $P_u$ is the variance of the regulated output $u$ (i.e. control performance measure) in the control system and $DI$ is the \emph{degree of instability} of the open-loop system defined as the product of open-loop unstable eigenvalues and is equal to the \emph{Bode sensitivity integral} (i.e. disturbance rejection measure). Here the tradeoffs mean that if one wishes to keep the first element in the pair small (such as low channel input power), the other element cannot be made arbitrarily large.  See Sec. \ref{sub:tradeoff} for more precise descriptions.  We call $DI$ the degree of anti-causality since it is associated with right-half plane (RHP) poles.  Note that references exist in addressing various aspects of fundamental limits; for an incomplete list, Van Trees \cite{vantrees} (pp. 501-511), de Bruijn, and Guo \emph{et al} \cite{guo:it05} (and therein references and subsequent works) discussed filtering versus smoothing as well as their relation to entropy and mutual information, Feng \emph{et al} \cite{fengfangkfinfo} examined the KF MMSE performance related to information theoretic measures, Iglesias and coauthors \cite{bode:iglesias,bode:zang_iglesias} studied BI and its information theoretic interpretation,  Seron \emph{et al} \cite{fundamental_filter_control} presented connections of the fundamental limitations between control and filtering, Martins and Dahleh \cite{nuno:tacsideinfo,nuno:tac} studied BI and entropy rates for systems over communication channels. See also \cite{tati:capI,sahai:main} and more discussions in Sec. \ref{subsec:fimcrbbi}.

Utilizing or motivated by the above mentioned equivalence relationship, we provide 1) New refinements to the Cover-Pombra capacity-achieving coding structure, including the complete characterization of the feedback generator; the necessity of KF in the CP structure; the orthogonality between future channel inputs and past channel outputs; the Gauss-Markov property of the transformed channel outputs; and the finite-dimensionality of the optimal message-carrying inputs.   2) Simple equivalence between generalized Schalkwijk-Kailath codes and the KF, which yields a convenient way to obtain a feedback communication scheme from an estimation problem. 3) \emph{Information theoretic characterization of KF}; that is, the KF is not only a device to provide 
sufficient statistics (which was shown in \cite{kavcic_it07}), but 
also a device to ensure the power efficiency and to recover the 
message optimally.  4) The \emph{necessity of MMSE estimation} in feedback communication problems over general additive noise channels with an average power constraint. Our results 1) - 3) hold for AWGN channels with intersymbol interference (ISI) 
where the ISI is modelled as a stable and minimum-phase FDLTI 
system; through the equivalence shown in 
\cite{tati:capI,kavcic_it07}, this channel is equivalent to a 
colored Gaussian channel with a rational noise power spectrum (which is assumed in a number of references) and 
without ISI.   The above results are mainly derived in the finite horizon, but we also show that the KF converges to a steady state as time goes to infinity, and the equivalence holds in the steady state system as well.  Note that, however, the infinite-horizon feedback capacity (or the stationary feedback capacity) problem is left open in this paper \footnote{We note that Kim in \cite{kim_allerton05} and further in \cite{kim06} claims the 
stationary conjecture is verified.  This leads to that stationary 
feedback capacity equals the asymptotic feedback capacity.}.

This paper is organized as follows. In Section \ref{sec:awgn:model}, a motivating example of feedback communication over an AWGN channel is presented. In Section \ref{sec:pre}, we 
describe the general Gaussian channel models. We then introduce the feedback capacity in finite horizon and the CP structure in Section \ref{app:cp}. In Section 
\ref{sec:kf}, we consider a general coding structure in 
finite-horizon which is closely related to the CP structure but allows us to easily see the necessity of the KF algorithm in feedback communication.  The presence of the KF links the feedback communication problem to an estimation 
problem and a control problem as shown in Section \ref{sec:dual}, and hence we rewrite the information 
rate and input power in terms of estimation theory quantities and control theory 
quantities and explore the connections; see Section \ref{subsec:crb}. More necessary conditions for the optimality of the coding structure are proposed in  Section \ref{sec:structureproperty}. Sections \ref{sec:kf} to 
\ref{sec:structureproperty} are focused on finite horizon. In Section 
\ref{sec:asym}, we extend the horizon to infinity and characterize 
the steady-state behavior.

\textbf{Notations:} We use underlines to specify vectors, and use 
boldface to specify matrices.  To ease the reading, all vectors in 
this paper are column vectors. We represent transpose by ${}'$.  We 
represent time indices by subscripts, such as $y_t$.  We denote by 
$\uy^T$ the collection $\{y_0,y_{1},$ $\cdots,$ $y_T\}$, and 
$\{y_t\}$ the sequence $\{y_t\}_{t=0}^\infty$.  We assume that the 
starting time of all processes is 0, consistent with the convention 
in dynamical systems but different from the information theory 
literature. We use $h(X)$ for the differential entropy of the random 
variable $X$.  For a random vector $\uy^T$, we denote its covariance 
matrix as $\bfK_{\uy}^{(T)}$. The norm $\|\uy\|$ is the Euclidean norm of the vector.  We 
denote $\calT_{xy}(z)$ as the transfer function from $x$ to $y$.  As a linear input-output relation (linear system) $\calZ(z)$ can be alternatively captured by a matrix, we represent the matrix associated with linear system $\calZ(z)$ by $\bcalZ(z)$ (boldface script $Z$). We 
denote ``defined to be" as ``$:=$".  We use $(\bfA,\uB,\uC', D)$ to 
represent system
\be \left\{ \ba{lll} \ux_{t+1} &=& \bfA \ux_t + \uB u_t \\
y_t &=& \uC' \ux_t + D u_t. \ea \right. \ee
Finally, in this paper, by ``capacity" we refer to the feedback 
capacity, if not specified otherwise.

\section{Motivating example: feedback capacity and optimal schemes for an AWGN channel} \label{sec:awgn:model}

To help the reader understand the intuition behind our study, we present a
simple example over an AWGN channel before we go into the Gaussian channels with memory. Below, we introduce a simple KF system (see Fig. \ref{fig:awgn:kf} (a)), followed by a straightforward rewrite of it (see Fig. \ref{fig:awgn:kf} (b)), which now has an interpretation as a feedback communication system. Finally we show that this feedback communication system is optimal as it is equivalent to the optimal SK scheme.  It motivates the further exploration of the connections among feedback communication, estimation, and feedback control.

\subsubsection{A Kalman Filter Problem} \label{subsubsec:awgnmot} 
\begin{figure}[h!]
\begin{center}
\subfigure[]
{\includegraphics[scale=0.46]{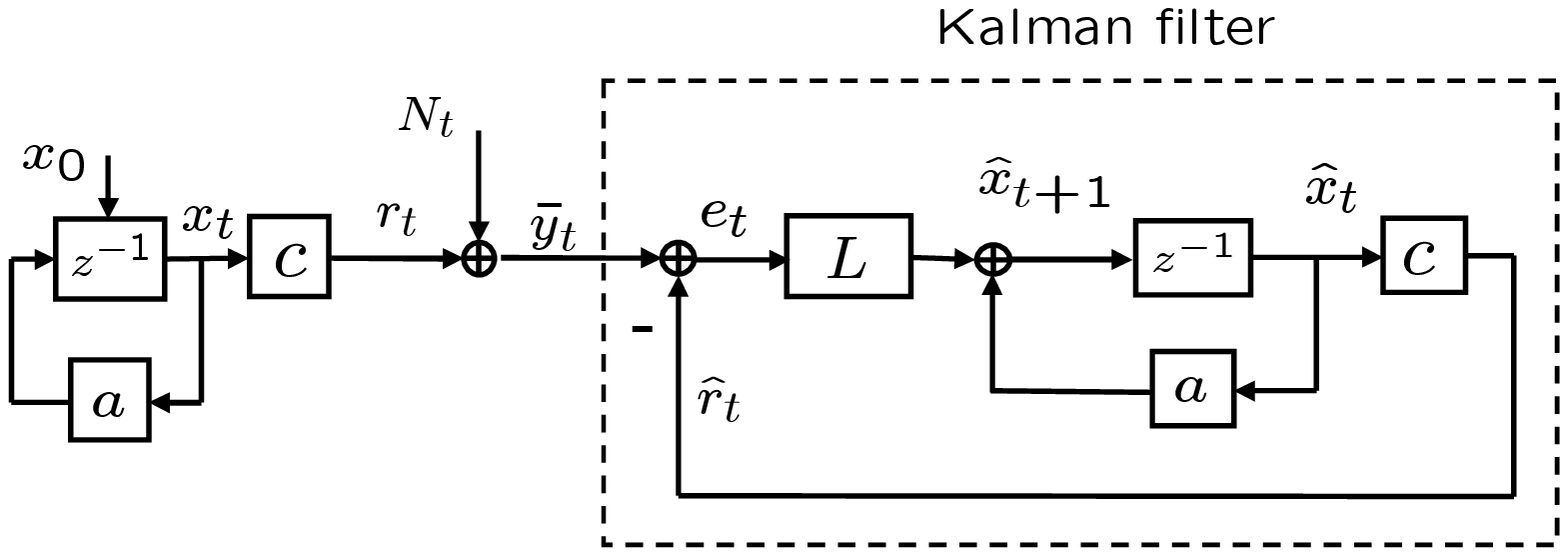}} \hspace*{-5pt}
\subfigure[]
{\includegraphics[scale=0.46]{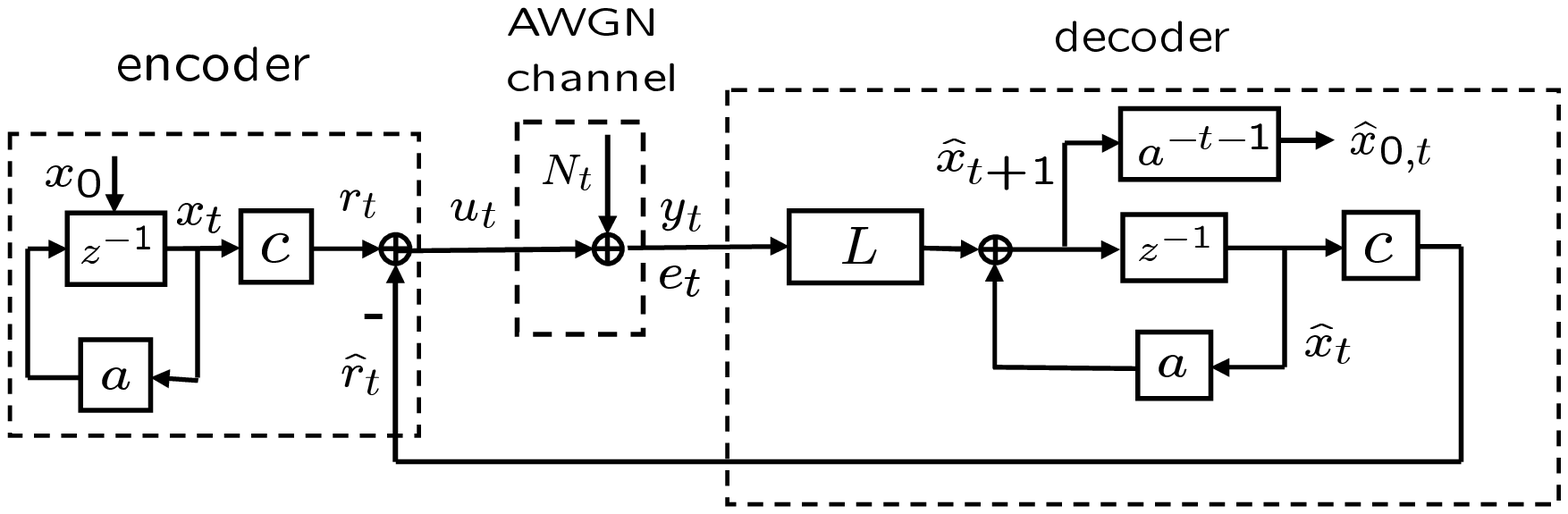}} 
 
\caption{\label{fig:awgn:kf} (a) A KF problem. (b) A KF-based coding structure.}
\end{center}
\end{figure}

Consider a standard KF problem for a first-order unstable LTI system
with noisy measurements:
\be \textnormal{to-be-estimated system: }\quad  \left\{\ba{lll} x_{t+1} &=& a x_t  \\
r_t&=& c x_t \\
\bar{y}_t &=& r_t +N_t, \label{dyn:1stsource} \ea \right.\ee
where $x_0$ is unknown, $a>1$ (namely the system is unstable), $a$ and $c$ are
known, and $N_t \iid \calN(0,1)$. The KF provides MMSE estimate of $\{x_t\}$ based on the noisy measurement process $\{\bar{y}_t\}$.  The (steady-state) \footnote{Though $\{\bar{y}_t\}$ is neither stationary nor even asymptotically stationary, a time-varying or time-invariant (steady-state) KF can be built to guarantee \emph{bounded} error covariance for
estimating $x_t$, and the difference between the time-varying one and time-invariant one vanishes as time increases, as
pointed out in Chapter 14 of \cite{kailath:book}. }
KF is described as (See Fig. \ref{fig:awgn:kf} (a) for the block diagram)

\be \textnormal{Kalman filter: }\quad \left\{ \ba{lll} \hat{x}_{t+1} &=& a \hat{x}_t + L e_t\\
\hatr_t & = & c \hat{x}_t \\
e_t &=& \bar{y}_t - c \hat{x}_t ,  \ea \right. \label{dyn:1stkf} \ee
where
\be L:= \frac{a \Si c}{1+c^2 \Si} \ee
is the asymptotic \emph{Kalman filter gain}, and $\Si$ is the asymptotic error covariance
for $\hat{x}_t$ (i.e. $\Si=\lim_{t \rightarrow \infty} \E (x_t-\hat{x}_t)(x_t-\hat{x}_t)' $), which is the positive solution to the discrete-time algebraic Riccati
equation (DARE)
\be \Si = a^2 \Si - \frac{a^2 c^2 \Si^2}{1+c^2\Si}. \ee
Solving the DARE, we obtain
\be  \Si = \disp \frac{a^2-1}{c^2}, \; L= \disp \frac{a^2-1}{ac} . \label{L:motiv} \ee

\subsubsection{KF-based Feedback Communication} Next, as illustrated in Fig. \ref{fig:awgn:kf} (b), we introduce a \emph{feedback communication coding scheme over an AWGN channel} by slightly changing the KF problem shown in Fig. \ref{fig:awgn:kf} (a).  Rather than closing the loop after
the AWGN $N_t$ (i.e. adding $(-\hatr_t)$ to $\bar{y}_t$), in Fig.
\ref{fig:awgn:kf} (b), the loop is closed before the AWGN $N_t$ (i.e. adding
$(-\hatr_t)$ to $r_t$). This does not change anything but the signals between the
two adders.  As indicated in Fig. \ref{fig:awgn:kf} (b), one can identify the encoder, the
AWGN channel, and the decoder, described in the following for time $t=0,1,\cdots$.
\be \textrm{AWGN channel: }\quad y_t = u_t + N_t, \label{awgn:channel} \ee
where $u_t$ is the channel input, $N_t \iid \calN(0,1)$ is the channel noise, and $y_t$ is the channel output. 
At time $t$, the encoder can access $\hatr_t$ (generated from 
$\uy^{t-1}$) via the noiseless feedback link:
\be \textrm{encoder dynamics: }\quad\left\{ \ba{lll} x_{t+1}&=&a x_{t} \\
r_t&=& c x_t \\
u_t&=& r_t - \hatr_t \ea \right. \label{awgn:enc} \ee
where $a$ and $c$ are encoder design parameters. The \emph{encoding procedure} is: Fix a set of $ M_T$ equally likely 
messages, then equally partition the interval
$[-\frac{1}{2}, \frac{1}{2}]$ into $M_T$ sub-intervals, and map the
sub-interval centers to the set of $M_T$ messages; this
is known to both the transmitter and receiver \textit{a priori}.  To transmit, let $x_{0}:=W$, the sub-interval center representing the to-be-transmitted message.  In other words, {the initial condition (at time 0) of the transmitter is the to-be-transmitted message}.
\be  \textrm{decoder dynamics: }\qquad\left\{ \ba{lll} \hat{x}_{t+1}&=&a \hat{x}_{t} +L y_t \\
\hatr_t&=& c \hat{x}_t \\
\hat{x}_{0,t}&=& a^{-t-1} \hat{x}_{t+1} ,\ea \right. \label{dyn:dec} \ee
and the \emph{decoding procedure} is to simply map $\hat{x}_{0,T}$ into the closest sub-interval center. 
 (Note that in Fig. \ref{fig:awgn:kf} (b), $y_t=e_t$.)

\begin{figure}[h!]
\begin{center}
{\includegraphics[scale=0.5]{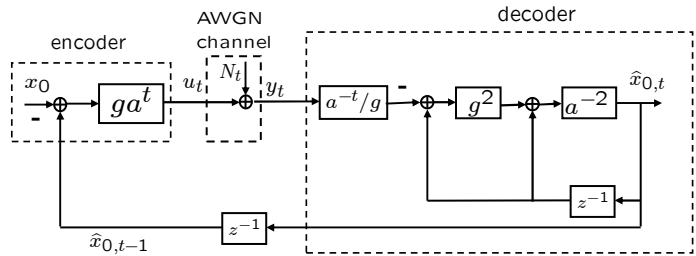}} 
\caption{\label{fig:sk} The SK coding scheme.}
\end{center}
\end{figure}

The \emph{objective} of the feedback communication problem is to, under
an average channel input power constraint
\be  \frac{1}{T+1} \E \| \uu^T \|^2  \leq \mathcal{P} \textnormal{ or }\lim _{T \rightarrow \infty}  \frac{1}{T+1} \E \| \uu^T \|^2  \leq \mathcal{P}
\label{awgn:powercon}\ee
with $\calP>0$ being the power budget, achieve 
\be C_{fb}(\calP)=C_{nf}(\calP)=\frac{1}{2} \log(1+\calP), \ee
where $C_{fb}(\calP)$ is the feedback capacity and $C_{nf}(\calP)$ is the
non-feedback capacity in either the finite horizon (time 0 to $T$) or infinite horizon (time 0 to $\infty$).
To attain this objective, one can fixed any coding length $(T+1)$ and
any $\ep>0$ (where $\ep$ is an arbitrarily small slack from the 
capacity $C_{fb}$).  Then let $a:=\sqrt{1+\calP}$,
$c \neq 0$ be arbitrary, $ M_T:=a^{(T+1)(1-\ep)}$, and follow the above-described encoding/decoding dynamics/procedures.
It can be shown that this communication scheme can transmit any message out of totally $ M_T $ 
messages with vanishing probability of error as $T \rightarrow \infty$ while satisfying the power constraint (\ref{awgn:powercon}).
Instead of proving the optimality directly, we may alternatively show that the coding scheme in Fig. \ref{fig:awgn:kf}(b) is a simple reformulation of the well-known SK coding scheme that has been shown to achieves the feedback capacity of the AWGN channel. To this aim, a slight variation of the original SK scheme proposed
in \cite{kailath2} is illustrated in Fig. \ref{fig:sk} \footnote{A few SK-type schemes and their variations are compared in \cite{liu:markov}.  The variation here performs the same operations every step, as opposed to the scheme in \cite{kailath2} in which the initialization step differs from later steps. See also \cite{elia_c5,gallager:sk}}.  In this figure, one can identify the encoder,
AWGN channel, decoder, and the feedback link with one-step delay.    

To see the connection between the two coding schemes, note that in the SK scheme, it holds that
\be \ba{lll} u_t&=& g a^t ( \hat{x}_{0,t-1} -x_0 ) \\
\hat{x}_{0,t} &=& \disp \frac{g^2+1}{a^2} \hat{x}_{0,t-1} - a^{-t-2} g y_t ; \ea \ee
and in the KF-based scheme, it holds that
\be \ba{lll} u_t&=& c a^t (x_0 - \hat{x}_{0,t-1}) \\
\hat{x}_{0,t} &=& \hat{x}_{0,t-1} + a^{-t-1} L y_t . \ea \ee
If we define
\be  g:=  \sqrt{a^2-1}, \; c:= -g , \ee
then both schemes generate identical channel inputs, outputs, and decoder
estimates respectively, and hence they are considered as equivalent.  The optimal choice of $g$
in the SK coding scheme indeed corresponds to the (optimal) KF
gain.  Thus, we conclude that \emph{the SK scheme essentially implements the KF algorithm}.  
In fact, more insights can be obtained from this AWGN example; see Chapter 3 of \cite{liu:phd}. These insights can be extended to the case of Gaussian channels with memory, which we now turn to.

\section{Channel model} \label{sec:pre}

In this section, we briefly describe two Gaussian channel models, 
namely the colored Gaussian noise channel without ISI and white 
Gaussian noise channel with ISI.

\subsection{Colored Gaussian noise channel without ISI} \label{subsec:color}

\begin{figure}[h!]
\center \subfigure[ 
]{{\scalebox{.5}{\includegraphics{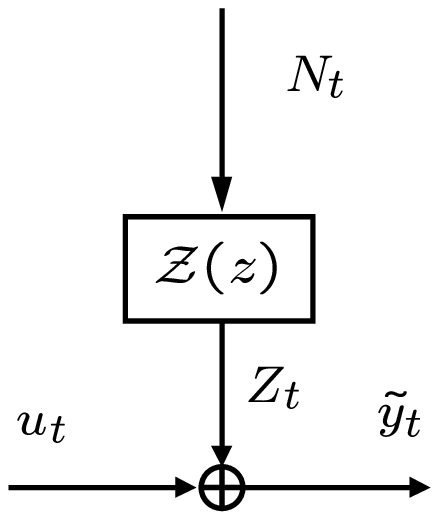}}}} \hspace{0in} 
\subfigure[ ]{{\scalebox{.5}{\includegraphics{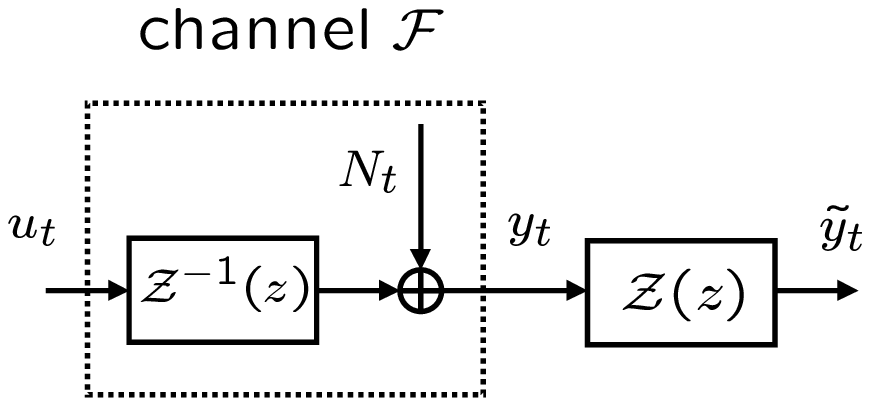}}}} 
\hspace{0in} \subfigure[ 
]{{\scalebox{.5}{\includegraphics{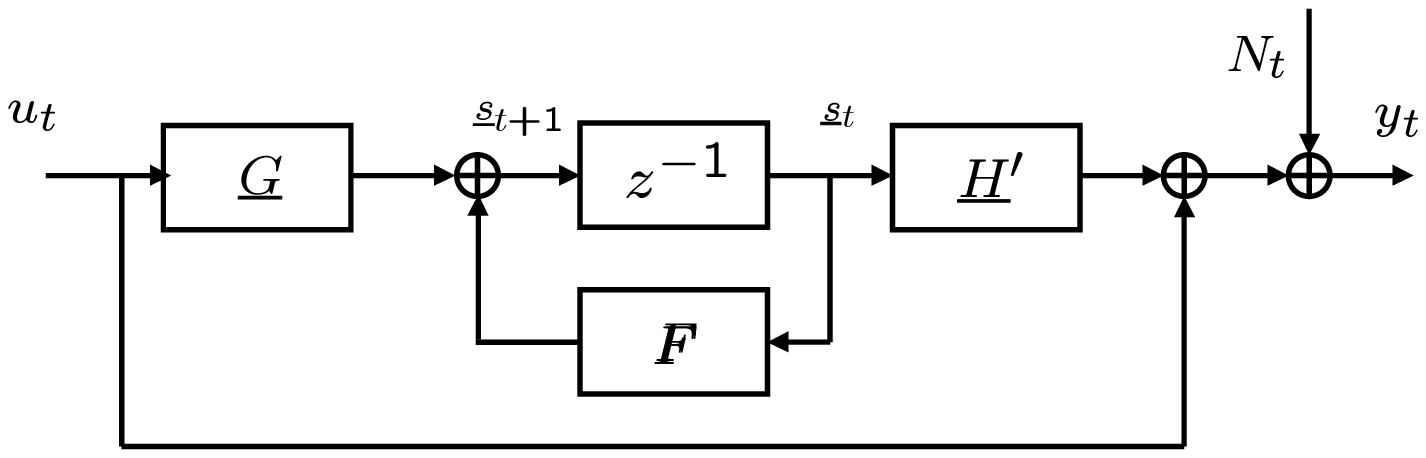}}}} \caption{ (a) A 
colored Gaussian noise channel without ISI.  (b) The induced ISI 
channel with AWGN. (c) State-space realization of channel $\calF$.} 
\label{fig:isicolor}
\end{figure}

Fig. \ref{fig:isicolor} (a) shows a colored Gaussian noise channel 
without ISI.  At time $t$, this discrete-time channel is described 
as
\be \tilde{y}_t = u_t + Z_t, \;\textnormal{ for } t=0,1,\cdots, 
\label{chan:color} \ee
where $u_t$ is the channel input, $Z_t$ is the channel noise, and 
$\tilde{y}_t$ is the channel output.  We make the following 
assumptions: The colored noise $\{Z_t\}$ is the output of a 
finite-dimensional stable and minimum-phase linear time-invariant 
(LTI) system $\calZ(z)$, driven by a white Gaussian process $\{N_t\}$ 
with zero mean and unit variance, and $\calZ(z)$ is at initial rest.  
We assume that the LTI system $\calZ(z)$ has order (or dimension) $m$ 
and $\calZ(\infty) \neq 0$ (i.e. $\calZ(z)$ is proper but 
non-strictly proper). We further assume, without loss of generality, that 
$\calZ(\infty)=1$; for cases where $g:=\calZ(\infty) \neq 1$, we can 
normalize $\calZ(z)$ using a scaling factor $1/g$. Then, the finite 
dimensionality of $\calZ(z)$ implies that $\calZ(z)$ admits the 
following \emph{transfer function representation}
\be \calZ(z)  = \frac{ z^m + f_{m-1} z^{m-1} + \cdots + f_1 z + f_0 
} {  z^m + (f_{m-1} + g_{m-1}) z^{m-1} + \cdots + (f_1 + g_1) z + 
(f_0 + g_0 )}, \label{calZtf} \ee
where $\{f_0, \cdots, f_{m-1}\}$ and $\{g_0, \cdots, g_{m-1}\}$ are 
such that $\calZ(z)$ is stable and minimum phase.  Define
\be \ba{lll} \calZ_z(z) &=& \disp
\frac{ z^m + f_{m-1} z^{m-1} + \cdots + f_1 z + f_0 }{z^m} \\
\calZ_p(z)  &=& \disp \frac{ z^m + (f_{m-1} + g_{m-1}) z^{m-1} + 
\cdots + (f_1 + g_1) z + (f_0 + g_0 )} {z^m}. \ea \label{calZpz} \ee
Then it holds that 
\be \calZ(z)  =  \frac{\calZ_z(z)}{\calZ_p(z)}, \label{calZfactor} 
\ee
that is, $\calZ_p (z)$ and $\calZ_z (z)$ contain the information 
about the poles and zeros of $\calZ(z)$, respectively. For future reference, we define 
\be \ba{lll} \uG_z &:=& [ f_{m-1}, \quad \cdots \quad, f_0 ]' \\
\uG_p &:=& [ f_{m-1}+g_{m-1}, \quad \cdots \quad, f_0+g_0 ]', \ea 
\label{HzHp} \ee
that is, $\uG_z'$ and $\uG_p'$ are the output matrices (vectors) for 
systems $\calZ_z(z)$ and $\calZ_p(z)$ (see Appendix \ref{appsub:represent} for relevant state-space representation concepts).

We can also represent the input-output relation based on \emph{time-domain operators (matrices)}.  For any block size (i.e. coding length) of $(T+1)$, we may 
equivalently generate $\uZ^T$ by
\be \uZ^T = \bcalZ_T \uN^T, \label{noise:color}\ee
where $\bcalZ_T$ is a $(T+1)\times(T+1)$ lower triangular Toeplitz 
matrix of the impulse response of $\calZ(z)$.  Since we have assumed 
that $\calZ(\infty)=1$, the diagonal elements of $\bcalZ_T$ are all 1.  Likewise, $\bcalZ_{z,T}$ 
and $\bcalZ_{p,T}$, the matrix 
representations of $\calZ_z(z)$ and $\calZ_p(z)$, are respectively given by
\be \ba{lll} \bcalZ_{z,T} &:=& \left[ \matrix{ 1 & 0  & 0 & \cdots & 
0 & 0 \cr f_{m-1} & 1 & 0 & \cdots & 0 & 0  \cr 
         f_{m-2} & f_{m-1} & 1 & \cdots & 0 & 0 \cr
         \vdots \cr 
         f_{0} & f_{1} & f_2 & \cdots & 0 & 0 \cr
         0 & f_{0} & f_1 & \cdots & 0 & 0 \cr
          \vdots \cr 
         0 & 0 & 0 & \cdots & f_{m-1} & 1   } \right] \\
    \bcalZ_{p,T} &:=& \left[ \matrix{ 1 & 0  & 0 & \cdots & 0 & 0 \cr f_{m-1}+g_{m-1} & 1 & 0 & \cdots & 0 & 0  \cr 
         f_{m-2}+g_{m-2} & f_{m-1}+g_{m-1} & 1 & \cdots & 0 & 0 \cr \vdots \cr
         f_{0} + g_0 & f_{1}+g_1 & f_2+g_2 & \cdots & 0 & 0 \cr
         0 & f_{0}+g_0 & f_1+g_1 & \cdots & 0 & 0 \cr
          \vdots \cr
           0 & 0 & 0 & \cdots & f_{m-1}+g_{m-1} & 1   } \right] , \ea    \label{def:ZzZp}
\ee
that is, $\bcalZ_{p,T}$ and $\bcalZ_{z,T}$ are lower triangular, 
Toeplitz, and banded with bandwidth $(m+1)$, corresponding to 
causal, LTI, $m$th order moving-average (MA-$m$) filters.  
Therefore, it holds that
\be \bcalZ_T = \bcalZ_{p,T}^{-1} \bcalZ_{z,T} = \bcalZ_{z,T} \bcalZ_{p,T}^{-1} . 
\label{eq:calZfactor}\ee

As a consequence of the above assumptions, $\{Z_t\}$ is 
asymptotically stationary.  Note that there is 
no loss of generality in assuming that $\calZ(z)$ is stable and 
minimum-phase (cf. Chapter 11, \cite{papoulis}).

\subsection{White Gaussian channel with ISI} \label{subsec:isi}

The above colored Gaussian channel induces a 
white Gaussian channel with ISI.  More precisely, notice that from 
(\ref{chan:color}) and (\ref{noise:color}), we have
\be \utildey^T = \bcalZ_T (\bcalZ_T^{-1} \uu^T + \uN^T ), \label{chan:color1} \ee
which we identify as a stable and minimum-phase ISI channel with 
AWGN $\{N_t\}$, see Fig. \ref{fig:isicolor} (b); $\bcalZ_T^{-1}$ is 
well defined since $\bcalZ_T$ is lower triangular with diagonal 
elements being $\calZ(\infty) = 1$.   Here $\calZ^{-1}(z)$ is 
also at initial rest. Note that $\bcalZ_T^{-1}$ is the matrix 
inverse of $\bcalZ_T$, equal to the lower-triangular Toeplitz matrix 
of impulse response of $\calZ^{-1}(z)$. For any fixed $\uu^T$ and 
$\uN^T$, (\ref{chan:color}) and (\ref{chan:color1}) generate the same channel output $\utildey^T$. 
\footnote{More rigorously, the mappings from $(u,N)$ to $\utildey$ 
are $T$-equivalent. For a discussion about systems representations 
and equivalence between different representations, see Appendix 
\ref{appsec:equiv}.} 

The initial rest assumption on $\calZ^{-1}(z)$ can be imposed in 
practice as follows.  First, before a transmission, drive the 
initial condition (which is enabled by the controllability 
requirement stated below) of the ISI channel to any desired value 
that is also known to the receiver a priori.  Then, after the 
transmission, remove the response due to that initial condition at 
the receiver.  Such an assumption is also used in 
\cite{kavcic_it07,tati:capI}. 

We can then write the \emph{minimal state-space representation} of 
$\calZ^{-1}(z)$ as $(\bfF,\uG,\uH',1)$, where $\bfF \in 
\mathbb{R}^{m}$ is stable, $(\bfF,\uG)$ is controllable, 
$(\bfF,\uH')$ is observable, and $m$ is the \emph{dimension} or 
\emph{order} of $\calZ^{-1}(z)$. Let us denote the channel from $u$ 
to $y$ in Fig. \ref{fig:isicolor}(b) as $\calF$, where
\be \uy^T := \bcalZ_T^{-1} \uu^T + \uN^T = \bcalZ_T^{-1} \utildey^T. 
\ee
The channel $\calF$ is described in \emph{state-space} as
\be  \textnormal{channel } \calF: \left \{\ba{lll} \us_{t+1}&=&\bfF \us_t + \uG u_t \\
y_t &=& \uH' \us_t +u_t +N_t, \ea \right. \label{channel:isi}\ee
where $\us_0=\underline{0}$; see Fig. \ref{fig:isicolor} (c).  Notice that 
channel $\calF$ is not essentially different than the channel from 
$u$ to $\tilde{y}$, since $\{\uy^t\}$ and $\{\tilde{\uy}^t\}$ causally 
determine each other.  Without loss of generality, we can choose $(\bfF,\uG,\uH',1)$ to 
have the following observable canonical form:
\be \ba{lll} \bfF := \left[ \ba{c|c} 
                                                                                \ba{c} -f_{m-1} \\ \vdots \\ -f_1 \ea 
                                                                                & \bfI _{m-1} \\
                                                                                \hline
                                                                                -f_0 
                                                                                & \uzero_{ (m-1) \times 1} 
                                                        \ea \right]
, \quad \uG :=\left[ \ba{cc} g_{m-1} \cr \vdots \cr g_0 \ea \right]  \\
\uH' := [ 1 \quad 0 \quad \cdots \quad 0 ]. \ea \label{FGH:ctrb0} 
\ee
In other words, it holds that $\calZ^{-1}(z) = \uH' (z \bfI - 
\bfF)^{-1} \uG +1$.  Note that we also have $\uG_p' = 
\uG' + \uG_z'$ (see (\ref{HzHp})).

We concentrate on the case $m \geq 1$; the case that $m$ is 0 (i.e., 
$\calF$ is an AWGN channel) was solved in \cite{kailath1,kailath2}.

\section{The feedback capacity in finite-horizon and the Cover-Pombra structure} \label{app:cp}

\subsection{The CP structure for the colored Gaussian noise channel and finite-horizon
capacity} \label{appsub:cpcolor}

We briefly review the CP coding structure for the colored Gaussian 
noise channel specified in Section \ref{subsec:color} (see 
\cite{cover-pombra-1989,cover}). 
Denote the covariance matrix of the colored Gaussian noise ${\uZ}^T$ as  
$\bfK_{\uZ}^{(T)}$, and let
\be \uu^T := \calB_T {\uZ}^T +\uv^T, \label{eq:cpu}\ee
where $\calB_T$ is a $(T+1)\times(T+1)$ strictly lower triangular 
matrix, $\uv^T$ is Gaussian with covariance $\bfK_{\uv}^{(T)} \geq 
0$ and is independent of ${\uZ}^T$  \footnote{This $\uv^T$ is called 
innovations in \cite{cover,kavcic_it07}; it should not be confused 
with the KF innovations in this paper.}.  Now 
the channel output is
\be \tilde{\uy}^T = \uu^T + \uZ^T = (\bfI+\calB_T ) {\uZ}^T +\uv^T.  \label{eq:cpy}\ee
Then $C_T$, the \emph{finite-horizon capacity}, is 
defined as the highest information rate that the CP structure can 
generate:
\be \ba{lll} C_{T}:= C_{T}(\calP) &:=& \disp \sup \frac{1}{T+1} I(\uv^T;{\utildey}^T) \\
&=&  \disp \sup \frac{1}{2(T+1)} \log \frac{\det \bfK_{\utildey} ^{(T)} } {\det \bfK_{\uZ}^{(T)}} \\
&=& \disp \sup \frac{1}{2(T+1)} \log \frac{ \det ((\bfI+\calB_T) 
\bfK_{\uZ}^{(T)}(\bfI+\calB_T)' + \bfK_{\uv}^{(T)} ) } {\det 
\bfK_{\uZ}^{(T)}}, \ea \label{cap:cpcolor} \ee
where the supremum is taken over all admissible $\bfK_{\uv}^{(T)}$ 
and $\calB_T$ satisfying the power constraint
\be P_T:=\frac{1}{T+1}\tr (\calB_T 
\bfK_{\uZ}^{(T)}\calB_T'+\bfK_{\uv}^{(T)}) \leq \calP. 
\label{constr:cpcolor} \ee

This finite-horizon capacity $C_T$ is the operational capacity as given by Theorem 1 of \cite{cover-pombra-1989} based on AEP and a random coding argument \footnote{One can also invoke Theorem 5.1 in \cite{tati:capI} and the equivalence between directed information and mutual information in this case to claim that $C_T$ is also the operation capacity.}.  Thus, we may focus only on the information rates in this paper and need not discuss coding in the operational sense.

To directly use the CP structure to construct a coding scheme is generally viewed as challenging for the following reasons. a) Its computation 
complexity grows faster than linearly with time ($O((T+1)^2)$ unknowns to be solved for each $T$), even though for each $T$ the search of $\bfK_{\uv}^{(T)}$ 
and $\calB_T$ can be posed as convex \cite{maxdet}. b) For each $T$ 
the optimal $\bfK_{\uv}^{(T)}$ and $\calB_T$ are not unique, (in 
fact there are an uncountable infinite number of optimizing 
solutions for each $T$, as can be easily seen from the $T=1$ 
case); moreover, the optimal solution to coding length $(T+1)$ does not 
necessarily contain a part that is optimal to coding length $T$. 
Hence the search of optimal $\bfK_{\uv}^{(T)}$ and $\calB_T$ for 
$T$ is not likely to suggest what the optimal 
coding scheme could be for any other time horizon. c) In 
\cite{cover-pombra-1989} the achievability of $C_T$ is proven using 
a random coding argument, but a specific practical code has not been proposed 
or applied to the CP structure.  Nevertheless, many insights can be obtained from the CP structure and it is also
the starting point of our development.

\subsection{The CP structure for the ISI Gaussian channel}

\begin{figure}[h!]
\begin{center}
{\scalebox{.5}{\includegraphics{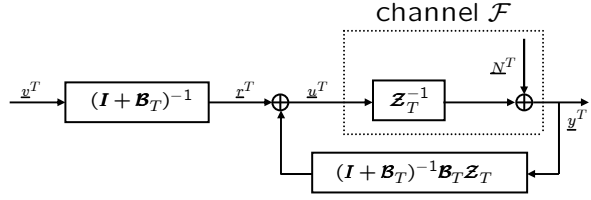}}} \caption{The block 
diagram of the CP structure for ISI Gaussian channel $\calF$.} 
\label{fig:cpisi}
\end{center}
\end{figure}

In light of the correspondence relation between the colored Gaussian noise channel 
and the ISI channel $\calF$, we can derive the CP coding structure 
for $\calF$, which is obtained from (\ref{eq:cpu}) by introducing a 
new quantity ${\ur}^T$ as
\be {\ur}^T : = (\bfI+\calB_T)^{-1} \uv^T. \ee
By ${\uZ}^T=\bcalZ_T \uN^T$ and $\tilde{y}^T=\bcalZ_T \uy^T$, we 
have
\be \ba{lll} \uu^T &=& \disp \calB_T \bcalZ_T \uN^T +(\bfI+\calB_T) {\ur}^T \\
\uy^T &=& \disp  \bcalZ^{-1}_T (\bfI+\calB_T) \bcalZ_T \uN^T + \bcalZ^{-1}_T (\bfI+\calB_T) {\ur}^T \\
&=& \disp  \bcalZ^{-1}_T (\bfI+\calB_T) (\bcalZ_T \uN^T + {\ur}^T) . 
\ea \label{eq:cpisi}\ee
This implies that, the channel input $\uu^T$ can be represented as
\be \uu^T =(\bfI+\calB_T)^{-1} \calB_T \bcalZ_T \uy^T + {\ur}^T, 
\label{input:cpisi}\ee
which leads to the block diagram in Fig. \ref{fig:cpisi}.
Then the capacity $C_T$ has the form:
\be \ba{lll} C_{T}(\calP) &= & \disp \sup \frac{1}{2(T+1)} \log \det \bfK_{\uy}^{(T)}  \\
&=& \disp \sup \frac{1}{2(T+1)}  \log  \det \left(\bcalZ^{-1}_T 
(\bfI+\calB_T) (\bcalZ_T \bcalZ_T'
+ \bfK_{\ur}^{(T)}) (\bfI+\calB_T)' \bcalZ_T^{-1}{}' \right) \\
&=& \disp \sup \frac{1}{2(T+1)}  \log  \det  (\bcalZ_T \bcalZ_T' + 
\bfK_{\ur}^{(T)}) \label{cap:cpisi} \ea \ee
where the supremum is over the power constraint
\be P_T:=\frac{1}{T+1}\tr (\calB_T \bcalZ_T \bcalZ_T' 
\calB_T'+(\bfI+\calB_T) \bfK_{\ur}^{(T)} (\bfI+\calB_T)' ) \leq 
\calP. \label{constr:cpisi} \ee
The capacity in this form is equivalent to 
(\ref{cap:cpcolor}).  Another form of the capacity based on the directed information, namely an input/output characterization,  can be shown as equivalent to the above form; see Appendix \ref{appsub:directI}.  One can also define the inverse function of $C_{T}(\calP)$ as $P_{T}(\calR)$, which is equal to the infimum power subject to a rate constraint 
\be \frac{1}{2(T+1)} \log \det \bfK_{\uy}^{(T)} \geq \calR . \label{constr:cpinverse} \ee

\section{Necessity of KF for optimal coding} \label{sec:kf}

In this section, we consider a finite-horizon feedback coding structure over channel $\calF$ denoted $\bbS:=\bbS(\calF)$, which is a variation of the CP structure. This variation is useful since: 1) searching 
over all possible parameters in the structure achieves $C_T$, that is, there is no loss of generality or optimality when
focusing on this structure only; 2) we can show that to ensure power 
efficiency (to be explained), structure $\bbS$ necessarily 
implements the KF algorithm.  This implies that our KF 
characterization leads to a refinement to the CP structure.  

\subsection{Coding structure $\bbS$} \label{sub:generalstructure}

\begin{figure}[h!]
\begin{center}
{\scalebox{.5}{\includegraphics{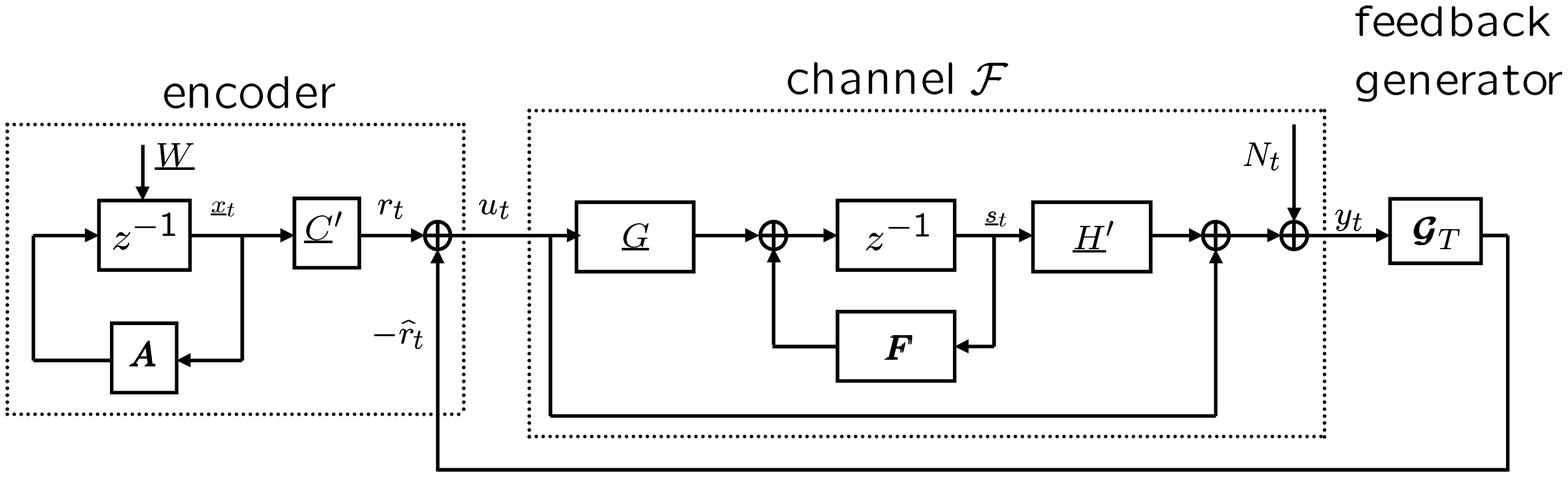}}} \caption{Coding structure $\bbS$ for channel $\calF$.} \label{fig:liu2}
\end{center}
\end{figure}

Fig. \ref{fig:liu2} illustrates the coding structure $\bbS$, 
including the encoder and the \emph{feedback generator}, which is a portion 
of the decoder.  (How the decoder produces the estimate of the decoded 
message will be considered shortly.)  Below, we fix the time horizon to span from time 0 to time $T$ and describe $\bbS$.

\textbf{Encoder:} The encoder follows the dynamics 
\be  \textnormal{Encoder:} \left \{ \ba{lll} \ux_{t+1} &=& \bfA \ux_t\\
r_t &=& \uC' \ux_t \\
u_t &=& r_t - \hatr_t . \ea \right. \label{enc1}\ee
where $\ux_0:=\uW\sim \calN(\uzero,\bfI_{n+1})$. We assume that the encoder dimension $(n+1)$ is a fixed integer 
satisfying $0 \leq n \leq T$; $\bfA \in \bbR^{(n+1)\times (n+1)}$; $\uC \in \bbR^{n+1}$; and the 
assumption (A1) holds:

\textbf{(A1)}: $(\bfA,\uC')$ is observable.  

We then let
\be \ba{llllll} 
\bfGa_T(\bfA,\uC)&:=&\bfGa_T&:=& [\uC,\bfA'\uC,\cdots,\bfA^{T}{}' \uC]'
        &\in \bbR^{(T+1) \times (n+1)}  \\
\bfK_{\ur}^{(T)}(\bfA,\uC)&:=&\bfK_{\ur}^{(T)}&:=&\E \ur^T \ur^T{}'
        &\in \bbR^{(T+1) \times (T+1)} . \ea \ee
Therefore, $\bfGa_n$ is the observability matrix for $(\bfA,\uC')$ 
and is invertible, $\bfGa_T$ has rank $(n+1)$, $\ur^T=\bfGa_T \uW$, and 
$\bfK_{\ur}^{(T)} = \bfGa_T \bfGa_T'$ with rank $(n+1)$.

\textbf{Feedback generator:} The feedback signal $(-\hatr_t)$ is 
generated through a feedback generator $\calG_T$, i.e.
\be -\underline{\hatr}^T= \calG_T \uy^T,  \label{fbgen} \ee
where $\calG_T \in \bbR^{(T+1)\times (T+1)}$ is a strictly 
lower triangular matrix, namely the output feedback is 
strictly causal. 

Throughout the paper, 
the above assumptions on the encoder/decoder are always assumed if 
not otherwise specified.  For future use, we compute the 
channel output as
\be \uy^T = (\bfI-\bcalZ_T^{-1} \calG_T) ^{-1} (\bcalZ_T^{-1} \ur^T 
+ \uN^T). \ee

\begin{definition}
Consider the coding structure $\bbS$ shown in Fig. \ref{fig:liu2}.  
Define the constraint capacity
\be \ba{lll} \displaystyle C_{T,n}:=  C_{T,n}(\calP)&:=& 
\displaystyle \sup _{\bfA \in
\bbR^{(n+1)\times (n+1)},\uC,\calG_T, (A1)} \frac{1}{T+1}I(\uW;\uy^T)\\
& &  ^{s.t. \: \E \| \uu^T \|^2/(T+1) \leq \mathcal{P} } \ea 
\vspace{-10pt} \ee
and define its inverse function as $P_{T,n}(\calR)$, that is,
\be \ba{lll} \displaystyle P_{T,n}:=  P_{T,n}(\calR)&:=& 
\displaystyle \inf _{\bfA \in
\bbR^{(n+1)\times (n+1)},\uC,\calG_T, (A1)} \frac{1}{T+1}\E \| \uu^T \|^2\\
& &  ^{s.t. \:  I(\uW;\uy^T)/(T+1) \geq \mathcal{R} } \ea \label{def:PTn}
\vspace{-10pt} \ee
\label{def:CTn}
\end{definition}

In other words, $C_{T,n}$ is the finite-horizon information capacity 
for a \emph{fixed encoder dimension} $(n+1)$, by searching over all 
admissible $\bfA$, $\uC$, and $\calG_T$ of appropriate dimensions.  The pair $(\calP,C_{T,n}(\calP))$ and the pair $(P_{T,n}(\calR),\calR)$ specify the \emph{optimal tradeoff} between the channel input power and information rate for the communication problem with fixed encoder dimension.

\subsection{Relation between the CP structure and the proposed structure $\bbS$}
\label{appsub:cp_relation}

The coding structure $\bbS$ over $\calF$ in Fig. \ref{fig:liu2} was motivated and is tightly associated with the CP 
structure over  the ISI Gaussian channel $\calF$ in Fig. \ref{fig:cpisi}.  Let $\uu^T(\bfK_{\ur}^{(T)},\calB_T)$ and $\uu^T(\bfA,\uC,\calG_T)$ denote the input sequences generated by encoders with $(\bfK_{\ur}^{(T)},\calB_T)$ and $(\bfA \in 
\bbR^{(T+1)\times(T+1)},\uC,\calG_T)$, respectively.

\begin{lemma} \label{lemma:ctn}
i) For any given pair $(\bfK_{\ur}^{(T)},\calB_T)$ with $\bfK_{\ur}^{(T)}>0$, there exists an admissible triple 
\newline $(\bfA \in 
\bbR^{(T+1)\times(T+1)},\uC,\calG_T)$ such that $\uu^T(\bfK_{\ur}^{(T)},\calB_T)= \uu^T(\bfA,\uC,\calG_T)$; for any given pair $(\bfK_{\ur}^{(T)},\calB_T)$ with $\bfK_{\ur}^{(T)} \geq 0$ but $\bfK_{\ur}^{(T)} \not > 0$,  there exists a sequence of admissible triples $\{(\bfA_i \in \bbR^{(T+1)\times(T+1)},\uC_i,\calG_{T,i})\}_{i=1}^{\infty}$ such that $\uu^T(\bfK_{\ur}^{(T)},\calB_T)= \lim _{i \rightarrow \infty}  \uu^T(\bfA_i,\uC_i,\calG_{T,i})$; 

ii) For any given triple $(\bfA \in \bbR^{(T+1)\times(T+1)},\uC,\calG_T)$, there is an admissible 
pair $(\bfK_{\ur}^{(T)},\calB_T)$ such that 
\newline $\uu^T(\bfK_{\ur}^{(T)},\calB_T)= \uu^T(\bfA,\uC,\calG_T)$;

iii)
\be C_T(\calP)=C_{T,T}(\calP) , \;  P_T(\calR)=P_{T,T}(\calR).\ee
\end{lemma}

\textbf{Proof: } See Appendix \ref{appsec:CPequivalent}.   \endproof

One advantage of considering the structure $\bbS$ is that we 
can have the flexibility of allowing $T \geq n$, which makes it possible to increase the 
horizon length to infinity without increasing the dimension of 
$\bfA$, a useful step towards the KF characterization 
of the feedback communication problem.

In what follows, several refinements to the coding structure $\bbS$
will be presented.

\subsection{The presence of the KF} \label{sub:nec4kf}

We first compute the mutual information in the aforementioned coding structure $\bbS$.

\begin{prop} \label{prop:I}
Consider the structure $\bbS$ in Fig. \ref{fig:liu2}. Let $0 \leq n \leq T$, $(\bfA, \uC)$ be observable with $\bfA \in 
\bbR^{(n+1)\times (n+1)}$ and $\calG_T$ be strictly lower triangular.  
Then 

i) It holds that
\be \ba{lll} I(\uW;\uy^T)&=& I(\ur^T;\uy^T) \\
& = & \disp I(\uu^T \rightarrow \uy^T) \\ \vspace{2pt} & = &  \disp  
\frac{1}{2} \log \det \bfK_{\uy}^{(T)}\\ \vspace{2pt} &=& \disp  
\frac{1}{2} \log \det (\bfI+ \bcalZ_T^{-1} \bfK_{\ur}^{(T)} 
\bcalZ_T^{-1'}) \\ \vspace{2pt} &=& \disp  \frac{1}{2} \log \det 
(\bfI+ \bcalZ_T^{-1} \bfGa_T \bfGa_T' \bcalZ_T^{-1'}); \ea \ee

ii) $I(\uW;\uy^T)$ is \emph{independent} of the feedback generator $\calG_T$.
\end{prop}

\textbf{Proof: } i)
\be  \ba{lll}  I(\uW;\uy^T) & = &h(\uy^T) - h(\uy^T|\uW) \\
&=& h(\uy^T) - h \left((\bfI-\bcalZ_T^{-1} \calG_T) ^{-1} (\bcalZ_T^{-1} \ur^T + \uN^T) | \uW \right) \\
&\eqa& \disp \frac{1}{2} \log \det (2 \pi e \bfK_{\uy}^{(T)}) - h(\uN^T)  \\
&\eqb& \disp I(\uu^T \rightarrow \uy^T) \\
 \vspace{2pt}&=&\disp  \frac{1}{2} \log \det \bfK_{\uy}^{(T)} \\   
&= & \disp  \frac{1}{2} \log \det (\bfI+ \bcalZ_T^{-1} 
\bfK_{\ur}^{(T)} \bcalZ_T^{-1'}) , \ea \ee
where (a) is due to $\ur^T=\bfGa_T \uW$, $\det (\bfA \bfB)=\det \bfA 
\det \bfB$, and $\det(\bfI-\bcalZ_T^{-1} \calG_T) ^{-1}=1$; and (b) 
follows from \cite{elia_c5} or a direct computation of $I(\uu^T 
\rightarrow \uy^T)$.  ii) It is clear from i) that $I(\uW;\uy^T)$ is 
independent of the feedback generator $\calG_T$, and depends only 
on $\bfK_{\ur}^{(T)}$, or equivalently on $(\bfA,\uC)$.
\endproof

\begin{remark} \rm Though simple, Proposition \ref{prop:I} has 
interesting interpretations and implications.  The first equality of 
i) shows that the mutual information between the message $W$ and 
channel output $\uy^T$ is completely preserved in the mutual information between the 
message-carrying signal $\ur^T$ and channel output $\uy^T$.  The second equality 
shows that the directed information (cf. \cite{tati:capI} and Appendix \ref{appsub:directI}) in this setup is 
equivalent to the message-output characterization based on the 
mutual information, which is convenient in many situations.  The 
third equality involves the output covariance  
matrix, a link towards the Bode waterbed effect and the fundamental concept of the KF 
innovations (to be explored in subsequent sections).  The rest of the proposition implies 
that, for the given channel $\calZ^{-1}, $\textit{fixed $(\bfA,\uC)$ leads to a fixed information rate regardless of the feedback generator}.  In fact, the mutual information may be interpreted as anti-causal and independent of the  the strictly causal feedback generator.   Hence \emph{the feedback 
generator $\calG_T$ has to be chosen to minimize the average channel input 
power} in order to achieve the capacity (recalling that the capacity problem can be expressed as
minimizing power while fixing the rate (\ref{def:PTn})), which necessitates
a KF.  Note that the infinite-horizon counterpart of this 
proposition was proven in \cite{elia_c5}.
\end{remark}

Next we solve the optimal feedback generator for a fixed $(\bfA,\uC)$, which is essentially a KF. Denote the \emph{optimal feedback generator} 
for a given $(\bfA,\uC)$ as $\calG_T^*(\bfA,\uC)$, namely
\be \calG_T^*(\bfA,\uC) :=    \arg \displaystyle \inf _{\calG_T} \displaystyle 
\frac{1}{T+1} \E \| \uu^T (\bfA,\uC,\calG_T)  \| ^2.\label{opt:calG2} \ee
By Proposition \ref{prop:I}, we can define, for a fixed 
$(\bfA,\uC)$, the information rate across the channel to be
\be R_T(\bfA,\uC):= \frac{I(\uW;\uy^T)}{T+1} .\ee

\begin{prop} \label{prop:kf} Consider coding structure $\bbS$ in Fig.
\ref{fig:liu2}. Fix any $0 \leq n \leq T$.  Then (recall the capacity definition $P_{T,n}(\calR)$ in (\ref{def:PTn}))

i) \be \ba{llcllcl}P_{T,n}(\calR) &= &   \displaystyle 
\inf_{\bfA \in \bbR^{(n+1)\times (n+1)},\uC }
 & \displaystyle \frac{1}{T+1}\E \| \uu^T(\bfA,\uC,\calG_T^*(\bfA,\uC)) \|^2 . \\
 & & ^{s.t. \: R_T(\bfA,\uC) \geq  \calR} & \ea
\label{opt:calG1}\ee

ii) The optimal feedback generator $\calG_T^*(\bfA,\uC)$ is given by
\be \calG_T^*(\bfA,\uC) = - \hatG_T^*(\bfA,\uC) (\bfI- \bcalZ_T^{-1} 
\hatG_T^*(\bfA,\uC)) ^{-1} , \label{eq:calGhatG}\ee
where $\hatG_T^*(\bfA,\uC)$ is the one-step prediction MMSE estimator 
(Kalman filter) of ${\ur}^T$ given the noisy observation $ \ubary^T 
:= \bcalZ_T^{-1} {\ur}^T+\uN^T $ (i.e. the optimal one-step prediction is $\hat{\ur}^T =\hatG_T^*(\bfA,\uC) \ubary^T  $), given by
\be \ba{lll} \hatG_T^*(\bfA,\uC) &:=&   \displaystyle \arg 
\min_{\hatG_T }  \displaystyle 
\frac{1}{T+1} \E ({\ur}^T-\hatG_T \ubary^T)({\ur}^T-\hatG_T 
\ubary^T)', \ea \label{opt:calG3}  \ee
where $\hatG_T$ is strictly lower triangular.

\end{prop}

\begin{figure}[h!]
\center \subfigure[ ]{\scalebox{.4}{\includegraphics{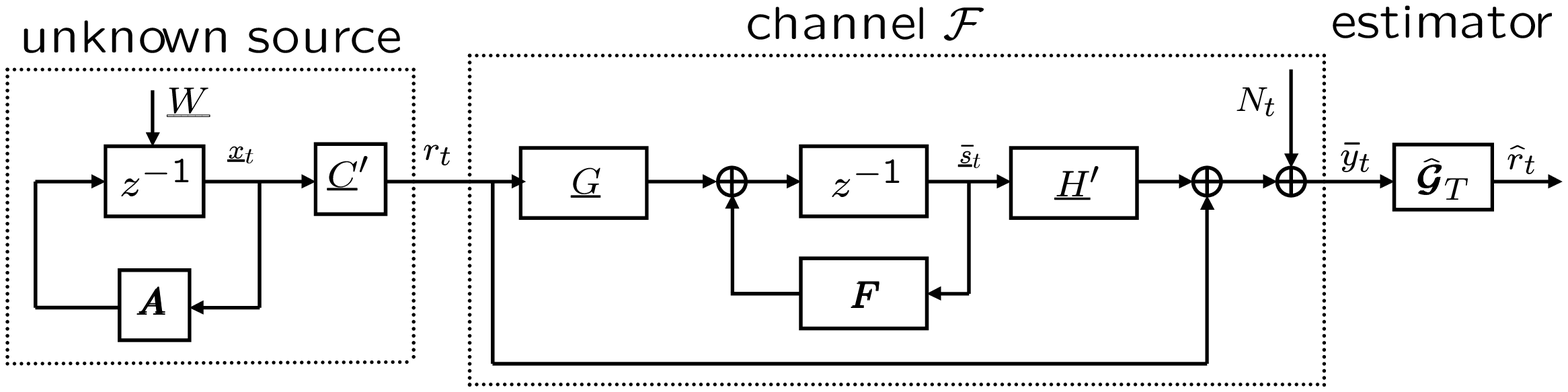}}}  
\subfigure[ ]{\scalebox{.4}{\includegraphics{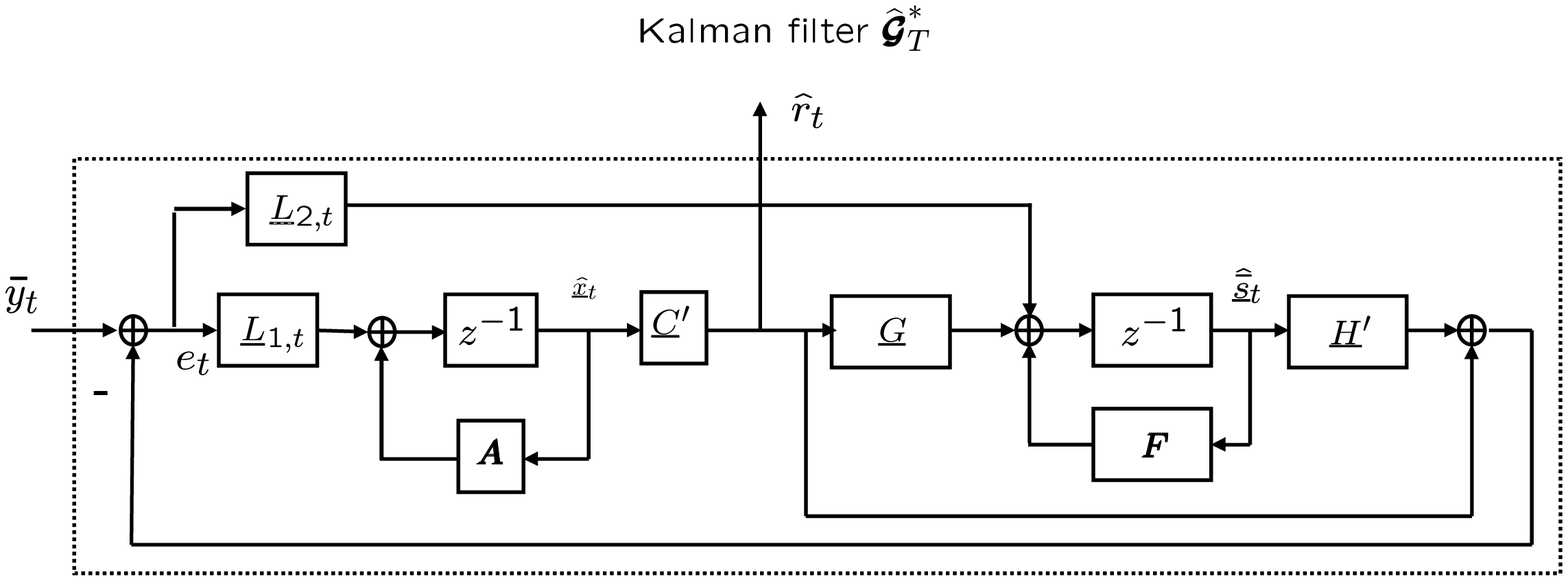}}} 
\subfigure[ ]{{\scalebox{.4}{\includegraphics{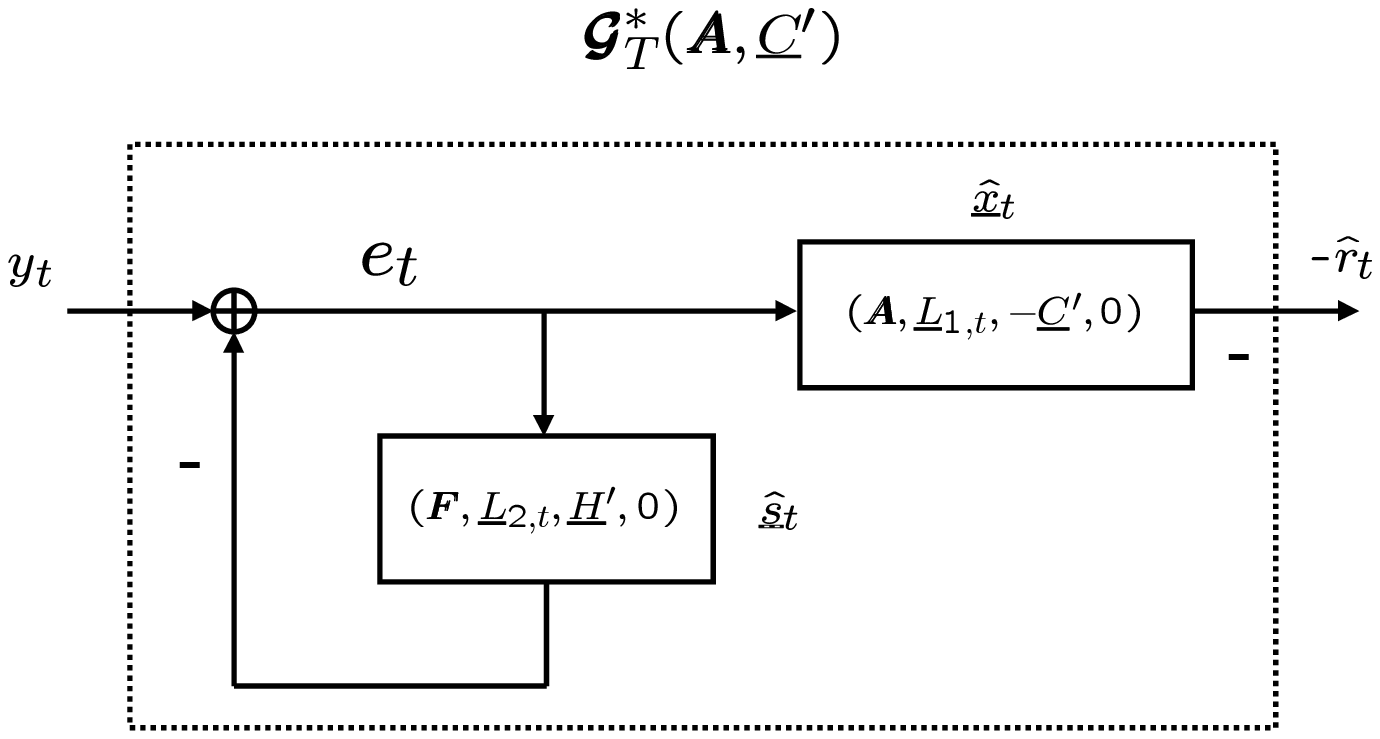}}}} 
\caption{ (a) An estimation problem over channel $\mathcal{F}$.  (b) 
The KF $\hatG_T^*(\bfA,\uC)$ for (a).  (c) The KF-based feedback generator $\calG_T^*(\bfA,\uC)$ in state space.   
$(\bfA,\uL_{1,t},-\uC,0)$ with $\hatx_t$ denotes a state-space 
representation with $\hatx_t$ being its state at time $t$, and initial condition
$\hatx_0$ being 0.} \label{fig:estim}
\end{figure}

Fig. 
\ref{fig:estim} (a) shows the associated estimation problem, (b)  
the KF $\hatG_T^*(\bfA,\uC)$ for (a), and (c) the state-space representation of the optimal 
feedback generator $\calG_T^*(\bfA,\uC)$ (see (\ref{def:bba}) and (\ref{ric:recur}) for 
$\uL_{1,t}$ and $\uL_{2,t}$).

\begin{remark} \rm Proposition \ref{prop:kf} reveals that, the minimization of channel input
power in a feedback communication problem is equivalent to the 
minimization of MSE in an estimation problem.  This equivalence 
yields a complete characterization (in terms of the KF
algorithm) of the optimal feedback generator $\calG_T^*(\bfA,\uC)$ 
for any given $(\bfA,\uC)$, as shown in Section \ref{sub:kfroles}. This proposition refines the CP structure as it shows that the CP structure necessarily contains a KF.
\end{remark}

\begin{remark} \rm Proposition \ref{prop:kf} i) implies that 
we may reformulate the problem of $C_{T,n}$ (or $P_{T,n}$) as a 
two-step problem: STEP 1: Fix $(\bfA,\uC)$ (and hence fix the 
rate), and minimize the input power by searching over all possible feedback generator $\calG$ for the fixed $(\bfA,\uC)$; STEP 2: Search over all possible 
$(\bfA,\uC)$ subject to the rate constraint of $R_T(\bfA,\uC) \geq 
\calR$.  Thus, \emph{one essential role of the feedback generator 
$\calG$} for any fixed $(\bfA,\uC)$ is to minimize the input power, which can be solved by considering the \emph{equivalent} optimal estimation problem in 
Fig. \ref{fig:estim} (a) whose solution is the KF. It also follows that $\E (u_t|y^{t-1})=0$, which implies no power waste due to a non-zero mean (cf. \cite{kavcic_it07}, Eq. (126)) and a center-of-gravity encoding rule (cf. \cite{schalkwijk:center}). The input generated by the KF-based feedback generator has the form 
\be u_t = r_t - E(r_t|\uy^{t-1}), \ee
which is related to the optimal input distributions obtained by e.g. \cite{kavcic_it07,ihara-1988-1,ihara-1988-2,ihara:book}.
\end{remark}

We also remark that the necessity of the KF in the 
optimal coding scheme is not surprising, given various indications 
of the essential role of KF (or minimum mean 
squared-error estimators or MMSE estimators; or cheap 
control, its control theory equivalence; or the sum-product 
algorithm, its generalization) in optimal communication designs.  
See e.g. 
\cite{kschischang01,forney03,kavcic_it07,elia_c5,mitter:kalman,kim06}.  
The study of the KF in the feedback communication problem 
along the line of \cite{mitter:kalman} may shed important insights 
on optimal communication problems and is under current 
investigation.

\textbf{Proof: }  i) Notice that for any fixed $(\bfA,\uC)$, $R_T (\bfA,\uC)$ 
is fixed.  Then from the definition of $P_{T,n}(\calR)$, we have
\be \ba{llcll}P_{T,n}(\calR) &= &   \displaystyle \inf_{\bfA,\uC, 
\calG_T}
 & \displaystyle \frac{1}{T+1}\E \| \uu^T(\bfA,\uC, \calG_T) \|^2 \\
 & & ^{s.t. \: R_T(\bfA,\uC) \geq  \calR} &  \\
 &= &   \displaystyle \inf_{\bfA,\uC,(A1)}
 & \disp \inf_{(\bfA,\uC) \textnormal{ \small fixed},(A1),\calG_T}
 \displaystyle \frac{1}{T+1}\E \| \uu^T(\bfA,\uC, \calG_T) \|^2 . \\
 & & ^{s.t. \: R_T(\bfA,\uC) \geq  \calR}  &   \ea
\ee
Then i) follows from the definition of $\calG^*_T(\bfA,\uC)$.

ii) Note that for the coding structure $\bbS$, it holds that
\be \uu^T={\ur}^T +(-\underline{\hatr}^T) ={\ur}^T+\calG_T 
\uy^T.\label{eq:input}\ee
Then, letting
\be \hatG_T:= -\calG_T(\bfI-\bcalZ_T^{-1} \calG_T)^{-1} \ee
and $\ubary^T:= \bcalZ_T^{-1} {\ur}^T+\uN^T$, we have $\calG_T 
\uy^T= -\hatG_T \ubary^T$. Therefore,
\be \ba{lll} \calG_T^*(\bfA,\uC) &=&   \displaystyle \arg 
\inf_{\calG_T} \displaystyle \frac{1}{T+1} \E ({\ur}^T+\calG_T
\uy^T)({\ur}^T+\calG_T \uy^T)' \\
&=&   \displaystyle \arg \inf_{\hatG_T} \displaystyle  \frac{1}{T+1} 
\E ({\ur}^T-\hatG_T \ubary^T)({\ur}^T-\hatG_T \ubary^T)'. \ea   \ee
The last equality implies that the optimal solution $\hatG_T^*$ is 
the strictly causal MMSE estimator (with one-step prediction) of 
${\ur}^T$ given $\ubary_T$; notice that $\hatG_T$ is strictly lower 
triangular. It is well known that such an estimator can be 
implemented recursively in state-space as a KF (cf. 
\cite{book:kay1,kailath:book}).  Finally, from the relation between 
$\calG_T$ and $\hatG_T$, we obtain (\ref{eq:calGhatG}).  The 
state-space representation of $\calG_T^*(\bfA,\uC)$, as illustrated 
in Fig. \ref{fig:estim} (c), can be obtained from straightforward 
computation, as shown in Appendix \ref{appsub:represent}.
\endproof

We remark that it is possible to derive a dynamic programming based 
solution (\cite{tati:capI}) to compute $C_{T,n}$, and if we further 
employ the Markov property in \cite{kavcic_it07} and the above 
KF-based characterization, we would reach a solution with 
complexity $O(T)$ for computing $C_{T,n}$ and $C_T$.  However, we do 
not pursue along this line in this paper as it is beyond the main scope of this paper.

\section{Connections among feedback communication, estimation, and feedback control} \label{sec:dual}

We have shown that in the coding structure $\bbS$, to ensure power 
efficiency for a fixed $(\bfA,\uC)$, one needs to design a 
KF-based feedback generator. The KF 
immediately links the feedback communication problem to estimation 
and control problems.  In this section, we present a \emph{unified 
representation} of the \emph{optimal} coding structure $\bbS^*$ (i.e., $\bbS^*$ is $\bbS$ but with $\calG$ being 
chosen as $\calG^*(\bfA,\uC)$), its estimation theory counterpart, 
and its control theory counterpart. Then in the next section we will establish 
relation among the information theory quantities, estimation 
theory quantities, and control theory quantities.

\subsection{Unified representation of feedback coding system, KF,
and cheap control} \label{subsec:dual}

\textbf{Coding structure $\bbS^*$}

The optimal feedback generator for a given $(\bfA,\uC)$ is solved in 
(\ref{eq:calGhatG}), see Fig. \ref{fig:estim} (c) for its structure.  
We can then obtain a state-space representation of the optimal feedback generator $\calG^*_T(\bfA,\uC)$, and describe the coding structure $\bbS^*$ which contains
$\calG^*_T(\bfA,\uC)$ as
\be \textnormal{coding structure } \: \bbS^* 
    \left\{ \ba{ll}
    \ba{lll} \ux_{t+1} &=& \bfA \ux_t \\
            r_t &=& \uC' \: \ux_t\\
            u_t &=& r_t - \hatr_t
    \ea  & \left. \ba{l}\\ \\ \ea \right\} \textnormal{encoder}           \\
    \ba{lll}  \us_{t+1} &=& \bfF \us_t + \uG u_t \\
                y_t &=& \uH' \: \us_t + u_t +N_t
        \ea  & \left. \ba{l}\\  \ea \right\} \textnormal{channel }\calF \\
    \ba{lll}  \hats_{t+1} &=& \bfF \hats_{t} + \uL_{2,t} e_t \\
                e_t &=& y_t - \uH' \: \hats_t \\
                \hatx_{t+1} &=& \bfA \hatx_{t} + \uL_{1,t} e_t \\
                -\hatr_t &=& -\uC' \: \hatx_t
        \ea  & \left. \ba{l}\\ \\ \\   \ea \right\} \textnormal{optimal feedback generator }\calG^*(\bfA,\uC)
                \ea \right. \label{dyn:coding}\ee
with $\ux_0=\uW$ unknown, $s_0=\hats_0=\uzero$, and $\hatx_0=\uzero$.  Here $\uL_{1,t} \in \bbR ^{n+1}$ and 
$\uL_{2,t} \in \bbR^m$ are the time-varying KF gains 
specified in (\ref{eq:Lt}).   See 
Appendix \ref{appsec:equiv} for the derivation of a state-space 
representation of $\calG^*_T(\bfA,\uC)$.

\textbf{The estimation system}

The estimation system in Fig. \ref{fig:estim} (a) and (b) consists 
of three parts: the unknown source ${\ur}^T$ to be estimated or 
tracked, the channel $\calF$ (without output feedback), and the 
estimator, which we choose as the KF $\hatG^*$; we assume 
that $(\bfA,\uC)$ is fixed and known to the estimator and hence the randomness in $\ur^T$ comes from the initial condition of $\ur^T$.  The system 
is described in state-space as
\be \textnormal{estimation system:}
\left\{
\ba{ll}
        \ba{lll}    \ux_{t+1} &=& \bfA \ux_t \\
                        r_t &=& \uC ' \: \ux_t
        \ea  & \left. \ba{l}\\  \ea \right\} \textnormal{unknown source}      \\
        \ba{lll}    \ubars_{t+1}  &=& \bfF \ubars_t + \uG r_t \\
                        \ubary_t &=& \uH' \ubars_t + r_t + N_t
        \ea  & \left. \ba{l}\\  \ea \right\} \textnormal{channel }\calF \\
        \ba{lll}    \hatx_{t+1} &=& \bfA \hatx_t + \uL_{1,t} e_t \\
                        \hatr_t &=& \uC' \: \hatx_t\\
                        \hat{\ubars}_{t+1} &=&  \bfF \hat{\ubars}_{t} + \uG \hatr_t + \uL_{2,t} e_t\\
                        e_t &=& \ubary_t - \uH' \: \hat{\ubars}_{t}  -\hatr_t
        \ea  & \left. \ba{l}\\ \\ \\ \\  \ea \right\} \textnormal{Kalman filter }\hatG^*(\bfA,\uC)
\ea
\right. \label{dyn:est}
\ee
with $\ux_0=\uW$, $\ubars_0=\hat{\ubars}_0= \uzero$, and 
$\hat{\ux}_0=\uzero$.  To write this in a more compact form, define
\be \ba{lll} 
\overline{\bbX}_t &:=&\left[ \matrix{\ux_{t} \cr \underline{\bar{s}}_{t}} \right] \\
\widehat{\bbX}_t &:=&\left[ \matrix{\hat{\ux}_{t} \cr \underline{\hat{\bar{s}}}_{t} }\right] \\
\bbA &:=&\left[ \ba{c|c} \bfA & \bfzero \\ \hline \uG \uC' & \bfF  \ea \right] \\
\bbC &:=&\left[ \matrix{ \uC \cr \uH } \right] \\
\uL_t &:=&\left[ \matrix{\uL_{1,t} \cr \uL_{2,t}} \right] . \ea 
\label{def:bbXhat}\ee
Then we have 
\be \textnormal{estimation system:}
\left\{
\ba{ll}
        \ba{lll}    \overline{\bbX}_{t+1} &=& \bbA \overline{\bbX}_t \\
                        \ubary_t &=& \bbC ' \overline{\bbX}_t + N_t
        \ea  & \left. \ba{l}\\  \ea \right\} \textnormal{unknown source and channel }\calF      \\
        \ba{lll}    \widehat{\bbX}_{t+1} &=& \bbA \widehat{\bbX}_t + \uL_{t} e_t \\
                        e_t &=& \ubary_t - \bbC \widehat{\bbX}_t.
        \ea  & \left. \ba{l}\\   \ea \right\} \textnormal{Kalman filter }\hatG^*(\bfA,\uC)
\ea
\right. \label{dyn:est_compact}
\ee
with $\overline{\bbX}_0=[\uW',\uzero']'$ and $\widehat{\bbX}_0= \uzero$.

It can be easily shown 
that $r_t$, $\hatr_t$, $e_t$, $\ux_t$, and $\hatx_t$ in 
(\ref{dyn:est}) and (\ref{dyn:coding}) are equal, respectively, and 
it holds that for any $t$,
\be \us_t - \hats_t = \ubars_t - \hat{\ubars}_t, \ee
which leads to the following unified representation as a control system.

\textbf{The unified representation: A cheap control problem}

\begin{figure}[h!]
\begin{center}
\subfigure[]
{\scalebox{.6}{\includegraphics{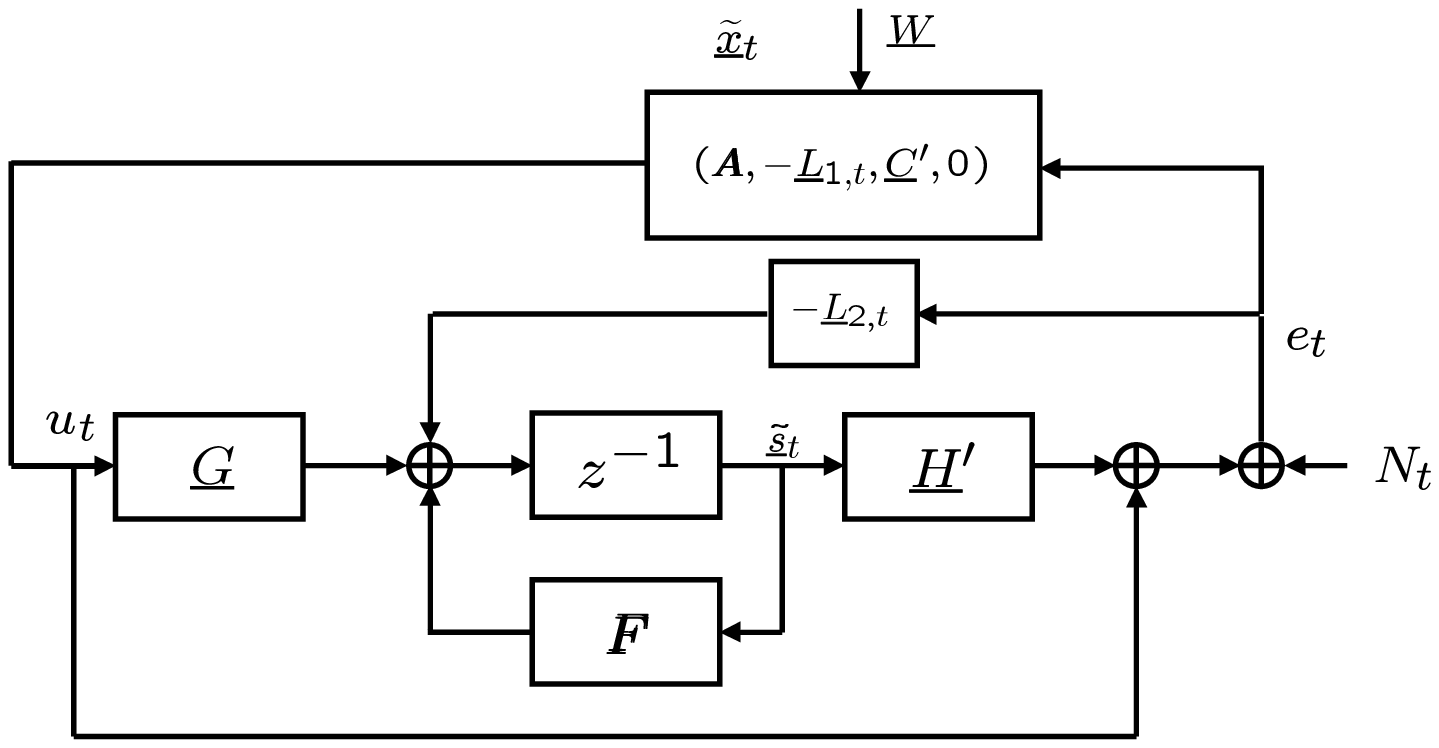}}}
\subfigure[]
{\scalebox{.5}{\includegraphics{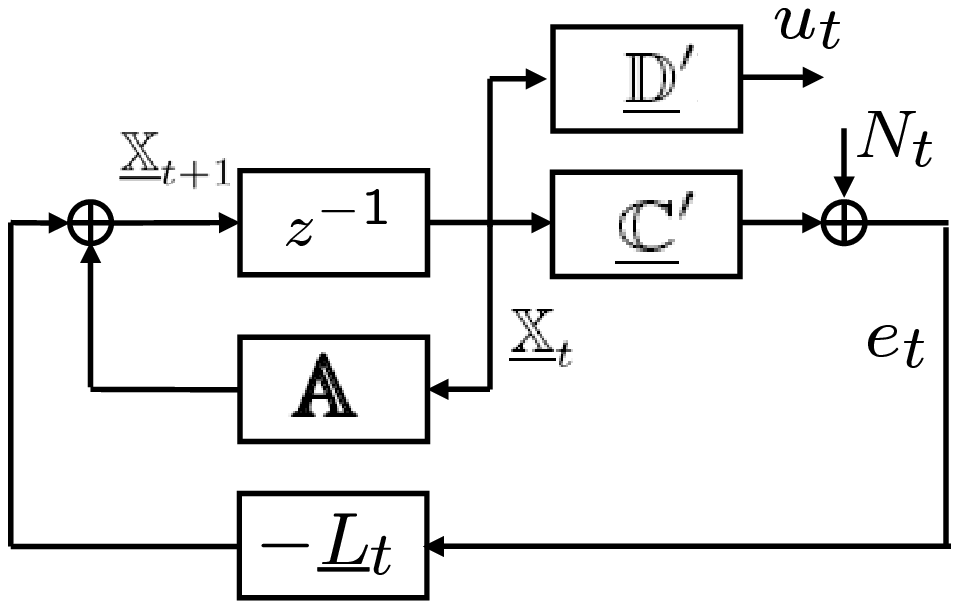}}}
 \caption{Two equivalent 
block diagrams for the cheap control system. In (a) the block 
$(\bfA,-\uL_{1,t},\uC,0)$ denotes the state-space representation 
with $\tilde{x}_t$ and $\uW$ being its states at time $t$ and at time 
0, respectively.} \label{fig:menergy}
\end{center}
\end{figure}

Define
\be \ba{lll} \tildex_t &:=& \ux_t - \hatx_t \\
\tildes_t &: =& \us_t - \hats_t = \ubars_t - \hat{\ubars}_t\\
\bbX_t &:=&\left[ \matrix{\tildex_{t} \cr \tildes_{t}} \right] = \overline{\bbX}_t - \widehat{\bbX}_t\\
\bbX_0 &:=&\left[ \matrix{\uW \cr \uzero} \right]\\
\bbD &:=&\left[ \matrix{ \uC \cr \uzero } \right] . \ea 
\label{def:bba}\ee
Note that $\bbX_t$ is the estimation error for $\bar{\bbX}_t$.  
Substituting (\ref{def:bba}) to (\ref{dyn:est}) and 
(\ref{dyn:coding}), we obtain that both systems become
\be \textnormal{control system:}\left\{ \ba{llll} \bbX_{t+1} &=& (\bbA - \uL_t \bbC') \bbX_t - \uL_t N_t=\bbA \bbX_t-\uL_t e_t & \textnormal{(state evolution)}\\
 e_t &=& \bbC' \: \bbX_t+N_t  & \textnormal{(noisy measurement)}\\
 u_t &=& \bbD' \: \bbX_t & \textnormal{(regulated output)} \ea \right. \label{dyn}\ee
See Fig. \ref{fig:menergy} for the block diagrams.  It is a control 
system where we want to minimize the power of the regulated output $u$ by appropriately 
choosing $\uL_t$.  More specifically, one may view $e_t$ as the noisy measurement which is also the input to the controller, $(-\uL_t)$ as the time-varying controller gain, $(-\uL_t e_t)$ as the controller's output which is also the input to the system with state $\bbX_t$.  The objective is to minimize $\E  \| \uu^T \|^2$; more formally we want to solve
\be \ba{lll} \displaystyle P_{T,n}(\bfA,\uC):=& 
\displaystyle \min _{\uL_0,\cdots,\uL_t} \frac{1}{T+1}\E \| \uu^T \|^2\\
&  ^{s.t. \:  (\ref{dyn}) } \ea \label{def:cheapcontrol}
\vspace{-10pt} \ee
in which $\bfA, \uC$ are given and $\uW$ unknown.  Note that the control effort is ``free'' as there is no direct penalty on the controller's output $(-\uL_t e_t)$.   This is a \emph{cheap control} problem, 
which is useful for us to characterize the steady-state solution and 
it is equivalent to the KF problem (see 
\cite{mincontrol:book}; also see \cite{liu:markov} for the discussion of cheap control and the closely related expensive control and minimum-energy control).

The signal $e_t$ in (\ref{dyn}) is the \emph{KF 
innovation} or simply \emph{innovation} \footnote{The innovation defined 
here is consistent with the Kalman filtering literature but different from that defined in 
\cite{cover-pombra-1989} or \cite{kavcic_it07}.}. One fact is that $\{e_t\}$ is 
a white process, that is, its covariance matrix $\bfK_{\ue}^{(T)}$ 
is a diagonal matrix. Another fact is that $\ue^T$ and $\uy^T$ 
determine each other causally, and we can easily verify that 
$h(\ue^T)=h(\uy^T)$ and $\det \bfK_{\uy}^{(T)}= \det 
\bfK_{\ue}^{(T)} $.  We remark that (\ref{dyn}) is the 
\emph{innovations representation} of the KF (cf. 
\cite{kailath:book}).

For each $t$, the optimal $\uL_t$ is determined as
\be \uL_t : = \left[ \ba{l}\uL_{1,t} \\ \uL_{2,t} \ea \right] := 
\frac{\bbA \Si_t \bbC}{ K_{e,t}}, \label{eq:Lt}\ee
where $\Si_t:= \E \bbX_t \bbX_t'$, $K_{e,t} := \E (e_t)^2  = \bbC' 
\Si_t \bbC+1$, and the error covariance matrix $\Si_t$ satisfies the 
Riccati recursion
\be \Si_{t+1} = \bbA \Si_t \bbA' - \frac{\bbA \Si_t \bbC \: \bbC' 
\Si_t \bbA' }{\bbC' \Si_t \bbC  +1} \label{ric:recur} \ee
with initial condition
\be \Si_0 :=\left[ \matrix{ \bfI_{n+1} & \bfzero \cr \bfzero & 
\bfzero } \right] , \label{ric:init}\ee
This completes the description of the optimal feedback generator for 
a given $(\bfA,\uC)$.

The existence of one unified expression for three different systems 
(\ref{dyn:est}), (\ref{dyn:coding}), and (\ref{dyn}) is because the 
first two are actually two different non-minimal realizations of the 
third.  The input-output mappings from $\uN^T$ to $\ue^T$ in the 
three systems are $T$-equivalent (see Appendix \ref{appsub:equiv}). 
Thus we say that the three problems, the optimal estimation problem, 
the optimal feedback generator problem, and the cheap 
control problem, are \emph{equivalent} in the sense that, if any one 
of the problems is solved, then the other two are solved. Since the 
estimation problem and the control problem are well studied, the 
equivalence can sometimes facilitate our study of the communication problem. 
Particularly, the formulation (\ref{dyn}) yields alternative 
expressions for the mutual information and average channel input 
power in the feedback communication problem, as we see in the next 
section.

\begin{figure}[h!]
\center{{\scalebox{.5}{\includegraphics{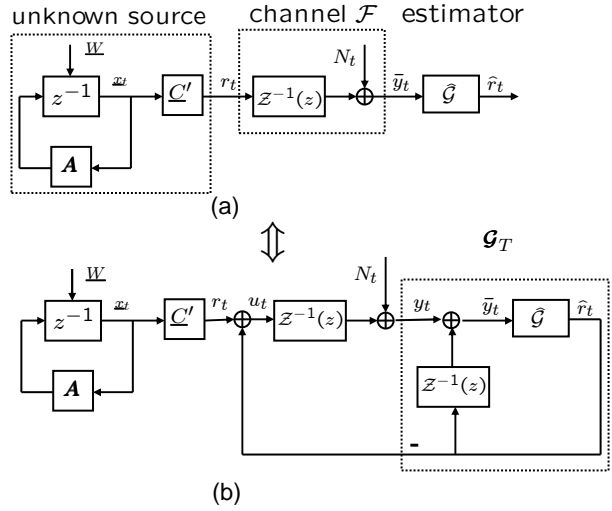}}}} 
\caption{ Relation between the estimation problem (a) and the 
communication problem (b).} \label{fig:dual}
\end{figure}

We further illustrate the relation of the estimation system and the 
communication system in Fig. \ref{fig:dual}, in which (b) is obtained from 
(a) by subtracting $\hatr_t$ from the channel input and adding 
$(\bcalZ_T^{-1} \hatr_t)$ back to the channel output, which does not 
affect the input, state, and output of $\hatG_T^*$.  It is clearly 
seen from the block diagram manipulations that \emph{the minimization of 
channel input power in feedback communication problem becomes the 
minimization of MSE in the estimation problem}.  This generalizes the observation we made regarding how to obtain a coding structure from a KF over an AWGN channel (as shown in Fig. \ref{fig:awgn:kf}) to more general Guassian channels. 

\subsection{Roles of the KF algorithm in feedback communication} \label{sub:kfroles}

We have seen that the KF algorithm is necessary to ensure the power efficiency in feedback communication.  Here we show that it is also needed to recover the transmitted signal $\ux_0:=\uW$.

The estimation of $\ux_0$ is a (an anti-causal) smoothing problem; more specifically, a fixed-point smoothing problem (cf. e.g. Ch. 10 of \cite{kailath:book}), whose solution is typically easily obtained by studying the innovations process of the KF used for prediction.  Note that $\overline{\bbX}_0:=[\ux_0 ',\us_0 ']:=[\uW',\uzero']'$, and hence the smoothed estimate for $\ux_0$ can be obtained by the smoothed estimate of $\bbX_0$ (the constraint that $\us_0:=\underline{0}$ should be automatically satisfied in the smoothing problem solution).  Denote $\widehat{ \bbX }_{0|t}:= \E(\overline{\bbX}_0|\uy^{t})$ and $\hat{ \ux }_{0|t}:= \E(\uW|\uy^{t})$.  The solution is given below.  Denote the closed-loop state transition matrices as $\pmb{\Phi} (t) := \bbA_{cl}(t-1)\bbA_{cl}(t-2)\cdots \bbA_{cl}(0)$ if $t>0$ and $\pmb{\Phi} (0) := \bfI$, where $\bbA_{cl}(t):=\bbA-\uL_t \bbC'$, and $\pmb{\phi} (t) := \bfA_{cl}(t-1)\bfA_{cl}(t-2)\cdots \bfA_{cl}(0)$ if $t>0$ and $\pmb{\phi} (0) := \bfI$, where $\bfA_{cl}(t):=\bfA-\uL_{1,t} \uC'$. (It holds that $\pmb{\phi} (t)$ is the upper left block of $\pmb{\Phi} (t)$.)  Then the smoothing equations are (see Problem 10.1 in \cite{kailath:book})
\be \ba{lll} \widehat{ \bbX }_{0,t} &=&  \widehat{ \bbX }_{0,t-1} + \pmb{\Sigma}_0 \pmb{\Phi}'(t) \bbC K_{e,t}^{-1} e_t \\
\hat{ \ux }_{0,t} &= & \hat{ \ux }_{0,t-1} + \pmb{\phi}'(t) \uC K_{e,t}^{-1} e_t, \ea \ee
which are based on the KF innovations.

The smoothed filter, in our special case of no process noise, can be alternatively obtained simply by invoking the invariance property of the MMSE estimation, if $\det \bfA \neq 0$ (as done in \cite{elia_c5}). To see this, notice that $\hatx_{t+1}$ is the MMSE estimate of $\ux_{t+1}$ with one-step prediction, i.e. $\hatx_{t+1} = \E(\ux_{t+1} | \uy^{t}) $.  Since $\ux_{t+1} = \bfA^{t+1} \uW$, it holds that 
\be \hat{ \ux }_{0,t} = \bfA^{-t-1} \hatx_{t+1} = \hat{ \ux }_{0,t-1} + \bfA^{-t-1} \uL_{1,t} e_t. \ee
The last equality, which specifies a recursive way to generate the smoothed estimate, is again based on the KF innovations \footnote{However, numerical problems may arise if $\bfA$ contains stable eigenvalues for large $t$.}. Similar equation holds for estimating $\overline{\bbX}_0$.  A by-product of the above reasoning is the following identities valid when $\det A \neq 0$:
\be \ba{lll} \uL_{t}=\bbA^{t+1}\pmb{\Phi}'(t) \bbC / K_{e,t} \\
\uL_{1,t}=\bfA^{t+1}\pmb{\phi}'(t) \uC / K_{e,t} . \ea   \ee

The estimation MSE error may be given by the following equations:
\be \ba{lll} \MMSE_{\uW,t} &:=& \E (\uW - \E(\uW|\uy^t))(\uW - \E(\uW|\uy^t))'\\
&=& \MMSE_{\uW,t-1} -  \pmb{\phi}'(t) \uC K_{e,t}^{-1} \uC'  \pmb{\phi}(t) \\ &=& \MMSE_{\uW,t-1} - K_{e,t} \bfA^{-t-1} \uL_{1,t} \uL_{1,t}' \bfA^{-t-1}{}' \\ &=& \bfA^{-t-1} \pmb{\Sigma}_{\ux,t+1} \bfA^{-t-1}{}', \ea   \ee
where the last equality hold only if $\bfA$ is invertible, and $\pmb{\Sigma}_{\ux,t+1}$ is the upper left $(n+1) \times (n+1)$ block of $\pmb{\Sigma}_{t+1}$.

\begin{remark} \rm We now have the \emph{complete characterization of the roles of KF algorithm in feedback communication}.  
The KF of an unknown process driven by its initial 
condition and observed through a Gaussian channel with memory, when reformulated in an appropriate form, is optimal in transmitting
information with feedback. The \emph{power efficiency} 
(i.e. the minimization of the channel input power) in communication 
is guaranteed by the strictly causal one-step prediction operation in Kalman 
filtering (i.e. the operation to generate $\E (r_t|\uy^{t-1})$ at time $t$); and 
the \emph{optimal recovery of the transmitted codeword} ( optimal in 
the MMSE sense) is guaranteed by the anti-causal smoothing operation in Kalman 
filtering (i.e. the operation to generate $\E (\ux_0|\uy^{t-1})$).  We 
may view this characterization as the optimality of KF 
in the sense of \emph{information transmission with feedback}, which 
is a complement to the existing characterization that KF 
is optimal in the sense of \emph{information processing} established 
by Mitter and Newton in \cite{mitter:kalman}.   It is also 
interesting to note that, though for different classes of channels, 
different optimal coding schemes have been derived along different 
directions, these schemes can be universally interpreted in terms of 
KF of appropriate forms; see \cite{liu:phd}.  
Thus, we consider that the KF acts as a ``unifier" for 
feedback communication schemes over various channels.  
\end{remark}

Finally, our study on the coding structure $\bbS$ also refines the CP 
structure.  Indeed, we conclude that the 
CP structure needs to have a KF inside.  We may further 
determine the optimal form of $\calB_T$. From (\ref{eq:g2b}) and 
(\ref{eq:calGhatG}), we have that
\be \calB_T^* = - \hatG_T^*(\bfA,\uC) \bcalZ_T^{-1} , \ee
where $\hatG_T^*(\bfA,\uC)$ is the KF given in (\ref{dyn:est}).
Therefore, to achieve $C_T$ in the CP structure, it is sufficient to 
search $(\bfK_{\uv}^{(T)},\calB_T)$ in the form of
\be \ba{lll} \bfK_v ^{(T)}&:=& (\bfI- \hatG_T^*(\bfA,\uC) \bcalZ_T^{-1} ) \bfGa_T(\bfA,\uC) \bfGa_T(\bfA,\uC)'  (\bfI- \hatG_T^*(\bfA,\uC) \bcalZ_T^{-1} ) '\\
\calB_T^* &:=& - \hatG_T^*(\bfA,\uC) \bcalZ_T^{-1} . \ea \ee

\section{Connections of fundamental limitations} \label{subsec:crb}
In this section, we discuss the connections of fundamental limitations.  These limitations involve the mutual information in the feedback communication system, the Fisher information, MMSE, and
CRB in the estimation system, and the Bode sensitivity integral in the feedback control system. We show that one limitation may be expressed in terms of the others, as a consequence of the equivalence established above.

\subsection{Fisher information matrix (FIM), CRB, and Bode-type sensitivity integral (sum)}
\label{subsec:fimcrbbi}
Let us first recall the general definitions of MMSE, Fisher 
information matrix (FIM), and CRB:
\be \MMSE_{\uW} := \E (\uW - \hatW) (\uW - \hatW)', \ee
where $ \hatW:= \E(\uW | \uy) $ is the MMSE estimator of $\uW$ based 
on noisy observation $\uy$;
\be \ba{lll} \calI_{\uW} &:=&  \disp \E \left( \frac{\partial \log
p_{\uW,\uy}(\uW,\uy)}{\partial \uW} \right)^2 \\
& =& \disp -\E \left( \frac{\partial ^2 \log 
p_{\uW,\uy}(\uW,\uy)}{\partial \uW^2} \right) \ea \ee
to be the (Bayesian) FIM, where $p_{\uW,\uy}(\uW,\uy)$ is the joint 
density of $\uW$ and $\uy$; and
\be \CRB_{\uW}:=\calI_{\uW}^{-1}  \ee
to be the (Bayesian) CRB~\cite{vantrees}.  Note that it always 
holds, as a \emph{fundamental limitation in estimation theory}, that
\be \MSE_{\uW} \geq \CRB_{\uW},\ee
regardless of how one designs the estimator \cite{vantrees}.  This 
inequality is referred to as the information inequality, Cramer-Rao 
inequality, or van Trees inequality \footnote{Some authors 
distinguish the Cramer-Rao inequality and van Trees inequality by 
restricting the former to be non-Bayesian and unbiased and the 
latter to be Bayesian and possibly biased.}.

The Bode sensitivity integral is a \emph{fundamental limitation in feedback control} (typically in steady state). Simply put, for any feedback design, the sensitivity of the output to exogenous disturbance cannot be made small uniformly over all frequencies since the sensitivity transfer function's power spectrum in log scale sums up (integrates) to be constant.  See Section \ref{sub:steady_eq} and \cite{fundamental_filter_control}.  A similar limitation holds in finite horizon as we now show.

As
\be \uy^T = (\bfI-\bcalZ_T^{-1} \calG_T) ^{-1} (\bcalZ_T^{-1} \ur^T 
+ \uN^T), \ee
the sensitivity of channel output $\uy^T$ to noise $\uN^T$ is $\mathcal{S}_T:=(\bfI-\bcalZ_T^{-1} \calG_T)^{-1}$.  It is then easily seen that, if the spectrum of $\mathcal{S}_T \mathcal{S}_T'$ is $\{\lambda_i\}_{i=1}^{T+1}$, then
\be \sum_{i=1}^{T+1} \log \lambda_i  = 0, \ee
which holds valid regardless of the choice of feedback generator $\calG_T$, including the case that there is no feedback (i.e. open loop).  Thus, the effect of noise $\uN^T$ cannot be made arbitrarily small in the measurements $\uy^T$, which may be viewed as a fundamental limitation of noise (or disturbance) suppression.

Since the noise $\uN^T$ is normalized, one may also define the sensitivity based on the spectrum of $\bfK_{\uy}^{(T)}$ or on the innovation process variance $K_{e,t}$
.  Let
\be \BI_T:=\frac{1}{2}\sum_{i=1}^{T+1} \log \lambda_i(\bfK_{\uy}^{(T)})  =  \sum _{t=0}^T \log  K_{e,t}, \label{def:BI} \ee
which is easily seen independent of any causal feedback and is the finite-horizon counterpart of the widely known Bode sensitivity integral of infinite-horizon.

\subsection{Expressions for mutual information and channel input power}

We have the following proposition.

\begin{prop} \label{prop:dual}

Consider the coding structure $\bbS$.  
For any fixed $0\leq n \leq T$ and observable $(\bfA,\uC')$ with 
$\bfA \in \bbR^{(n+1) \times (n+1)}$, it holds that

i) \be \ba{lll}I(\uW;\uy^T)& =& \BI_T
        = \frac{1}{2} \sum _{t=0}^T \log  K_{e,t}\\
& =&  \disp \frac{1}{2} \sum _{t=0}^T \log  (\bbC' \Si_t \bbC+1 )\\  
\vspace{2pt} & =&  \disp \frac{1}{2} \log \det \MMSE_{\uW,T}^{-1} \\  
\vspace{2pt} & =&  \disp \frac{1}{2} \log \det \calI_{\uW,T} \\   
&=&  \disp \frac{1}{2} \log \det \textnormal{CRB}_{\uW,T} ^{-1}; \ea 
\ee

ii) \be \ba{lll}P_{T,n}(\bfA,\uC) &=&  \disp \frac{1}{T+1} \sum 
_{t=0} ^T \bbD'
\Si_t \bbD \\
&=&  \disp \frac{1}{T+1} \trace (\PMMSE_{\ur,T})\\
&=&  \disp \frac{1}{T+1} \sum _{t=0}^T \uC' \bfA^t \MMSE_{\uW,t} 
\bfA^t{}'\uC , \ea \ee
where $\MMSE_{\uW,T}$ is the minimum MSE of $\uW$ at time $T$, 
$\CMMSE_{\ur,T}$ is the causal minimum MSE of ${\ur}^T$ at time $T$, 
$\calI_{\uW,T}$ is the Bayesian Fisher information matrix of $\uW$ 
at time $T$ for the estimation system (\ref{dyn:est}), and 
$\textnormal{CRB}_{\uW,T}$ is the Bayesian CRB of $\uW$ at time $T$.

\end{prop}

Note that $\PMMSE_{\ur,T}:=\E (\hat{\ur}^T - \ur^T)(\hat{\ur}^T - \ur^T)'$, in which $\hat{\ur}^T = [\hat{r}_0,\cdots,\hat{r}_T]$ contains the (strictly causal) estimates with one-step prediction $\hat{r}_t:=\uC' \E(\ux_t|\uy^{t-1})$ for $t=0,\cdots,T$.

\begin{remark} \rm This proposition connects the mutual information to the Bode sensitivity integral of the associated control problem and to the innovations process, Fisher information, (minimum) MSE, and CRB of the associated 
estimation problem.  Note that any mutual information larger than the value given above is not possible regardless of how one designs the feedback generator, and how much mutual information we may obtain is limited by the control problem fundamental limitations and by how
well the estimation can be done and hence by the Fisher
information, MMSE, and CRB.  Thus the fundamental limitation in feedback communication is linked to 
the fundamental limitations in control and estimation.
\end{remark}

This proposition also shows that the spectrum of the output covariance matrix or the innovation variance cannot be made large or small uniformly, which may be viewed as the finite-horizon, time domain counterpart of the Bode sensitivity integral in the steady state and frequency domain.  Notice that so far the estimation problem and control problem do not rely on asymptotic notions such as stability (stability was used to establish the Bode-Shannon connections between feedback communication and feedback stabilization in steady state \cite{elia_c5}).

As a side note, if one defines the complementary sensitivity as $\mathcal{T}_T:=\bcalZ_T^{-1} \calG_T(\bfI-\bcalZ_T^{-1} \calG_T)^{-1}$, it still holds that $\mathcal{S}_T-\mathcal{T}_T = \bfI$, which resembles the fundamental algebraic tradeoff in the steady state and frequency domain (cf. \cite{fundamental_filter_control}).

\textbf{Proof: } i) First we simply notice that $h(\uy^T)=h(\ue^T)$, and 
$K_{e,t} = \bbC' \Si_t \bbC+1$.  Next, to find MMSE of $\uW$, note 
that in Fig. \ref{fig:estim} (a)
\be \ubary^T = \bcalZ_T^{-1} \bfGa_T \uW+\uN^T \ee
and that $\uW \sim \calN(\uzero,\bfI)$, $\uN^T \sim 
\calN(\uzero,\bfI)$.  Thus, by \cite{book:kay1} we have
\be \MMSE_{\uW,t} = (\bfI + \bfGa_T' \bcalZ_T^{-1} {}' \bcalZ_T^{-1} 
\bfGa_T)^{-1}  = \bfI - \bfGa_T' ( \bcalZ_T \bcalZ_T ' + \bfGa_T' \bfGa_T )^{-1} 
\bfGa_T, \ee
yielding
\be \ba{lll} \det \MMSE_{\uW,t} &=& \det (\bfI + \bcalZ_T^{-1} \bfGa_T
\bfGa_T' \bcalZ_T^{-1} {}' )^{-1} \\
&=&\det (\bfI + \bcalZ_T^{-1} \bfK_{\ur}^{(T)}\bcalZ_T^{-1} {}' 
)^{-1} = \det (\bfK_{\ur}^{(T)} + \bfK_{\uZ}^{(T)} )^{-1}. \ea \ee
Besides, from Section 2.4 in \cite{vantrees} we can directly compute 
the FIM of $\uW$ to be $(\bfI + \bfGa_T' \bcalZ_T^{-1} {}' 
\bcalZ_T^{-1} \bfGa_T)$.  Then i) follows from Proposition 
\ref{prop:I} and (\ref{dyn}).

ii) Since $u_t=\bbD' \: \bbX_t = \uC' \: \tildex_t =r_t-\hatr_t$ and 
$\E \tilde{x}_t \tilde{x}_t' =\bfA^t \MMSE_{\uW,t} \bfA^t{}'$, we 
have $\E (u_t)^2=\bbD ' \Si_t \bbD= \uC' \E \tildex_t \tildex_t' \uC 
= \E (r_t-\hatr_t)^2$, and then ii) follows. \endproof

\subsection{Connections of the fundamental tradeoffs}  \label{sub:tradeoff}

The above fundamental limitations are based on \emph{one fixed} $(\bfA,\uC)$ with $\bfA \in 
\mathbb{R}^{n \times n}$.  Searching over \emph{all} admissible $(\bfA,\uC)$ with $\bfA \in 
\mathbb{R}^{n \times n}$ for all $n\leq T$, one can obtain the optimal tradeoffs for feedback communication, estimation, and feedback control, as well as the corresponding relation among these tradeoffs. Note that the linear scheme with $(\bfA,\uC)$ can attain the optimal tradeoffs as we have established in the feedback communication system (see Proposition \ref{prop:kf}), and hence the optimal tradeoffs obtained by searching over all admissible $(\bfA,\uC)$ are indeed the optimal tradeoffs over all (possibly nonlinear, provided relevant quantities are well defined) feedback communication designs, estimator designs, and feedback control designs.
These fundamental tradeoffs are elaborated below.

The fundamental tradeoff in the feedback communication problem over the channel $\calF$ for finite-horizon from time 0 to time $T$ is the capacity $C_{T,T}(\calP)$ (or $P_{T,T}(\calR)$, see Definition \ref{def:CTn}) in the form of the optimal power-rate pair. (As indicated by Proposition \ref{prop:kf}, searching over all admissible $(\bfA,\uC)$ achieves the capacity.)
That is, we have:

\noindent \textbf{(T1) Optimal Feedback Communication Tradeoff}: Given the channel $\calF$ with one-step delayed output feedback and an average channel input power $\calP$, the achievable information rate $R_T(\pmb{f},\calP)$ cannot be higher than a constant $C_{T,T}(\calP)$ for any feedback communication design $\pmb{f}$; here $R_T(\pmb{f},\calP):=\frac{1}{T+1} I\left(\uu^T(\pmb{f}) \rightarrow \uy^T(\pmb{f}) \right)$ is the information rate with feedback design $\pmb{f}$ such that $\frac{1}{T+1} \E \| \uu^T(\pmb{f}) \|^2 \leq \calP $. 

\noindent Alternatively

\noindent \textbf{(T1') Optimal Feedback Communication Tradeoff}: Given the channel $\calF$ with one-step delayed output feedback and an information rate $\calR$, the achieable average channel input power $P_T(\pmb{f},\calR)$ cannot be lower than a constant $P_{T,T}(\calR)$ for any feedback communication design; here $P_T(\pmb{f},\calR):=\frac{1}{T+1} \E \| \uu^T(\pmb{f}) \|^2$ is the average channel input power with feedback design $\pmb{f}$ such that $\frac{1}{T+1} I\left(\uu^T(\pmb{f}) \rightarrow \uy^T(\pmb{f}) \right) \geq \calR $.

Note that the average input power depends on the strictly causal feedback from the channel output; the information rate, however, is independent of the causal feedback, may be achieved by anti-causally processing the channel outputs $\uy^T$, and hence can be used as a measure of anti-causality of the system.

A fundamental tradeoff for the estimation problem over the channel $\calF$ is the causal estimation performance versus anti-causal estimation performance.  Assume a process $\ur^T$ is passed through the channel $\calF$ and generates measurements $\ubary^T$.  Let $\uW:=\mathbf{R}^{-1}\ur^T$, where $\mathbf{R}:=\left(\bfK_{\ur}^{(T)}\right)^{1/2}$ if $\bfK_{\ur}^{(T)}$ is of full rank; otherwise $\mathbf{R}$ is such that $\bfK_{\uW}$ is of full rank with rank$(\bfK_{\uW})=$rank$(\bfK_{\ur}^{(T)})$ and $\bfK_{\uW}=\bfI$.  That is, $\uW$ may be viewed as the to-be-estimated, normalized signal that completely determines the process $\ur^T$.  Therefore we have a linear model $\ubary^T = \bcalZ_T^{-1} \mathbf{R} \uW+\uN^T$.  Again one can define innovation as $e_t:=\bar{y}_t-\E(\bar{y}_t|\ubary^{t-1})$ for each $t$.

\noindent \textbf{(T2) Optimal Estimation Tradeoff}: Given the channel $\calF$ and the time-averaged one-step prediction MMSE 
\be \PMMSE_r:=\frac{1}{T+1}\sum_{t=0}^T\left(r_t-\E(r_t|\ubary^{t-1})\right)^2, \ee
the decay rate of the anti-causal, smoothing MMSE 
\be \frac{1}{2(T+1)} \log \det \MMSE_{\uW}^{-1}=-\frac{1}{2(T+1)} \log \det \E\left(\uW-\E(\uW|\ubary^{T})\right)\left(\uW-\E(\uW|\ubary^{T})\right)' \ee
cannot be larger than a constant, and the average of innovations variance in log scale $\frac{1}{2(T+1)} \sum_{t=0}^T \log K_{e,t}$ cannot be larger than a constant, for any one-step predictor design and smoother design.

\noindent Alternatively

\noindent \textbf{(T2') Optimal Estimation Tradeoff}: Given the channel $\calF$ and the decay rate of the anti-causal, smoothing MMSE $\frac{1}{2(T+1)} \log \det \MMSE_{\uW}^{-1}$ (or the average of innovations variance in log scale $\frac{1}{2(T+1)} \sum_{t=0}^T \log K_{e,t}$), the time-averaged one-step prediction MMSE $ \PMMSE_r$ cannot be smaller than a constant, for any one-step predictor design and smoother design.

Note that the prediction MMSE depends on causality, while the smoothing MMSE is anti-causal and independent of the causal processing (if any) done by the estimator.  That is, this tradeoff is concerned with prediction versus smoothing tradeoff, or more fundamentally, the causality versus anti-causality tradeoff.

A fundamental tradeoff for the cheap control problem over the channel $\calF$ is the control performance (regulated output variance, in this case the variance of the channel input signal) versus the Bode integral (or the disturbance rejection measure, degree of anti-causality, as defined in (\ref{def:BI})). View the channel input $u_t(\pmb{f})$ as the regulated output with control design $\pmb{f}$, $y_t(\pmb{f})$ be the associated channel output, and
\be \BI_T(\pmb{f}):=\frac{1}{2}\sum_{i=1}^{T+1} \log \lambda_i(\bfK_{\uy}^{(T)}(\pmb{f}))  . \ee

\noindent \textbf{(T3) Optimal Feedback Control Tradeoff}: Given the channel $\calF$ and the average regulated output variance $\frac{1}{T+1} \sum_{t=0}^T \E (u_t(\pmb{f}))^2$, the Bode integral $\BI_T(\pmb{f})$ cannot be larger than a constant for any control design $\pmb{f}$.

\noindent Alternatively

\noindent \textbf{(T3') Optimal Feedback Control Tradeoff}: Given the channel $\calF$ and the Bode integral,  the average regulated output variance cannot be smaller than a constant for any control design $\pmb{f}$.

Note that this specifies the relation between the control performance achievable via causal feedback and the anti-causality of the system (that is, the Bode sensitivity integral or disturbance rejection measure which is independent of causal feedback).

To summarize, we have seen that \emph{all three tradeoffs are essentially the fundamental tradeoff between causality and anti-causality}, which manifests itself in the three different but closely related problems. The causal entities, e.g. the channel inputs in feedback communication, one-step prediction in estimation, and regulated output in control, are closed-loop entities generated in a causal, progressive way by the causal feedback, and hence vary as the causal feedback varies.  On the other hand, the anti-causal entities, e.g. the information rate (and the decoded message) in communication, the smoothed estimate in estimation, and the BI in control, are invariant regardless of whether the systems are in open-loop or closed-loop or how the closed-loop is done.    It is worth noting the various discussions involving causal versus anti-causal operations and filtering versus smoothing in the literature; see \cite{vantrees, guo:it05} and therein references.

In contrast, the power versus rate tradeoff in communication problems without output feedback cannot be interpreted as causality versus anti-causality tradeoff, nor can the tradeoff in the corresponding estimation problems.  To see this, we again assume the linear Gaussian model $\ubary^T = \bcalZ_T^{-1} \mathbf{R} \uW+\uN^T$.  One can see that the channel input power is related to the unknown's prior covariance (i.e. covariance matrix of channel input $\mathbf{R} \uW$), whereas the mutual information $I(\mathbf{W};\ubary^T)=\frac{1}{2} \log \det \MMSE_{\uW}^{-1}$ is related to the posterior covariance (cf. Theorem 10.3, \cite{book:kay1}).  Thus, in communication without output feedback, the power versus rate tradeoff may be translated into the tradeoff between the unknown's prior covariance and posterior covariance (or more generally the tradeoff between the unknown's prior and posterior distributions). %
  Note it is easily verified that these two tradeoffs coincide in the AWGN channel case as one might expect.%

\section{Necessary conditions for the optimality of the finite-horizon coding structure $\bbS$} 
\label{sec:structureproperty}

We discuss in this 
section a few useful properties of the coding structure $\bbS^*$ with 
the optimal feedback generator. The first two properties, i.e., the 
orthogonality between future channel inputs and previous channel 
outputs, and the Gauss-Markov property of the transformed channel 
outputs, are direct consequences of the KF.  Naturally, 
they can be viewed as necessary conditions for optimality of the 
feedback communication scheme as we have proven the necessity of 
the KF for optimality.  The third property, the 
finite-dimensionality of the optimizing $\ur^T$, yet another 
necessary condition for optimality, is a joint consequence of the 
KF structure and the waterfilling requirement for 
optimality for the finite-dimensional channel $\calF$. Finally, we show that the MMSE one-step predictor is necessary for achieving the feedback capacity of general additive channels with an average power constraint, followed by an extension of the orthogonality property over such channels.

\subsection{Necessary condition for optimality: Orthogonality condition}

First, we show that the coding structure $\bbS^*$ satisfies a necessary condition for optimality 
discussed in \cite{kim04} \footnote{This was later referred to as 
the orthogonality condition in \cite{kim06}, based on which a Kalman 
filter structure is identified.  It was also discussed in 
\cite{liu:allerton,kavcic_it07}.}.  The condition says that, the channel input 
$u_t$ needs to be orthogonal to the past channel outputs 
$\uy^{t-1}$. This is intuitive since to ensure the fastest transmission, 
the transmitter should not (re-)transmit any information that the 
receiver has already obtained, thus the transmitter needs to remove 
any correlation with $\uy^{t-1}$ in $u_t$ (to this aim, the 
transmitter has to access the channel outputs through feedback). This property, albeit a rather natural/simple consequence due to the Kalman filter, can yield interesting results, see e.g. \cite{kim06}.

\begin{prop} \label{prop:projection}
In system (\ref{dyn:coding}), for any $0 \leq \tau < t$, it holds 
that $\E u_t e_\tau=0$ and $\E u_t y_\tau=0$.  Equivalently, 
matrices $\E \uu^T \uy^T{}'$ and $\E \uu^T \ue^T{}'$ are upper 
triangular for any $T$.
\end{prop}

The justification of this proposition follows simply from the 
famous Projection Theorem (for MMSE estimators, the estimation error at a time
is orthogonal to all available measurements, see e.g. 
\cite{luenberger,kailath:book,book:kay1}) which holds for the  
KF.  Here note that $u_t$ is in fact the one-step 
prediction error (i.e. $u_t=\uC' \tildex_{t}$ where $\tildex_{t}$ is the estimation error for $\ux_t$ with one-step prediction using the estimator $\E(\ux_t|\uy^{t-1})$).  We 
also provide an alternative proof based on the state-space model in the 
appendix.

\textbf{Proof: } See Appendix \ref{appsub:structureproperty}.  \endproof

\subsection{Gauss-Markov property of the transformed output process}  

In this subsection, we show that the process $\ucy^T$, a 
transformation of the output process $\uy^T$ or $\utildey^T$, is a 
Gauss-Markov (GM) process.  In particular, it is an MA-$m$ Gaussian 
process.  This is a generalization of the result obtained in 
\cite{ordentlich}, which states that if the channel has an MA-$m$ 
Gaussian noise process and has no ISI, a necessary condition for 
optimality is that the channel output needs to be an MA-$m$ Gaussian 
process; see Corollary IV.1 in \cite{kim06} for the detailed 
statement and proof of the result of \cite{ordentlich}. This result 
has been generalized in \cite{kim06}, that is, if the channel has an 
$m$th order autoregressive moving-average (ARMA-$m$) Gaussian 
process and has no ISI, a necessary condition for optimality is that 
the channel output needs to be an ARMA-$m$ Gaussian process; see 
Proposition VII.1 in \cite{kim06}.  Our result here, on the other 
hand, is concerned with an transformed output which is sometimes 
simpler to deal with.

Recall the relevant definitions in (\ref{def:ZzZp}) and (\ref{eq:calZfactor}) of Section \ref{sec:pre}, and define the transformed output process 
\be \ucy^T := \bcalZ_{z,T}  \uy ^T . \label{def:ycheck} \ee
From (\ref{eq:calZfactor}), it holds that
\be \ucy^T = \bcalZ_{z,T} ( \bcalZ_T^{-1} \uu^T + \uN^T ) = 
\bcalZ_{p,T} \uu^T + \bcalZ_{z,T} \uN^T.\ee
This implies that, $\cy_{t+m+1}$ is a linear combination of 
$\uu_{t+1}^{t+m+1}$, and $\cy_t$ is a linear combination of 
$\uu_{t-m}^t$, since $\bcalZ_{p,T}$ is banded (and lower triangular) 
with bandwidth $(m+1)$.  But the Projection Theorem yields that 
$\uu_{t+1}^{t+m+1}$ is independent of $\cy_{t}$, so $\cy_{t+m+1}$ is 
independent of $\cy_{t}$.  Repeat this argument and we can show that 
$\ucy^T$ is a banded process, i.e., an MA-$m$ process.  More 
formally, we have

\begin{prop} \label{prop:maout}
In system (\ref{dyn:coding}), it holds that the transformed output 
process $\ucy^T$ is an MA-$m$ Gaussian process, or equivalently 
\be \bfK_{\ucy}^{(T)} := \E \ucy^T \ucy^T {}' \label{def:kycheck} 
\ee
is banded with bandwidth $(2m+1)$, i.e., $\bfK_{\ucy}^{(T)} (i,j) = 
0$ if $|i-j| \geq m+1 $.

\end{prop} 

\textbf{Proof: } See Appendix \ref{appsub:structureproperty}.  \endproof

As a result of this proposition, we see that $\utildey^T$ is an 
ARMA-$m$ process, as claimed in \cite{kim06}.

\begin{figure}[h!]
\center {{\scalebox{.5}{\includegraphics{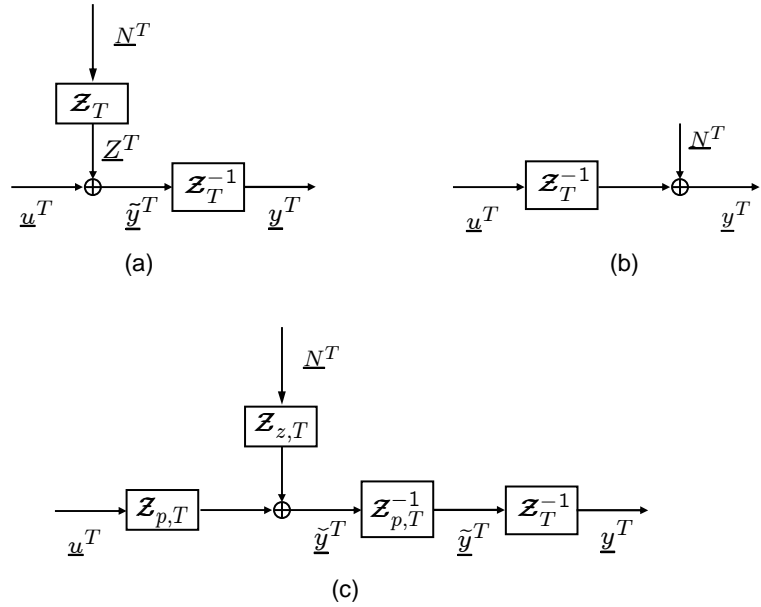}}}} 
\caption{ (a) A colored Gaussian noise channel without ISI.  This 
formulation may be directly used to study the waterfilling property 
of the optimal solution.   (b) An equivalent ISI channel with AWGN.  
This formulation may be easily used to study the KF 
properties of the optimal solution. (c) Another equivalent channel 
model with both ISI and colored noise, but the ISI and colored noise 
filter are both MA-$m$ filters.  This formulation may be used to 
study the finite-dimensionality of the channel input/output 
processes.  Note that $\bcalZ_T$ can be realized as $(\bfF - \uG 
\uH', -\uG, \uH', 1)$, $\bcalZ_T^{-1}$ as $(\bfF , \uG, \uH', 1)$, 
$\bcalZ_{p,T}$ as $(\bfF_z, \uG_p, \uH', 1)$, $\bcalZ_{p,T}^{-1}$ as 
$(\bfF+\uG_z \uH', \uG_p, \uH', 1)$, and $\bcalZ_{z,T}$ as $(\bfF_z, 
\uG_z, \uH', 1)$} \label{fig:ISI3ways}
\end{figure}

The different forms of channel outputs, i.e. $\uy^T$, $\utildey^T$, and $\ucy^T$, causally determine each other; see Fig. 
\ref{fig:ISI3ways} for their relations.  Fig. \ref{fig:ISI3ways} (a) shows the ISI-free 
colored Gaussian noise channel with a direct channel output 
$\utildey^T$ and a transformed output $\uy^T$.  Since this channel 
has no ISI, the optimal effective input process must waterfill the 
effective noise spectrum and hence $\utildey^T$ is the waterfilling 
output for the optimal scheme.  Fig. \ref{fig:ISI3ways} (b) shows 
the ISI channel corrupted by AWGN, with a channel output $\uy^T$.  
Since the channel noise is white, it may be easy to directly apply the KF algorithm.  Fig. \ref{fig:ISI3ways} (c) 
is an ISI channel corrupted by a colored Gaussian noise with a 
channel output $\ucy^T$, but both the ISI filter and the filter 
generating the colored noise are MA-$m$ filters.  It may be easily 
used to establish that $\ucy^T$ is an MA-$m$ process.  These 
formulations are $T$-equivalent and can be easily converted from one to 
another.

\subsection{Finite dimensionality of the optimizing $\ur^T$}  

We now show that, to achieve the finite-horizon feedback 
capacity $C_{T,n}$, the covariance matrix of the feedback-free, message-carrying process $\ur^T$ can 
have rank at most $(m+1)$, where $m$ is the order of the channel 
$\calZ(z)$.  This is an extension of the finite-rankness property by Ordentlich (c.f. \cite{ordentlich,kim06}) for a Gaussian channel with an MA-$m$ noise process to a Gaussian channel with an ARMA-$m$ noise process.

\begin{prop} \label{prop:fdft}
For system (\ref{dyn:coding}), the optimal $\bfK_{\ur}^{(T)}$ that 
solves $C_{T,n}$ as defined in (\ref{def:CTn}) has rank at most 
$(m+1)$.

\end{prop} 

The proof of this proposition is based on Lemma \ref{lemma:Cfhft_dimm} below.  This lemma deals with a special class of $m$th order channel $\bcalZ_T$, that is, any $\bcalZ_T$ such that $(f_0 + g_0)=0$ (see Sec. \ref{sec:pre} for notations).  In other words, $\bcalZ_{p,T}$ is in fact an MA-$(m-1)$ model.  For this class of channels, it is easy to extend the idea of Ordentlich (c.f. \cite{ordentlich}) and prove that the optimal $\bfK_{\ur}^{(T)}$ has rank at most $m$.  Then the proposition can be proven by approaching any arbitrary $\bcalZ_T$ by elements in the special class of channels based on certain continuity properties.

\begin{lemma} \label{lemma:Cfhft_dimm}
For system (\ref{dyn:coding}) with $(f_0 + g_0)=0$, the optimal $\bfK_{\ur}^{(T)}$ that 
solves $C_{T,n}$ as defined in (\ref{def:CTn}) has rank at most 
$m$.
\end{lemma}

See Appendix \ref{appsub:fdft} for the proofs of the lemma and the proposition.

\subsection{Necessity of the MMSE predictor for general channels with feedback}

The necessasity of the Kalman filter in achieving the optimality for the channel $\calF$ under an average power constraint can be easily extended.  Assume an arbitrary additive channel 
\be \uy^T = \mathbf{H} \uu^T + \uZ^T \label{ch:add} \ee
with an average power constraint $\E \|\uu^T\|^2 \leq (T+1)\calP$, where $\uZ^T$ is an arbitrary additive noise process.  Assuming one-step delayed channel output feedback,  then such a channel needs to contain an MMSE one-step predictor in order to achieve the feedback capacity.

\begin{prop} \label{prop:mmse} Let $\tilde{u}_t:=u_t - \E (u_t|\uy^{t-1})$ and $\underline{\tilde{y}}^T := \mathbf{H} \tilde{\uu}^T + \uZ^T$.  Then $I(\uu^T \rightarrow \uy^T) = I(\tilde{\uu}^T \rightarrow \tilde{\uy}^T)$ and $\E \uu^T{}' \uu^T \geq \E \tilde{\uu}^T{}' \tilde{\uu}^T$.
\end{prop}

\textbf{Proof: } Note that $\E (u_t|\uy^{t-1})$ can be generated and added back to the channel output at the receiver side and hence the directed information across the channel or mutual information from the message to channel outputs is the same using either $\uu^T$ or $\tilde{\uu}^T$ as the channel inputs.  The average power of using $\tilde{\uu}^T$ is no larger since it has minimum variance.
\endproof

Simple as it is, this necessary condition for optimality is rather universal. A corollary is that in the optimal feedback coding scheme the current channel input $u_t$ is independent of all past channel outputs $\uy^{t-1}$ by the Projection Theorem, an extension of Proposition \ref{prop:projection}. Moreover, since $\E \tilde{\uu}^T = 0$ by the law of total expectation, it is a center-of-gravity encoding rule (cf. \cite{schalkwijk:center,kavcic_it07}). It is also straightforward to see that if the channel output feedback delay is $d$ steps, then an MMSE $d$-step predictor is needed for optimality.

\section{Asymptotic analysis of the feedback system} \label{sec:asym}

By far we have completed our analysis in finite-horizon.  We have 
shown that the optimal design of encoder and decoder must contain a 
KF, and connected the feedback communication problem to 
an estimation problem and a control problem. Below, we briefly consider the 
steady-state communication problem, by studying the limiting 
behavior ($T$ going to infinity) of the finite-horizon solution 
while fixing the encoder dimension to be $(n+1)$. The infinite-horizon capacity problem will not be considered in this paper. Here and 
hereafter, we make the following assumption unless otherwise 
specified: 

\textbf{(A2)}: $(\bfA,\uC')$ is observable, and none of the 
eigenvalues of $\bfA$ are on the unit circle or at the locations of 
the eigenvalues of $\bfF$.

\subsection{Convergence to steady-state}  \label{subsec:asym}

The time-varying KF in (\ref{dyn}) converges to a 
steady-state, namely (\ref{dyn}) is \emph{stabilized} in 
closed-loop: The distributions of $u_t$, $e_t$, and $y_t$ will 
converge to steady-state distributions, and $\Si_t$, $\uL_t$, 
$\calG^*_t(\bfA,\uC)$, $\hatG_t^*$, and $K_{e,t}$ will converge to 
their steady-state values.  That is, asymptotically (\ref{dyn}) 
becomes an LTI system
\be \textnormal{steady-state:}\left\{ \ba{lll} \bbX_{t+1}
&=& (\bbA - \uL  \: \bbC') \bbX_t - \uL N_t=\bbA \bbX_t-\uL e_t\\
 e_t &=& \bbC' \: \bbX_t+N_t \\
 u_t &=& \bbD' \: \bbX_t, \ea \right. \label{dyn:steady}\ee
where
\be \uL : = \frac{\bbA \Si \bbC}{ K_{e}}, \ee
$K_{e} = \bbC' \Si \bbC +1$, and $\Si$ is the unique stabilizing 
solution to the Riccati equation
\be \Si = \bbA \Si \bbA' - \frac{\bbA \Si \bbC \: \bbC' \Si \bbA' 
}{\bbC' \Si \bbC  +1}. \label{ric:eq} \ee

This LTI system is sometimes easy to analyze (e.g., it allows transfer 
function based study) and to implement.  For instance, the 
cheap control (cf. \cite{liu:markov} and \cite{mincontrol:book}) of an LTI system 
claims that the transfer function from $N$ to $e$ is an 
\emph{all-pass} function in the form of
\be \calT_{Ne} (z) = \prod _{i=0}^k \frac{z-\la_i}{z-\la_i^{-1}} 
\label{allpass}\ee
where $\la_0,\cdots,\la_k$ are the \emph{unstable eigenvalues} of 
$\bfA$ or $\bbA$ (noting that $\bfF$ is stable).  Note that this is 
consistent with the whiteness of innovations process $\{e_t\}$.

The existence of steady-state of the KF is proven in the 
following proposition. Notice that (\ref{dyn}) is a \emph{singular} 
KF since it has no process noise; the convergence of such 
a problem was established in \cite{gallivan_riccati05}.

\begin{prop} \label{prop:dare} Consider the Riccati recursion (\ref{ric:recur}) and
the system (\ref{dyn}).   Assume (A2) and that $\la_0,\cdots,\la_k$ 
are the \emph{unstable eigenvalues} of $\bfA$.

i) Starting from the initial condition given in (\ref{ric:init}), 
the Riccati recursion (\ref{ric:recur}) generates a sequence 
$\{\Si_t\}$ that converges to $\Si_\infty$, the unique stabilizing 
solution to the Riccati equation (\ref{ric:eq}), and $\Si_\infty$ 
has rank $(k+1)$.

ii) The time-varying system (\ref{dyn}) converges to the unique 
steady-state as given in (\ref{dyn:steady}).

\end{prop}

\textbf{Proof: } See Appendix \ref{appsub:convg}. \endproof

\subsection{Steady-state quantities} \label{sub:steady_eq}

Now we fix $(\bfA,\uC)$ and let the horizon $T$ in  
the coding structure $\bbS^*$ go to infinity.  Let $\mathcal{H}(e)$ be the 
entropy rate of $\{e_t\}$, 
\be DI(\bfA):=\prod _{i=0}^k |\lambda_i| \label{def:di} \ee
be the \emph{degree of instability} or the \emph{degree of anti-causality} of $\bfA$, and $S (e^{j 2\pi 
\theta}) := Y (e^{j 2\pi \theta}) / N (e^{j 2\pi \theta})$ be the spectrum of the sensitivity function of system 
(\ref{dyn:steady}) (cf. \cite{elia_c5}). Then the limiting result of 
Proposition \ref{prop:dual} is summarized in the next proposition.

\begin{prop} \label{prop:steady}

Consider the coding structure $\bbS^*$. For 
any $n \geq 0$ and $(\bfA,\uC')$ with $\bfA \in \bbR^{(n+1) \times 
(n+1)}$ satisfying (A2),

i) The asymptotic information rate is given by
\be \ba{lll}  R_{\infty,n}(\bfA,\uC) &: =& \displaystyle \lim  _{T
\rightarrow \infty} \frac{1}{T+1} I(\uW;\uy^T) \\
&=&  \displaystyle  \mathcal{H}(e) - \frac{1}{2} \log 2 \pi e \\
&=&  \displaystyle \frac{1}{2} \log K_{e}\\
&=& \displaystyle \log DI(\bfA) \\
&=& \displaystyle \displaystyle \int_{-\frac{1}{2}} ^{\frac{1}{2}}
\log |S (e^{j 2\pi \theta})| d \theta \\
& =& \displaystyle \frac{1}{2}  \log  (\bbC' \Si \bbC+1 )\\
 &=& \displaystyle \lim _{T \rightarrow \infty}\frac{\log \det \calI_{\uW,T}}{2(T+1)} \\
 &=& \displaystyle - \lim _{T \rightarrow \infty}\frac{\log \det \MSE_{\uW,T}}{2(T+1)} \\
&=& \displaystyle -\lim _{T \rightarrow \infty}\frac{\log  \det  
\CRB_{\uW,T}}{2(T+1)} . \ea \ee

ii) The average channel input power is given by
\be \ba{lll} P_{\infty,n}(\bfA,\uC) & :=& \disp \lim _{T \rightarrow 
\infty} \frac{1}{T+1} \E 
\| \uu^T \| ^2 \\
&=& \displaystyle \bbD' \Si \bbD . \ea \ee

\end{prop}

\begin{remark} \rm Proposition \ref{prop:steady} links the asymptotic 
information rate to 
the degree of anti-causality and Bode sensitivity integral 
(\cite{elia_c5}) for the control system, to the entropy rate and steady-state variance of the innovations process,  asymptotic increasing rate of the Fisher 
information, and the asymptotic decay rate of smoothing MSE or of CRB for the estimation system.
Note that the Bode sensitivity integral is the fundamental 
limitation of the disturbance rejection (control) problem, and the 
asymptotic decay rate of CRB is the fundamental limitation of the 
recursive estimation problem.  Hence, the fundamental limitations in 
feedback communication, control, and estimation coincide. More specifically, the asymptotic information rate cannot be made higher or lower than a constant regardless of the feedback generator choice; the disturbance rejection measure cannot be made smaller  than a constant regardless of the feedback controller design; the decay rate of the estimate error cannot be made faster  than a constant regardless of the estimator design; and the constant is the logarithm of the degree of anti-causality of $\bfA$.
\end{remark}

\begin{remark} \rm It is straightforward to extend the finite-horizon connections between the fundamental tradeoffs for feedback communication, estimation, and feedback control to infinite horizon.  As the limits exist, quantities in fundamental tradeoffs (T1) through (T3) given in Section \ref{sub:tradeoff} are well defined in infinite horizon and the corresponding relationship still holds.  Note it is more obvious to see that the Bode integral is associated with anti-causality since it equals the logarithm of the degree of anti-causality of $\bfA$.
\end{remark}

\textbf{Proof: } Proposition \ref{prop:dare} leads to that, the limits of the 
results in Proposition \ref{prop:dual} are well defined.  Then
\be \ba{lll}  R_{\infty,n}(\bfA,\uC) & =& \displaystyle \lim  _{T
\rightarrow \infty} \frac{1}{2(T+1)} \sum_{t=0}^T \log K_{e,t} \\
& =& \displaystyle \lim  _{T
\rightarrow \infty} \frac{1}{2}  \log K_{e,T} \\
&=&  \displaystyle \mathcal{H}(e) -\frac{1}{2} \log 2\pi e, \ea \ee
where the second equality is due to the Cesaro mean (i.e., if $a_k$ 
converges to $a$, then the average of the first $k$ terms converges 
to $a$ as $k$ goes to infinity), and the last equality follows from 
the definition of entropy rate of a Gaussian process (cf. 
\cite{cover}).

Now by (\ref{allpass}), $\{e_t\}$ has a flat power spectrum with 
magnitude $DI(\bfA)^2$. Then $R_{\infty,n}(\bfA,\uC)=\log DI(\bfA)$.  
The Bode integral of sensitivity follows from \cite{elia_c5}.  The 
other equalities are the direct applications of the Cesaro mean to 
the results in Proposition \ref{prop:dual}.
\endproof

Proposition \ref{prop:steady} implies that the presence of stable 
eigenvalues in $\bfA$ does not affect the rate (see also 
\cite{elia_c5}). Stable eigenvalues do not affect 
$P_{\infty,n}(\bfA,\uC)$, either, since the initial condition 
response associated with the stable eigenvalues can be tracked with 
zero power (i.e. zero average MSE).  Therefore, we conclude that the 
presence of stable eigenvalues in $\bfA$ does not affect either the 
rate $R_{\infty,n}(\bfA,\uC)$ or the power $P_{\infty,n}(\bfA,\uC)$. 
We have thus seen that the communication problem is essentially a 
problem of tracking an anti-causal source over a communication 
channel (\cite{sahai:phd,elia_c5,sahai:main}).

\begin{coro} \label{coro:stable_eig}
Suppose that $(\bfA,\uC)$ with $\bfA \in \bbR^{(n+1) \times (n+1)}$ 
satisfies (A2).  Suppose further that $\bfA$ has $(k+1)$ unstable 
eigenvalues denoted $\la_0, \cdots, \la_k$ where $0 \leq k \leq 
(n+1)$.  Then there exists an observable pair $(\bfA_k,\uC_k')$ with 
$\bfA_k \in \bbR^{(k+1) \times (k+1)}$ being anti-stable such that 
$R_{\infty,n}(\bfA,\uC) = R_{\infty,k}(\bfA_k,\uC_k)$ and 
$P_{\infty,n}(\bfA,\uC) = P_{\infty,k}(\bfA_k,\uC_k)$.
\end{coro}

\textbf{Proof: }  See Appendix \ref{appsub:stable_eig}.\endproof

\section{Conclusions and future work}

In this paper, we proposed a perspective that integrates information transmission (communication), information processing (estimation), and information utilization (control).  We identified and explored fundamental limitations in feedback communication, estimation, and feedback control over Gaussian channels with memory.  Specifically, we established a certain equivalence of a feedback communication system, an estimation system, and a feedback control system.  We demonstrated that a simple reformulation of the Kalman filter becomes the celebrated Schalkwijk-Kailath codes, and the well-studied Cover-Pombra structure necessarily contains a Kalman filter in order to be optimal.  We characterized the roles of Kalman filtering in an optimal feedback communication system as to ensure power efficiency and to optimally recover the transmitted codewords.  We showed that the fundamental limitations/tradeoffs in these three systems also coincide: The power versus rate tradeoff in feedback communication, the causal prediction versus smoothing tradeoff in estimation, and the control performance versus Bode integral tradeoff in control, are equivalent and in essence, all of them are the causality versus anti-causality tradeoffs.  We also presented a coding scheme achieving the finite-horizon feedback capacity of the Gaussian channel.  The scheme is based on the Kalman filtering algorithm, and provides refinements and extensions to the Cover-Pombra coding structuure and Schalkwijk-Kailath codes.

Our new perspective has been recently generalized in 
\cite{liu:phd} to uniformly address the fundamental limits 
of several classes of feedback communication problems, and we 
envision that this perspective can generate a new avenue for 
studying more general feedback communication problems, such as 
multiuser feedback communications.  Our ongoing research includes 
extending our proposed scheme to address the optimality of more 
feedback communication problems (such as single-user MIMO systems with 
output feedback, multi-user MIMO systems with output feedback).  We also anticipate that the perspective and the 
approaches developed in this paper be extended and help to build a 
theoretically and practically sound paradigm that unifies 
information, estimation, and control.

\appendices

\section{Systems representations and equivalence} \label{appsec:equiv}

The concept of system representations and the equivalence between 
different representations are extensively used in this paper.  In 
this subsection, we briefly introduce system representations and the 
equivalence. For more thorough treatment, see e.g. 
\cite{oppen:signals,chen:book,dahleh:book}.

\subsection{Systems representations} \label{appsub:represent}

Any discrete-time linear system can be represented as a linear 
mapping (or a linear operator) from its input space to output space; 
for example, we can describe a single-input single-output (SISO) 
linear system as
\be \uy^t = \pmb{\calM}_t \uu^t \ee
for any $t$, where $\pmb{\calM}_t \in \bbR^{(t+1) \times (t+1)}$ is 
the matrix representation of the linear operator, $\uu^t \in 
\bbR^{t+1} $ is the stacked input vector consisting of inputs from 
time 0 to time $t$, and $\uy^t \in \bbR^{t+1} $ is the stacked 
output vector consisting of outputs from time 0 to time $t$.  For a 
(strictly) causal SISO LTI system, $\pmb{\calM}_t$ is a (strictly) 
lower triangular Toeplitz matrix formed by the coefficients of the 
impulse response.  Such a system may also be described as the 
(reduced) transfer function, whose inverse $z$-transform is the 
impulse response; by a (reduced) transfer function we mean that its 
zeros are not at the same location of any pole.

A causal SISO LTI system can be realized in state-space as
\be \left\{ \ba{lll} \ux_{t+1} &=& \bfA \ux_t + \underline{B} u_t \\
y_t &=& \uC' \ux_t + D u_t, \ea \right. \label{state-space}\ee
where $\ux_t \in \bbR^l$ is the state, $u_t \in \bbR$ is the input, 
$y_t \in \bbR$ is the output, $\bfA$ is the state matrix, $\underline{B}$ is the input matrix (vector), $\uC$ is the output matrix (vector), and $D$ is the direct feedthrough term. We call $l$ the \emph{dimension} 
or the \emph{order} of the realization.  The state-space 
representation (\ref{state-space}) may be conveniently denoted as 
$(\bfA,\underline{B},\uC', D)$.   Note that in the study of 
input-output relations, it is sometimes convenient to assume that 
the system is relaxed or at initial rest (i.e. zero input leads to 
zero output), whereas in the study of state-space, we generally 
allow $\ux_0 \neq \uzero$, which is not at initial rest.  For 
multi-input multi-output (MIMO) systems, linear time-varying 
systems, etc., see \cite{chen:book,dahleh:book}.

The state-space representation of an causal FDLTI system $\calM(z)$ 
is not unique.  We call a realization $(\bfA,\underline{B},\uC', D)$ 
\emph{minimal} if $(\bfA,\underline{B})$ is controllable and 
$(\bfA,\uC')$ is observable. All minimal realizations of $\calM(z)$ 
have the same dimension, which is the minimum dimension of all 
possible realizations.  All other realizations are called 
\emph{non-minimal}.  The transfer function for the state-space representation $(\bfA,\underline{B},\uC', D)$ is $\uC' (z \bfI - \bfA) ^{-1} \underline{B} + D$.

\textbf{Example: Derivation of state-space representation of 
$\calG_T^*(\bfA, \uC)$}

We demonstrate here how we can derive a realization of a system.  
Consider $\calG_T^*(\bfA,\uC)$ in (\ref{eq:calGhatG}) in Section 
\ref{sec:kf}, which is given by
\be \calG_T^*(\bfA,\uC) = - \hatG_T^* (\bfI- \bcalZ_T^{-1} 
\hatG_T^*) ^{-1} , \label{eg:calG} \ee
where the state-space representations for $\hatG_T^*(\bfA,\uC)$ and 
$\bcalZ_T^{-1}$ are illustrated in Fig. \ref{fig:dual} (b) and Fig. 
\ref{fig:isicolor} (c).  This result shows that the block diagram in 
Fig. \ref{fig:estim} (c) is indeed the dynamics of $\hatG_T^*$, as 
claimed in Proposition \ref{prop:kf} iii).

\begin{figure}[h!]
\begin{center}
{\scalebox{.5}{\includegraphics{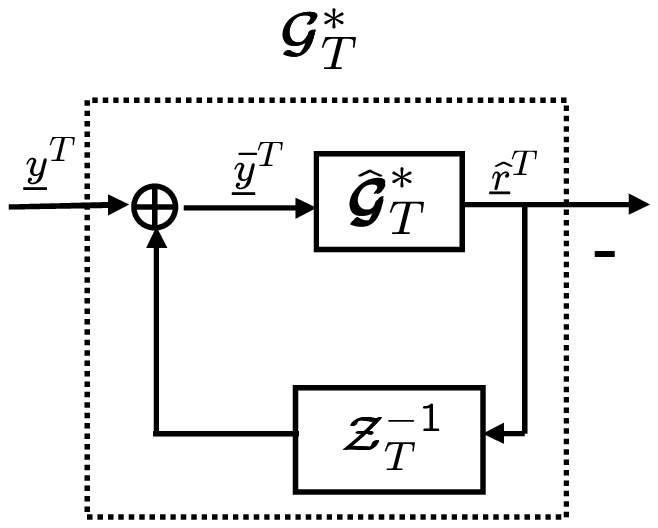}}} \caption{$\calG^*$ is 
a feedback connection of $\hatG^*$ and $\bcalZ^{-1}$.} 
\label{fig:calG}
\end{center}
\end{figure}

Since (\ref{eg:calG}) suggests a feedback connection of $\hatG^*$ 
and $\bcalZ^{-1}$ as shown in Fig. \ref{fig:calG}, we can write the 
state-space for $\calG^*$ as
\be 
\left\{ \ba{ll}
            \ba{lll}    \hatx_{t+1} &=& \bfA \hatx_t + \uL_{1,t} e_t \\
                     \hatr_t &=& \uC' \hatx_t\\
                      \hat{\ubars}_{t+1} &=&  \bfF \hat{\ubars}_{t} + \uG \hatr_t + \uL_{2,t} e_t\\
                        e_t &=& \ubary_t - \uH' \hat{\ubars}_{t}  -\hatr_t 
            \ea 
            & \left. \ba{l}\\ \\ \\ \\  \ea \right\} \textnormal{Kalman filter }\hatG_T^*(\bfA,\uC)      \\
            \ba{lll}            \us_{a,t+1} &=& \bfF \us_{a,t} + \uG \hatr_t \\
                                \ubary_t &=& y_t + \uH' \us_{a,t} + \hatr_t 
            \ea
             & \left. \ba{l}\\ \\ \ea \right\} \bcalZ_T^{-1}      
        \ea
            \right. \ee
Letting $\hats_t:=\hat{\ubars}_{t} - \us_{a,t}$, the above reduces 
to
\be \left\{
\ba{lll}    \hatx_{t+1} &=& \bfA \hatx_t + \uL_{1,t} e_t \\
                     \hatr_t &=& \uC' \hatx_t\\
  \hats_{t+1} &=& \bfF \hats_{t} + \uL_{2,t} e_t \\
                e_t &=& y_t - \uH' \hats_t  . \ea \right. \ee
This is the dynamics shown in Fig. \ref{fig:estim} (c).  Note that 
the above reduction of realization is allowed since it preserves the 
``$T$-equivalence", see the next subsection.

\textbf{Example: State-space representation of an inverse of a system}

Given a linear system with input $u$ and output $y$ and represented as $(\bfA,\underline{B},\uC', 1)$ in state space, it can be inverted if it is both stable and minimum-phase.  The inverse system that maps $y$ back to $u$ can be realized in state space as $(\bfA-\underline{B}\uC',-\underline{B},\uC', 1)$.  

\textbf{Example: State-space representations of $\bcalZ_z$ and $\bcalZ_p$}

It is easily shown that $\bcalZ_z(z)$ can be realized as $(\bfF_z, \uG_z, \uH', 1)$, $\bcalZ_p(z)$ can be realized as $(\bfF_z, \uG_p, \uH', 1)$, and $\bcalZ_z ^{-1}(z)$ can be realized as $(\bfF, -\uG_z, \uH', 1)$, where $\bfF_z := \bfF + \uG_z \uH'$.  

\subsection{Equivalence between representations} \label{appsub:equiv}

\begin{definition} \label{def:equiv}

i) Two FDLTI systems represented in state-space are said to be 
\emph{equivalent} if they admit a common transfer function (or a 
common transfer function matrix) and they are both stabilizable and 
detectable.

ii) Fix $0 \leq T < \infty$.  Two linear mappings $\calM_{i,T}: \: 
\bbR^{ q(T+1) } \rightarrow \bbR^{ p(T+1) }$, $i=1,2$, are said to be \emph{$T$-equivalent} if for any $\uu^T 
\in \bbR^{q(T+1)}$, it holds that
\be \calM_{1,T}(\uu^T) = \calM_{2,T}(\uu^T).\ee

\end{definition}

We note that i) is defined for FDLTI systems, whereas ii) is for 
general linear systems. i) implies that, the realizations of a 
transfer function are not necessarily equivalent. However, if we 
focus on all realizations that do not ``hide" any unstable modes, 
namely all the unstable modes are either controllable from the input 
or observable from the output, they are equivalent; the converse is 
also true.  ii) concerns about the \emph{finite-horizon} 
input-output relations only. Since the states are not specified in 
ii), it is not readily extended to infinite horizon: Any unstable 
modes ``hidden" from the input and output will grow unboundedly 
regardless of input and output, which is unwanted.

\textbf{Example: $T$-equivalence between the estimation system 
(\ref{dyn:est}) and coding structure $\bbS$ (\ref{dyn:coding})}

To show the $T$-equivalence, it is sufficient to show that for each 
$t$, the sets of signals $r_t$, $\hatr_t$, $e_t$, $\ux_t$, and 
$\hatx_t$ in (\ref{dyn:est}) and (\ref{dyn:coding}) are equal, 
respectively.  To this aim, first note that for $t=0$, the sets 
signals are equal, respectively, and that $\us_0 - \hats_0 = 
\ubars_0 - \hat{\ubars}_0$. Assume that for $t \leq \tau$, the sets 
of signals are equal, respectively, and that $\us_\tau - \hats_\tau 
= \ubars_\tau - \hat{\ubars}_\tau$.  Now use induction.  Apparently, 
$r_{\tau+1}$ and $\ux_{\tau+1}$ generated by (\ref{dyn:est}) and 
(\ref{dyn:coding}) are equal, respectively.  Then
\be \ba{lll} \us_{\tau+1} - \hats_{\tau+1} 
    & = & \bfF (\us_\tau - \hats_\tau) + \uG (r_\tau - \hatr_\tau) - \uL_{2,\tau} e_\tau \\
        &=& \ubars_{\tau+1} - 
\hat{\ubars}_{\tau+1}, \ea \ee
and $e_{{\tau+1}}$ from both (\ref{dyn:est}) and (\ref{dyn:coding}) 
equals
\be \uH' (\ubars_\tau - \hat{\ubars}_\tau) + (r_\tau - \hatr_\tau) + 
N_{\tau+1}. \ee
Thus we have proven the $T$-equivalence.

Likewise, we can show that the estimation system (\ref{dyn:est}), 
feedback communication system (\ref{dyn:coding}), and control system 
(\ref{dyn}) are $T$-equivalent.

\textbf{Examples}

As we mentioned in Section \ref{subsec:isi}, for any $\uu^T$ and 
$\uN^T$, Fig. \ref{fig:isicolor} (a) and (b) generate the same 
channel output $\utildey^T$.  That is, the mappings from 
$(\uu^T,\uN^T)$ to $\utildey^T$ for the two channels are identical, 
and both are given by
\be \utildey^T = \bcalZ_T (\bcalZ_T^{-1} \uu^T + \uN^T ). \ee
Thus, we say the two channels are $T$-equivalent.

\section{Input/output characterization of finite-horizon 
information capacity: Directed information} \label{appsub:directI}

\begin{definition} \label{def:di_I}
The directed information from $\uu^T$ to $\uy^T$ is defined as
\be I(\uu^T \rightarrow \uy^T):= \sum_{t=0}^T I(\uu^t;y_t|\uy^{t-1}). 
\label{directedinfo}\ee
\end{definition}

See \cite{tati:capI} for details.  One important feature about the directed information is that it is 
an input-output counterpart of the mutual information, which is especially useful to deal with channels within loops.

\begin{prop} \label{prop:CTsekhar} (Tatikonda and Mitter) It holds that
\be C_T(\calP) = \sup _{\uu^T} \frac{1}{T+1}I(\uu^T \rightarrow 
\uy^T) , \label{opt:c_t}\ee
where the supremum is over all possible feedback-dependent Gaussian 
input distributions satisfying the power constraint
\be  \frac{1}{T+1} \E \uu^T{}' \uu^T \leq \mathcal{P}, 
\label{eq:pc}\ee
and in the form
\be u_t=\underline{\gamma}_t ' \uu^{t-1} + \underline{\eta}_t ' 
\uy^{t-1} + \xi_t \label{u_sekhar}\ee
for any $\underline{\gamma}_t \in \bbR^{t}$, $\underline{\eta}_t \in 
\bbR^{t}$, and zero-mean Gaussian random variable $ \xi_t \in \bbR$ 
independent of $\uu^{t-1}$ and $\uy^{t-1}$.
\end{prop}

This proposition follows directly from the following lemma.

\begin{lemma} \label{lemma:input}
The CP structure for the ISI Gaussian channel $\calF$ shown in Fig. 
\ref{fig:cpisi} can generate any Gaussian channel input process 
$\{u_t\}$ in the form of (\ref{u_sekhar}) and vice versa.
\end{lemma}

\textbf{Proof: } Note that any input generated by the scheme in Fig. 
\ref{fig:cpisi} has the form of
\be \uu^t=\calB_t \bcalZ_t \uN^t + \uv^t = \calB_t \bcalZ_t 
\uy^t-\calB_t \uu^t + \uv^t, \ee
leading to
\be \uu^t = (\bfI + \calB_t)^{-1} \calB_t \bcalZ_t \uy^t +( \bfI + 
\calB_t)^{-1} \uv^t,  \label{u3} \ee
where $\uv^t$ is independent of $\uN^t$ and hence $\uZ^t$.

On the other hand, from (\ref{u_sekhar}), we have
\be \uu^t = \pmb{\bar{\gamma}}_t \uu^t + \pmb{\bar{\eta}}_t \uy^t + 
\underline{\xi}^t,  \ee
leading to
\be \uu^t = (\bfI - \pmb{\bar{\gamma}}_t)^{-1}  \pmb{\bar{\eta}}_t 
\uy^t +(\bfI- \pmb{\bar{\gamma}}_t)^{-1}  \underline{\xi}^t, 
\label{u4} \ee
where $\pmb{\bar{\gamma}}_t \in \bbR^{(t+1) \times (t+1)}$ is the 
strictly lower triangular matrix formed by $\underline{\gamma}_0 
',\cdots, \underline{\gamma}_t '$, and $\pmb{\bar{\eta}}_t\in 
\bbR^{(t+1) \times (t+1)}$ is the strictly lower triangular matrix 
formed by $\underline{\eta}_0 ',\cdots,\underline{\eta}_t '$.  Since 
for any $\tau \leq t$, $ \xi_\tau$ is independent of $\uu^{\tau-1}$ 
and $\uy^{\tau-1}$, $ \xi_\tau$ is independent of $\uN^{\tau-1}$.  
By causality $ \xi_\tau$ is independent of $N_{\tau}$, $N_{\tau+1}$, 
$\cdots$.  Therefore, $\underline{\xi}^t$ is independent of 
$\uN^{t}$ and hence $\uZ^t$.  Then the lemma follows by comparing 
(\ref{u3}) and (\ref{u4}). 
\endproof

\textit{Proof of Proposition \ref{prop:CTsekhar}:} This proposition 
follows trivially from the observation that, for any Gaussian input 
of form (\ref{input:cpisi}), it holds that 
\be \ba{lll} I(\uu^T \rightarrow \uy^T) &=& \sum_{t=0}^T \left( 
h(y_t|\uy^{t-1}) - h(y_t|\uu^t, \uy^{t-1}) \right)  \\
&=& h(\uy^T) - \sum_{t=0}^T h(N_t|\uu^t, \uy^{t-1}) \\
&=& h(\uy^T) - h(\uN^T)  \\
&=& I(\ur^T; \uy^T) . \ea \ee
\endproof

Note that the above proof can be easily used to show that $I(\uu^T 
\rightarrow \uy^T)$ equals $I(W;\uy^T)$ and 
$I(\underline{\xi}^T;\uy^T)$, where $W$ is the message.  That is, 
the directed information from the input signal to output signal 
(both signals are inside the feedback loop and causally affecting each 
other) effectively captures the mutual information between the 
message $W$ (or a message-carrying signal outside the feedback loop, 
such as $\underline{\xi}^T$ or $\ur^T$) and the output signal.  
Therefore, directed information has the advantage of capturing 
the capacity without the need to identify the message, see e.g. 
\cite{kavcic_it07}.

\section{Corresponding relation between the CP structure and coding structure $\bbS$} \label{appsec:CPequivalent}

i) Assume $\bfK_{\ur}^{(T)} > 0$ first. For any fixed 
$(\bfK_{\ur}^{(T)},\calB_T)$ in the CP structure, define in $\bbS$ that
\be \ba{lll} \calG_T &:=&(\bfI+\calB_T)^{-1} \calB_T \bcalZ_T \\
\bfA &:=& \bfGa_a^{-1} \left[ \begin{tabular}{c|c}0 & $\bfI_T$ \cr 
\hline
* & * \end{tabular} \right] \bfGa_a  :=  \bfGa_a^{-1} \bfA_a \bfGa_a  \;\; \in \bbR^{(T+1)\times(T+1)}\\
\uC &:=& \bfGa_a \left[ \matrix{1 &0& \cdots & 0} \right]' :=  \bfGa_a \ue_1 ,\ea \ee
where $\bfGa_a :=  (\bfK_{\ur}^{(T)})^{\frac{1}{2}} = \bfGa_a '$ is a positive definite square root, * can be 
any number, and $\bfA_a$ and $\ue_1$ are defined in obvious ways (note that this vector $\ue_1$ should not be confused with the Kalman filter innovation $e_t$).  Then it is easily verified that $\calG_T$ is strictly 
lower triangular and $(\bfA,\uC')$ is observable \footnote{$\bfA$ can also be chosen to be such that its
eigenvalues are not on the the unit circle and not at the locations of 
$\bfF$'s eigenvalues, as Assumption (A2) requires in order to guarantee convergence in Section \ref{sec:asym}.}.  
One can compute that the observability matrix for $(\bfA,\uC')$ is in fact $\bfGa_a$, that is, 
\be \left [ \matrix{ \uC' \cr \uC'\bfA \cr \uC' \bfA ^2 \cr \cdots \cr \uC'\bfA ^ T } \right ]
 = \left [ \matrix{\ue_1 ' \cr \ue_1 ' \bfA_a \cr  \ue_1 ' \bfA_a ^2  \cr \cdots \cr \ue_1'\bfA_a ^ T } \right ] \bfGa_a = \bfGa_a, 
 \ee
where the last equality is due to the structures of $\bfA_a$ and $\ue_1$.  Thus, $(\bfA,\uC')$ can generate process $\ur^T$ with covariance matrix $\bfK_{\ur}^{(T)}$. Then by (\ref{input:cpisi}), we know that for any given 
$(\bfK_{\ur}^{(T)},\calB_T)$ with $\bfK_{\ur}^{(T)} > 0$, we can 
find an admissible $(\bfA,\uC,\calG_T)$ generating the same channel input $\uu^T$ as $(\bfK_{\ur}^{(T)},\calB_T)$ does.

Now consider the case that $\bfK_{\ur}^{(T)} \geq 0$ but 
$\bfK_{\ur}^{(T)}$ is not positive definite.  Consider the positive 
definite sequence $\{\bfK_{\ur}^{(T)} + \frac{1}{i} \bfI 
\}_{i=1}^\infty$.  Therefore, for each pair $(\bfK_{\ur}^{(T)} + \frac{1}{i} \bfI,\calB_T)$, we 
can find an admissible triple $(\bfA_i,\uC_i,\calG_{T,i})$ corresponding 
to it per above construction.  It is easily shown that the sequence of triples generate a sequence of inputs that converge to $\uu^T(\bfK_{\ur}^{(T)},\calB_T)$.   Note that, however, power constraint or rate constraint given in Definition \ref{def:CTn} is not considered here and hence may not hold unless some further constraint on the sequence is imposed.

ii) Conversely, for any fixed admissible $(\bfA,\uC,\calG_T)$ with $\bfA\in 
\bbR^{(T+1)\times(T+1)}$, we can obtain an admissible 
$(\bfK_{\ur}^{(T)},\calB_T)$ as
\be \ba{lll} \\
\calB_T &:=& \calG_T \bcalZ_T^{-1}(\bfI-\calG_T \bcalZ_T^{-1}) ^{-1} \\
\bfK_{\ur}^{(T)} &:=& \bfGa_T(\bfA,\uC) \bfGa_T(\bfA,\uC)', \ea 
\label{eq:g2b}\ee
which generates identical channel input $\uu^T$ as 
$(\bfA,\uC,\calG_T)$ does.

iii) By the continuity of the mutual information and power, the limits of the power sequence and the mutual information sequence generated by 
$\uu^T(\bfA_i,\uC_i,\calG_{T,i})$ are equal to the power and mutual information 
generated by $\uu^T(\bfK_{\ur}^{(T)},\calB_T)$, respectively. 
Then note that $P_{T,T}(\calR)$ is the infimum power over all admissible 
$(\bfA,\uC,\calG_T)$ with $\bfA \in \bbR^{(T+1)\times(T+1)}$ according to Definition \ref{def:CTn}, and that $P_{T}(\calR)$ is the infimum power over all admissible 
$(\bfK_{\ur}^{(T)},\calB_T)$ according to (\ref{cap:cpcolor}), subject to rate constraints in (\ref{constr:cpinverse}) and (\ref{def:PTn}) respectively, which implies that to show iii), it is sufficient to show that the equivalent inputs (or equivalent input sequences) hold the rate constraint in Definition \ref{def:CTn}.  By the construction of the sequence of admissible triples $\{(\bfA_i,\uC_i,\calG_{T,i})\}_{i=1}^{\infty}$, it is straightforward to see that the rate constraint (i.e. $I(\uW;\uy^T)/(T+1) \geq \mathcal{R}$) can be satisfied, yielding $P_T(\calR)=P_{T,T}(\calR)$ and hence $C_T(\calP)=C_{T,T}(\calP)$.  

On the other hand, one can directly prove $C_T(\calP)=C_{T,T}(\calP)$ without resorting to $P_{T,T}(\calR)$.  To this aim, use an arbitrarily small reduction $\epsilon>0$ from the power budget $\calP$, that is, consider only those $(\bfK_{\ur}^{(T)},\calB_T)_\epsilon$ such that $\E \| \uu^T \|^2/(T+1) \leq \mathcal{P}-\epsilon$.  As $\epsilon$ vanishes, those $(\bfK_{\ur}^{(T)},\calB_T)_\epsilon$ can arbitrarily approach $C_T(\calP)$ or $C_{T,T}(\calP)$.  Now fix $(\bfK_{\ur}^{(T)},\calB_T)_\epsilon$, and construct an admissible sequence $(\bfA_i,\uC_i,\calG_{T,i})_\epsilon$ as ii) does.  Clearly, for sufficiently large $i$, the input $\uu^T(\bfA_i,\uC_i,\calG_{T,i})_\epsilon$ satisfies the power constraint $\E \| \uu^T \|^2/(T+1) \leq \mathcal{P}$.  That is, for any rate achievable by $\uu^T(\bfK_{\ur}^{(T)},\calB_T)_\epsilon$, it can be arbitrarily approached by $\uu^T(\bfA_i,\uC_i,\calG_{T,i})_\epsilon$ satisfying the power constraint.  Then the result follows.

\section{Proof of properties of the coding structure $\bbS^*$}
 \label{appsub:structureproperty}

\subsubsection{Proof of Proposition \ref{prop:projection}}
Here we show that the coding structure $\bbS^*$, in the form of 
(\ref{dyn}), satisfies the necessary condition for optimality as 
presented in Proposition \ref{prop:projection}.

Since $\{y_t\}$ is interchangeable with the innovations process 
$\{e_t\}$, in the sense that they determine each other causally and 
linearly, it suffices to show that $\E u_t e_\tau=0$. Note that
\be u_t = \bbD \bbX_t =\bbD ' \bbA \bbX_{t-1} - \bbD ' \uL_{t-1} 
e_{t-1}, \ee
and thus
\be  \ba{lll}\E u_t e_{t-1} &=& \E \bbD ' \bbA \bbX_{t-1} e_{t-1} - 
\bbD '
\uL_{t-1} \bfK_{e,t-1} \\
&\eqa& \E \bbD ' \bbA \bbX_{t-1}  \bbX_{t-1}' \bbC + \E \bbD ' \bbA 
\bbX_{t-1}  N_{t-1} - \bbD ' \bbA \Si_{t-1} \bbC \\&=& \bbD ' \bbA 
\Si_{t-1} \bbC + 0 - \bbD ' \bbA \Si_{t-1} \bbC =0, \ea \ee
where (a) follows from (\ref{dyn}) and (\ref{eq:Lt}). Similarly we 
can prove $\E u_t e_\tau=0$ for any $\tau < t-1$.

\subsubsection{Proof of Proposition \ref{prop:maout}}

We first prove a simple technical lemma that is useful in the 
following development.  It says that the product of a banded, lower 
triangular matrix with bandwidth $(m+1)$ and an upper triangular 
matrix is banded in its lower triangular part with bandwidth 
$(m+1)$.

\begin{lemma} \label{lemma:banded}
Suppose $\bfA \in \bbR^{n \times n}$ is banded and lower triangular 
with bandwidth $(m+1)$, i.e., $\bfA (i,k) = 0$ if $i < k$ or $i > 
(k+m)$.  Suppose $\bfB \in \bbR^{n \times n}$ is upper triangular, 
i.e., $\bfB (k,j) = 0$ if $k > j$.  Then $\bfC := \bfA \bfB$ is 
banded in its lower triangular part with bandwidth $(m+1)$, i.e., 
$\bfC (i,j) = 0$ if $i > j+m$.
\end{lemma}

\textbf{Proof: } Simply note that $\bfC (i,j) = \sum_{k=1} ^n \bfA (i,k) \bfB 
(k,j) $.  So if $i > j+m$, then $\bfA (i,k) \bfB (k,j) = 0$ for any 
$k$.  \endproof

Now we go back to Proposition \ref{prop:maout}.  To show this 
proposition, it is sufficient to show that $\bfK_{\ucy}^{(T)}$ is 
banded in its lower triangular part with bandwidth $(m+1)$, since 
$\bfK_{\ucy}^{(T)}$ is symmetric.  By Lemma \ref{lemma:banded}, we 
only need to show that $\bfK_{\ucy}^{(T)}$ can be written as the sum 
of products of banded, lower triangular matrices (of bandwidth 
$(m+1)$) with upper triangular matrices.  To this aim, some algebra 
shows that
\be \ba{lll} \E \uN^T \uu^T {}' & =& \E \uN^T ( (\bfI - \calG_T \bcalZ_T^{-1} ) \ur^T - \hatG_T \uN^T )'  = -\hatG_T' \\
\bcalZ_{p,T} \E \uu^T \ucy^T {}' &=& \bcalZ_{p,T} \bfK_{\uu}^T \bcalZ_{p,T}' + \bcalZ_{p,T} \E \uu^T \uN^T{}'  \bcalZ_{z,T}' \\
\bfK_{\ucy}^{(T)} &=& \bcalZ_{p,T} \bfK_{\uu}^T \bcalZ_{p,T}' + \bcalZ_{z,T} \bcalZ_{z,T}' + \bcalZ_{p,T} \E \uu^T \uN^T{}'  \bcalZ_{z,T}' - \bcalZ_{z,T} \hatG_T ' \bcalZ_{p,T}' \\
 &=& \bcalZ_{p,T} \E \uu^T \ucy^T {}' + \bcalZ_{z,T} \bcalZ_{z,T}' -  \bcalZ_{z,T} \hatG_T ' \bcalZ_{p,T}'. \ea \ee
As $\bcalZ_{z,T}$ is lower triangular, $\E \uu^T \ucy^T {}'$ is 
upper triangular.  Therefore, on the right-hand-side of the last 
equality, $\bcalZ_{p,T}$ and $\bcalZ_{z,T}$ are banded and lower 
triangular with bandwidth $(m+1)$, and $\E \uu^T \ucy^T {}'$, 
$\bcalZ_{z,T}'$, and $\hatG_T ' \bcalZ_{p,T}'$ are upper triangular.  
Then the result follows.

As an alternative proof or a verification of the above result, let us consider the mapping $\calM_{\ue,\ucy}$ from 
$\ue^T$ to $\ucy^T$ (incorporating the feedback loop).  Since 
\be \bfK_{\ucy}^{(T)} = \calM_{\ue,\ucy} \bfK_{\ue}^{(T)} 
\calM_{\ue,\ucy}'  \ee
and $\bfK_{\ucy}^{(T)} >0$, $\calM_{\ue,\ucy}$ is lower triangular 
and uniquely defined (cf. \cite{kailath:book} for relevant 
discussions of innovation processes, QR factorization, and Cholesky 
factorization).  It is sufficient to show that $\calM_{\ue,\ucy}$ is banded and lower 
triangular with bandwidth $(m+1)$.  To this aim, we characterize $\calM_{\ue,\ucy}$ in state space.  Note that the 
state-space representation from $\uy^{T}$ to $\ue^{T}$ is $(\bfF - 
\uL_{2,t} \uH', \uL_{2,t}, -\uH', 1)$ and hence the one from $\ue^T$ 
to $\uy^T$ is $(\bfF, -\uL_{2,t}, -\uH', 1)$ (see Appendix \ref{appsub:represent}), and that the 
state-space representation from $\uy^{T}$ to $\ucy^{T}$ is $(\bfF 
_z, \uG_z, \uH', 1)$ (i.e., this is the state-space for the transfer 
function $\calZ_z(z)$).  Hence, we obtain the state-space from 
$\ue^T$ to $\ucy^T$ is
\be \left( 
        \left( \matrix{ \bfF & \bfzero \cr -\uG_z \uH' & \bfF_z }\right),
        \left( \matrix{ -\uL_{2,t} \cr \uG_z  }\right),
        ( -\uH', \uH'),
        1
         \right)
         ; \ee
where $\bfF_z := \bfF + \uG_z \uH'\in \bbR^{m \times m}$ is a nilpotent matrix, that 
is, $\bfF_z^m = \bfzero$ and only the $(1,m)$th entry of 
$\bfF^{m-1}_z$ is non-zero (equal to 1).  The above state-space 
realization is not minimal.  Simple computation shows that the above 
$2m$th order representation is equivalent to the $m$th order 
representation $(\bfF_z, \uG_{e,t}, \uH', 1 )$ where 
$\uG_{e,t}:=\uG_z + \uL_{2,t}$.  From the relation between the 
state-space representation and the impulse response, it holds that 
the $(i,j)$th entry in $\calM_{\ue,\ucy}$ is $\uH ' \bfF_z^{i-j-1} 
\uG_{e,j-1}$, if $i>j$.  Notice that $\calM_{\ue,\ucy} (i,i)=1$.  
Because $\bfF_z $ is nilpotent matrix, we have
$\calM_{\ue,\ucy} (i,j)=0$ if $(i-j-1) \geq m$, namely, the lower 
triangular part of $\calM_{\ue,\ucy}$ is banded with bandwidth 
$(m+1)$.  Thus, $\bfK_{\ucy}^{(T)}$ 
is banded with bandwidth $(m+1)$.

\subsection{Proof of Proposition \ref{prop:fdft}}  \label{appsub:fdft}

We follow the following steps to prove the lemma.  First, by 
considering the equivalent open loop, non-feedback communication 
problem, we show that in order for an $\ur^T$ to achieve $C_{T,n}$, 
it must hold that the effective channel input $(\bfI- \calG_T 
\bcalZ_T^{-1}) ^{-1}   \ur^T$ waterfills the effective channel noise 
$(\bfI- \calG_T \bcalZ_T^{-1}) ^{-1} \bcalZ_T \uN^T$.  Second, we 
show this would yield that, if the optimizing $\bfK_{\ur}^{(T)}$ has 
rank $k$, then the optimal channel output covariance matrix 
$\bfK_{\utildey}^{(T)}$ has its smallest (positive) eigenvalue, 
denoted $\lambda_0$, repeated exactly $k$ times.  Therefore, 
$(\bfK_{\utildey}^{(T)} - \lambda_0 \bfI)$ has rank $(T+1-k)$.  This 
in turn results in that 
\be \bfM:= \bfK_{\ucy}^{(T)} - \lambda_0 \bcalZ_{p,T} \bcalZ_{p,T}' 
\label{mat:rankcheck} \ee
has rank $(T+1-k)$.  However, since $\bfK_{\ucy}^{(T)}$ and 
$\bcalZ_{p,T} \bcalZ_{p,T}'$ are banded with bandwidth $(2m+1)$, it 
can be shown that $\bfM$ has rank at least $(T+1-m)$.  
So $k$ has to be no larger than $m$.  The details follow.

First, we prove that the effect channel input needs to waterfill the 
effective channel noise to achieve $C_{T,n}$.  Since
\be \utildey^T = \bcalZ_T \uy^T =  (\bfI- \calG_T \bcalZ_T^{-1}) 
^{-1}  ( \ur^T +\bcalZ_T \uN^T) \ee
and
\be \bfK_{\uu}^{(T)} = (\bfI- \calG_T \bcalZ_T^{-1}) ^{-1}   
\bfK_{\ur}^{(T)} (\bfI- \calG_T \bcalZ_T^{-1}) ^{-1} {}'  + (\bfI- 
\calG_T \bcalZ_T^{-1}) ^{-1} \calG_T \calG_T ' (\bfI- \calG_T 
\bcalZ_T^{-1}) ^{-1} {}' : = \bfP_r + \bfP_Z\ee
where $\bfP_r$ and $\bfP_Z$ are defined in obvious way, the 
optimization problem for $C_{T,n}(\calP)$ can be recast as
\be \ba{lll} \displaystyle  C_{T,n}(\calP)&=& \displaystyle \sup 
_{\bfA \in \bbR^{(n+1)\times (n+1)},\uC,\calG_T} 
    \frac{1}{T+1} \log \det (\bfI- \calG_T \bcalZ_T^{-1}) ^{-1}  
    ( \bfK_{\ur}^{(T)} +\bcalZ_T \bcalZ_T'  ) (\bfI- \calG_T \bcalZ_T^{-1}) ^{-1} {}'.\\
& &  ^{s.t. \: \trace (\bfP_r + \bfP_Z)/(T+1) \leq \mathcal{P} } \ea 
 \ee
If the optimizing $\calG_T *$ is plugged into the above optimization 
problem, noticing the resulting $P_Z$ is independent of the choice 
of $\bfK_{\ur}^{(T)}$, we end up with the following optimization 
problem
\be \ba{lll} \displaystyle  C_{T,n}(\calP)&:=& \displaystyle \sup 
_{\bfA \in \bbR^{(n+1)\times (n+1)},\uC,\calG_T, (A1)} 
    \frac{1}{2(T+1)} \log \det (\bfI- \calG_T \bcalZ_T^{-1}) ^{-1}  
    ( \bfK_{\ur}^{(T)} +\bcalZ_T \bcalZ_T'  ) (\bfI- \calG_T \bcalZ_T^{-1}) ^{-1} {}' .\\
& &  ^{s.t. \: \trace (\bfP_r)/(T+1) \leq \mathcal{P} } \ea 
 \ee
This may be viewed as a finite-horizon \emph{non-feedback} capacity 
problem, in which the effective input to the channel without feedback is 
$(\bfI- \calG_T \bcalZ_T^{-1}) ^{-1}   \ur^T$, and the effective 
channel noise is $(\bfI- \calG_T \bcalZ_T^{-1}) ^{-1} 
\bcalZ_T \uN^T$.  This idea has been used in 
\cite{klein:phd,liu:allerton,kim06}.  Thus, in order to give rise to 
the maximum mutual information between the effective input and the 
channel output $\utildey^T$, it is necessary to have the effective 
input to waterfill the effective noise.

The implication of the waterfilling argument is that, if the 
(effective) input covariance matrix has rank $k$, then the output 
covariance matrix must have its smallest eigenvalue repeated exactly 
$k$ times.  This is intuitively suggested by the name 
``waterfilling".  In other words, if the waterfilling level (cutoff 
value) is $\lambda_0$, then there are $k$ eigenvalues of 
$\bfK_{\utildey}^{(T)}$ that are equal to $\lambda_0$ and are the 
consequence of ``waterfilling" by $k$ positive eigenvalues of 
$(\bfI- \calG_T \bcalZ_T^{-1}) ^{-1} \bfK_{\ur}^{(T)}(\bfI- \calG_T 
\bcalZ_T^{-1}) ^{-1}{}'$ or equivalently $\bfK_{\ur}^{(T)}$, and the 
rest of the eigenvalues are strictly larger than $\lambda_0$ that 
remain unchanged after the waterfilling. Hence, if the optimizing 
$\bfK_{\ur}^{(T)}$ has rank $k$, then the resulting 
$(\bfK_{\utildey}^{(T)} - \lambda_0 \bfI)$ has rank $(T+1-k)$.  
Since $\bcalZ_{p,T}$ has full rank, it holds that
\be \bfM:=\bfK_{\ucy}^{(T)} - \lambda_0 \bcalZ_{p,T} \bcalZ_{p,T}' = 
\bcalZ_{p,T} (\bfK_{\utildey}^{(T)} - \lambda_0 \bfI) \bcalZ_{p,T}' 
\label{mat:rankcheck1} \ee
has rank $(T+1-k)$.  Since $\bfK_{\ucy}^{(T)}$ and 
$\bcalZ_{p,T} \bcalZ_{p,T}'$ are banded with bandwidth no larger than $(2m+1)$,  
we have that $\bfM$ is banded with bandwidth $(2m+1)$.  By $(f_0+g_0)=0$, it holds that $\bcalZ_{p,T}(j+m,j) = 0 $ for any $j$.  Therefore, $\bfM(j+m,j)=\bfK_{\ucy^{(T)}}(j+m,j)$.
However, from the proof of Proposition \ref{prop:maout}, we have
\be 
\bfK_{\ucy}^{(T)} = \bcalZ_{p,T} \E \uu^T \ucy^T {}' + \bcalZ_{z,T} \bcalZ_{z,T}' -  \bcalZ_{z,T} \hatG_T ' \bcalZ_{p,T}',\ee
which leads to that $\bfK_{\ucy^{(T)}}(j+m,j) = f_0 \neq 0$ for any $j$; notice that $\hatG_T$ is strictly lower triangular and $\bcalZ_{p,T}(j+m,j) = 0 $.  Then the banded structure of $\bfM$ implies that the rank of $\bfM$ is at least $(T+1-m)$.  This 
immediately follows that $k$ has to be no larger than $m$.
This proves Lemma \ref{lemma:Cfhft_dimm}.

Now we go back to Proposition \ref{prop:fdft}.  For any $m$th order channel $\calZ(z)$, consider the following perturbation that leads to an $(m+1)$st order channel:
\be \ba{lll} \calZ_\ep(z) &:=& \frac{\calZ_{z,\ep}(z)}{\calZ_p(z)} \\&:=&
\disp \frac{(1-\ep)( 1 + f_{m-1}  z^{-1} + 
\cdots + f_1  z^{-m+1} + f_0  z^{-m} ) - \ep^i z^{-m-1}  } 
{ 1 + (f_{m-1} + g_{m-1}) z^{-1} + 
\cdots + (f_1 + g_1) z^{-m+1} + (f_0 + g_0 ) z^{-m}}, \ea \label{calZep} \ee
where $\ep>0$ and $i$ is an integer to be determined.  In other words, $\calZ_{z}(z)$ is perturbed to be $\calZ_{z,\ep}(z)$.
Consequently, $\bcalZ_{p,\ep,T} = (1-\ep) \bcalZ_{p,T} -\ep^i \bfJ $, where $\bfJ$ is the down-shift matrix with $\bfJ(j+m+1,j)=1$ for any $j$ and all other entries equal to 0.

The number $i$ is chosen such that for sufficiently small $\ep>0$, we have $\bcalZ_{z,T}\bcalZ_{z,T}' > \bcalZ_{z,\ep} \bcalZ_{z,\ep}'$.  Such an $i$ always exists.  To see this, note that the difference of the two covariance matrices is
\be \ba{lll} && \ep (2 - \ep) \bcalZ_{z,T}\bcalZ_{z,T}' + \ep^i  (1-\ep) ( \bfJ \bcalZ_{z,T}' +  \bcalZ_{z,T}\bfJ' )  - \ep^{2i}  \bfJ  \bfJ '  \\
&>& \ep (2 - \ep) \bcalZ_{z,T}\bcalZ_{z,T}' + \ep^i  (1-\ep) ( \bfJ \bcalZ_{z,T}' +  \bcalZ_{z,T}\bfJ' - \bfJ  \bfJ ' ) .\ea \ee
As $\bcalZ_{z,T}\bcalZ_{z,T}'$ is positive definite and $( \bfJ \bcalZ_{z,T}' +  \bcalZ_{z,T}\bfJ'  - \bfJ  \bfJ ')$ is symmetric, the above difference admits simultaneous diagonalization, which transforms both terms to be diagonal.  That is, there exists $\bfS$ non-singular such that the congruence transformation using $\bfS$ leads to
\be \ep (2 - \ep) \bfD_1 + \ep^i  (1-\ep) \bfD_2  \ee
with both $\bfD_1$ and $\bfD_2$ are diagonal and independent of $\ep$.  Suppose $\rho:= \max_{j} |\bfD_2(j,j)/\bfD_1(j,j)|$.  Then for any $i$ such that
\be i > 1 + \frac{ \log \frac{2-\ep}{\rho (1-\ep)} }{ \log \ep } , \ee
the above difference is positive definite.  However, for sufficiently small $\ep$, the right-hand-side of the inequality approaches 1, so it is sufficient to choose $i:=2$, independent of $\ep$.     Hence $\bfK_{\bcalZ}^{(T)} > \bfK_{\bcalZ_{\ep}}^{(T)}$.  Similarly we can show 
\be \bfK_{\bcalZ_{\ep_1}}^{(T)} > \bfK_{\bcalZ_{\ep_2}}^{(T)} \label{ineq:eporder} \ee
if $\ep_1 < \ep_2$.

Next we show that as $\ep>0$ approaches zero, $P_T(\calR,\bcalZ_{\ep,T})$ admits a limit which is no larger than  $P_T(\calR,\bcalZ_{T})$.  Due to $\bfK_{\bcalZ}^{(T)} \geq \bfK_{\bcalZ_{\ep}}^{(T)}$, the feasible set $\phi_{\bcalZ,T, \calR}$ for $\bcalZ_{T}$ is strictly contained in $\phi_{\bcalZ_\ep,T, \calR}$.
Using the ordering in (\ref{ineq:eporder}) and the capacity inequality proven in \cite{kim04}, it is seen that $P_T(\calR,\bcalZ_{\ep,T})$ is no larger than $P_T(\calR,\bcalZ_{T})$ and is non-decreasing as $\ep$ approaches zero from above and hence the limit exists.

Consider an arbitrarily small slack $\delta>0$ of rate, i.e., consider $P_T(\calR - \delta,\bcalZ_{T})$.  For any $\delta>0$, there exist $\ep>0$ such that the feasible set $\phi_{\bcalZ_\ep,T, \calR}$ is contained in $\phi_{\bcalZ,T, \calR-\delta}$.  To see such an $\ep$ always exists, note that
\be  \det ( I + \bfGa_T ( \bfK_{\bcalZ_{\ep}}^{(T)} ) ^{-1} \bfGa_T' ) = \det ( I + \bfGa_T ( \bfK_{\bcalZ}^{(T)} ) ^{-1} \bfGa_T' + \Delta_\ep ),
\ee
where $\Delta_\ep \geq 0 $ and vanishes as $\ep$ tends to zero.  Thus if 
\be  \frac{1}{2} \log \det ( I + \bfGa_T ( \bfK_{\bcalZ_{\ep}}^{(T)} ) ^{-1} \bfGa_T' ) \geq \calR
\ee
then 
\be  \frac{1}{2} \log \det ( I + \bfGa_T ( \bfK_{\bcalZ}^{(T)} ) ^{-1} \bfGa_T' ) \geq \calR - \delta
\ee
for small enough $\ep$.

Interconnecting the optimizing $\ur_{\ep,T}$ and $\calG_{\ep,T}$ (obtained for $P_T(\calR,\bcalZ_{\ep,T})$) \footnote{To be more rigorous, an arbitrarily small slack to $P_T(\calR,\bcalZ_{\ep,T})$ may be needed since this optimization problem is an infimization problem as opposed to a minimization problem.  The idea in the proof can be easily adopted when the slack is used and the same result holds.} with $\bcalZ_T$, we see that the interconnection satisfies the rate constraint for $P_T(\calR - \delta,\bcalZ_{T})$.  The consumed power becomes 
\be
\trace [ (\bfI- \calG_{\ep,T} \bcalZ_{\ep,T}^{-1} ) 
^{-1} ( \bfK_{\ur,\ep}^{(T)} + \calG_{\ep,T} \bfK_{\bcalZ}^{(T)} \calG_{\ep,T}' ) (\bfI- \calG_{\ep,T} \bcalZ_{\ep,T}^{-1} {}' ) ], 
\ee
which is greater than 
\be P_T(\calR,\bcalZ_{\ep,T}) = \trace [ (\bfI- \calG_{\ep,T} \bcalZ_{\ep,T}^{-1}) 
^{-1} ( \bfK_{\ur,\ep}^{(T)} + \calG_{\ep,T} \bfK_{\bcalZ,\ep}^{(T)} \calG_{\ep,T}' ) (\bfI- \calG_{\ep,T} \bcalZ_{\ep,T}^{-1} {}' ) ] 
\ee
but the difference dependent on $( \bfK_{\bcalZ}^{(T)} - \bfK_{\bcalZ,\ep}^{(T)} )$ vanishes as $\ep$ goes to zero.  Consequently, as $\ep$ goes to zero, the consumed power of the sequence of interconnections converges to $P_T(\calR,\bcalZ_{\ep,T})$, no larger than $P_T(\calR,\bcalZ_{T})$.  However, each optimizing $\bfK_{\ur,\ep}^{(T)}$ has rank no larger than $(m+1)$ by Lemma \ref{lemma:Cfhft_dimm}. Therefore, $P_T(\calR,\bcalZ_{T})$ can be achieved by a sequence of $\bfK_{\ur,\ep}^{(T)}$ with rank no larger than $(m+1)$.  Thus we have proved the proposition.

\section{Proof of Proposition \ref{prop:dare}: Convergence to steady-state} \label{appsub:convg}

We show that system (\ref{dyn}) converges to a 
steady-state, as given by (\ref{dyn:steady}).  To this aim, we first 
transform the Riccati recursion into a new coordinate system, then 
show that it converges to a limit, and finally prove that the limit 
is the unique stabilizing solution of the Riccati equation. The 
convergence to the steady-state follows immediately from the 
convergence of the Riccati recursion.

Consider a coordinate transformation given as
\be \tbbA:= \pmb{\Psi} \bbA \pmb{\Psi}^{-1}:=\left[\matrix{\bfA & 0 
\cr 0 & \bfF}\right], \;\;\tbbC:=  \pmb{\Psi}^{-1}{}' \bbC = \left[ 
\matrix{\uC + \bfphi' \uH \cr \uH} \right], \;\; \tbbD:=  
\pmb{\Psi}^{-1}{}' \bbD = \left[ \matrix{\uC \cr \uzero} \right], 
\;\;\tSi_t:= \pmb{\Psi} \tSi_t \pmb{\Psi}', \label{trsf:conv} \ee
where
\be \pmb{\Psi}:=\left[\matrix{\bfI_{n+1} & \bfzero \cr -\pmb{\psi} & 
\bfI_m}\right]  \textrm{ (i.e. } \pmb{\Psi}^{-1} = 
\left[\matrix{\bfI_{n+1} & \bfzero \cr \pmb{\psi} & \bfI_m}\right] 
\textrm{ )}, \label{trsf:phi} \ee
and $\pmb{\psi}$ is the unique solution to the Sylvester equation
\be \bfF \pmb{\psi} - \pmb{\psi} \bfA = -\uG \: \uC'. \label{eq:syl} 
\ee
Note that the existence and uniqueness of $\pmb{\psi}$ is guaranteed 
by the assumption on $\bfA$ that $\lambda_i (-\bfA) + \lambda_j 
(\bfF) \neq 0$ for any $i$ and $j$ (see Section 
\ref{sub:generalstructure}).

Assume $k=n$ for the rest of the proof; i.e., $\bfA$ is anti-stable. 
For the case $k < n$, we can further transform $\tbbA$, $\tbbC$, and 
$\tSi$ into $\cbbA$, $\cbbC$, and $\cSi$ such that 
\be \cbbA = \diag [\bfA_+ , \bfA_- , \bfF], \label{trsf:stable_eig} 
\ee
where $\bfA_+ \in \bbR^{(k+1) \times (k+1)}$ is anti-stable and 
$\bfA_-$ is stable; then the following argument can be easily 
modified for the case $k<n$.

we can further decompose $\bfA$ into block-diagonal form and 
incorporate the stable block of $\bfA$ into $\bfF$, and the proof 
follows similarly; note that $\tbbA$ does not have any eigenvalues 
on the unit circle.  

The transformation defined in (\ref{trsf:conv}) transforms $\bbA$ 
into a block-diagonal form with the unstable and stable eigenvalues 
in different on-diagonal blocks, and transforms the initial 
condition $\Si_0$ to
\be \tSi_0:=\pmb{\Psi} \left[\matrix{\bfI_{n+1} & \bfzero \cr 
\bfzero & \bfzero}\right] \pmb{\Psi} ' = \left[\matrix{\bfI & 
-\pmb{\psi}' \cr -\pmb{\psi}  & \pmb{\psi} \pmb{\psi}' }\right]. \ee
Therefore, the convergence of (\ref{ric:recur}) with initial 
condition $\Si_0$ is equivalent to the convergence of
\be \tSi_{t+1} = \tbbA \tSi_t \tbbA' - \frac{\tbbA \tSi_t \tbbC \: 
\tbbC' \tSi_t \tbbA' } {\tbbC' \tSi_t \tbbC +1} \label{ric:recurNew} 
\ee
with initial condition $\tSi_0$.  By \cite{gallivan_riccati05}, 
$\tSi_t$ would converge if
\be \det \left( \left[\matrix{\bfzero & \bfzero \cr \bfzero & 
\bfI_m}\right] - \tSi_0 \left[\matrix{\bfI_{n+1} & \bfzero \cr 
\bfzero & \bfX_{22}} \right] \right) \not = 0, \ee
where $\bfX_{22}$ is the negative semi-definite matrix to the 
discrete-time Lyapunov equation
\be \bfX_{22} = \bfF \bfX_{22} \bfF' - (\uC+\pmb{\psi}' \uH ) (\uC+ 
\pmb{\psi}' \uH )'. \ee
Notice that $(\uC + \pmb{\psi}' \uH )$ is the upper $(n+1)\times 1$ 
block in $\tbbC$. Since
\be \ba{lll} \det \left( \left[\matrix{\bfzero & \bfzero \cr \bfzero 
& \bfI}\right] - \left[ \matrix{\bfI & -\pmb{\psi}' \cr -\pmb{\psi}  
& \pmb{\psi} \pmb{\psi}' } \right] \left[\matrix{\bfI & \bfzero \cr 
\bfzero & \bfX_{22}}\right] \right )
&=& \det \left( \left[\matrix{-\bfI & \pmb{\psi}'\bfX_{22}  \cr 
\pmb{\psi} & \bfI-\pmb{\psi} \pmb{\psi}' \bfX_{22}}\right]
\right)  \\
&=& \det (-\bfI)  \det \left( \bfI - \pmb{\psi} \pmb{\psi}'\bfX_{22} + \pmb{\psi} \pmb{\psi}' \bfX_{22} \right) \\
&\neq & 0, \ea \ee
we conclude that $\tSi_t$ converges to a limit $\tSi_\infty$.

This limit $\tSi_\infty$ is a positive semi-definite solution to
\be \tSi_\infty = \tbbA \tSi_\infty \tbbA' - \frac{\tbbA \tSi_\infty 
\tbbC \tbbC' \tSi_\infty \tbbA' } {\tbbC' \tSi_\infty \tbbC  +1} . 
\label{ric:eqNew} \ee
By \cite{kailath:book}, (\ref{ric:eqNew}) has a unique stabilizing 
solution because $(\tbbA,\tbbC')$ is observable (noting that $(\bbA, 
\bbC')$ is observable) and $\tbbA$ does not have any eigenvalues on 
the unit circle.  Therefore, $\tSi_\infty$ is this unique 
stabilizing solution, which can be computed from (\ref{ric:eqNew}) 
as (see also \cite{gallivan_riccati05})
\be \tSi_\infty = \left[ \matrix{\tSi_{11} & \bfzero \cr \bfzero & 
\bfzero} \right] \ee
where $\tSi_{11}$ is the positive-definite solution to a 
reduced-order Riccati equation
\be \tSi_{11} = \bfA \tSi_{11} \bfA' - \frac{\bfA \tSi_{11} (\uC' + 
\uH' \pmb{\psi})' (\uC' +\uH' \pmb{\psi}) \tSi_{11} \bfA' } {(\uC' + 
\uH' \pmb{\psi}) \tSi_{11} (\uC' + \uH' \pmb{\psi})' +1} . 
\label{ric:reduced} \ee
and has rank equal to the number of anti-stable eigenvalues of 
$\tbbA$ (cf. \cite{gallivan_riccati05}).  Thus, $\Si_t$ converges to
\be \Si_\infty = \left[ \matrix{\tSi_{11} & \tSi_{11}\pmb{\psi}' \cr 
\pmb{\psi}\tSi_{11} & \pmb{\psi}\tSi_{11}\pmb{\psi}'} \right] \ee
with rank equal to the number of anti-stable eigenvalues of $\tbbA$.

ii) Immediately from i).

\section{Proof of Corollary \ref{coro:stable_eig}} \label{appsub:stable_eig}

Consider the coordinate transformations used in the proof of 
Proposition \ref{prop:dare} that transform $\bbA$, $\bbC$, and $\Si$ 
into $\tSi$ into $\cbbA$, $\cbbC$, and $\cSi$.  Note that the block 
in $\cSi_\infty$ (i.e. the solution to the Riccati equation defined 
by $\cbbA$ and $\cbbC$) associated with the $\bfA_-$ block is zero.  
By Proposition \ref{prop:steady}, in the new coordinates the rate 
and power due to the $\bfA_-$ block are both zero, and hence in the 
original coordinates the rate and power due to the stable 
eigenvalues of $\bfA$ are both zero.  Then we remove the dimensions 
corresponding to $\bfA_-$ in $\cbbA$, $\cbbC$, $\cSi$, and the 
coordinate transformation matrix.  It is easy to check that this 
leads to a pair of reduced order $(\bfA_k,\uC_k)$ with $\bfA_k$ 
anti-stable and satisfying $R_{\infty,n}(\bfA,\uC) = 
R_{\infty,k}(\bfA_k,\uC_k)$ and $P_{\infty,n}(\bfA,\uC) = 
P_{\infty,k}(\bfA_k,\uC_k)$.

\vspace{.25in} \hspace*{2.5in} \textbf{ACKNOWLEDGEMENTS} 
\vspace{.15in}

The authors would like to thank Anant Sahai, Sekhar Tatikonda, 
Sanjoy Mitter, Zhengdao Wang, Murti Salapaka, Shaohua Yang, 
Donatello Materassi, and Young-Han Kim for useful discussion.

\vspace{-15pt} \hfill \markright{\textsf{References}} \small
\bibliographystyle{unsrt}
\bibliography{bib070722}

\end{document}